\documentclass[pra,aps,amsmath,amssymb,amsfonts,twocolumn,nofootinbib,longbibliography,superscriptaddress]{revtex4-2}
\usepackage{amssymb}
\usepackage{bm,mathrsfs}
\usepackage{graphicx}
\usepackage{subfigure}
\usepackage{epsfig}
\usepackage{amsmath,bbm}
\usepackage{amsfonts,amssymb}
\usepackage{times}
\usepackage{verbatim}
\usepackage{diagbox}
\usepackage[sort&compress]{natbib}
\usepackage{amsmath}
\usepackage[colorlinks=true,citecolor=blue,linkcolor=blue,urlcolor=blue]{hyperref}
\usepackage[usenames]{color}

\newcommand{\bra}[1]{\langle #1|}
\newcommand{\ket}[1]{|#1\rangle}
\newcommand{\inner}[2]{\langle #1|#2\rangle}

\usepackage{float}

\begin{document}

\title{Entangling two Dicke states in a periodic modulated quantum system}
\author{Wuji Zhang}
\affiliation{Center for Quantum Sciences and School of Physics, Northeast Normal University, Changchun 130024, China}

\author{Ruifang Wu}
\affiliation{Center for Quantum Sciences and School of Physics, Northeast Normal University, Changchun 130024, China}

\author{Chunfang Sun}
\affiliation{Center for Quantum Sciences and School of Physics, Northeast Normal University, Changchun 130024, China}

\author{Chunfeng Wu}
\affiliation{Science, Mathematics and Technology, Singapore University of Technology and Design, 8 Somapah Road, Singapore 487372, Singapore}

\author{Gangcheng Wang}
\email{wanggc887@nenu.edu.cn}
\affiliation{Center for Quantum Sciences and School of Physics, Northeast Normal University, Changchun 130024, China}

\date{\today}
\begin{abstract}
We propose a theoretical approach for entangling two Dicke states in a periodic modulated quantum system. By considering two qubit ensembles that are nonuniformly coupled to a common resonator, we can derive an effective Hamiltonian whose energy levels depend nonlinearly on the excitation number of each qubit ensemble. More simplified effective Hamiltonian can be obtained by selecting appropriate driving parameters and initial state. Based on the dynamic evolution of the effective Hamiltonian, we can selectively achieve Dicke state transitions and generate entangled Dicke states controllably. For a special case, we can obtain ensemble-ensemble entangled states by performing a projective even-odd cat measurement. By implementing Gaussian soft temporal modulation, we can effectively suppress off-resonant contributions in the interaction and enhance the fidelity of target states. Furthermore, by utilizing the Holstein-Primakoff transformation, we study the resonator-ensemble coupling system in the thermodynamic limit and investigate the generation of entangled magnon states. Additionally, we propose a scheme of creating magnon NOON states through frequency modulation and study the influence of decoherence on the fidelity of target states.

\end{abstract}

\maketitle
\section{Introduction}
\label{SecI}
Quantum entanglement, as a valuable resource for quantum information processing \cite{2017Silveira,2019Zhu,2020Qiao,2022Contreras,2022Morvan,2022Eddins}, has garnered significant attention due to its pivotal role in testing the fundamentals of quantum mechanics \cite{2016Alsina,2017Su,2021Alsina,2022Orito}. Recently,  it has demonstrated a wide range of potential applications across various systems, encompassing atomic systems \cite{2019Gonz,2017Wong}, photons \cite{2022Seidelmann,2022Chaisson,2016Wangxilin}, superconducting systems \cite{2021Tzemos,2022Kounalakis,Wang2018,2016Houzhibo,2011Monz,2017Song}, trapped ions \cite{2022Kranzl,2022Guo}, and waveguides \cite{2022Song}. However, for large numbers of atoms, these schemes that relying on global control of the spin ensemble typically require intricate measurement-based feedback, high-fidelity control, and extensive preparation times, resulting in costly entangling gate operations between them \cite{PhysRevA.99.023822,PhysRevA.70.022106,PhysRevLett.125.190403}. Consequently, strategies for optimizing such schemes remain a significant challenge that requires attention \cite{PhysRevA.102.042610}.

Dicke states \cite{2019Liu,PhysRevA.85.022322,RevModPhys.90.035005}, originally introduced in the context of superradiance phenomenon \cite{PhysRev.93.99,PhysRevA.107.063713}, represent a special type of multiparticle entangled state. Due to their favorable properties, they hold promise for providing a reliable approach to address this issue in quantum sensing \cite{2007Xiao,2009Prevedel,2010Shao,2009Prevedel,PhysRevLett.67.661}. Compared with other states typically used in quantum sensing, Dicke states have been demonstrated to possess greater robustness to diverse sources of noise, such as qubit dephasing, qubit damping, and fluctuations in spin number \cite{PhysRevApplied.15.044026,PhysRevA.99.023808,https://doi.org/10.1002/qute.202200173}. It is noteworthy that utilizing entangled states has the potential to enhance measurement sensitivity beyond the standard quantum limit, in principle \cite{PRXQuantum.3.020308}. Considering the challenges involved in achieving precise control over individual spin qubits, global control over a collective of spins facilitates the realization of optimal sensitivity through entangled Dicke states \cite{Zhang_2014,2015Apellaniz}. On the other hand, periodic driving has significant advantages in enhancing controllability of systems. This technique has been widely utilized in various areas of quantum physics, such as the investigation of exotic quantum phenomena in artificial materials \cite{Wang:20}, analysis of stochastic processes \cite{JUNG1993175}, floquet engineering \cite{PhysRevA.95.023615} and so on. Through periodic driving, one can effectively promote selective resonant transitions while suppressing undesired transitions within the targeted Hilbert subspace. As a result, a broad spectrum of controllable and selective interactions can be achieved, enabling enhanced controllability and selectivity \cite{2022Yu}.

The concept of selective interaction plays an important role in the preparation of controllable quantum entangled states \cite{2020CongL,2020Shao,2020Neuman,Mu:20}. This approach exploits the non-linearity of energy levels to generate desired entangled states. In Refs. \cite{2017Chunfeng,2022Yu}, the authors investigated the generation of Dicke states by means of selective interactions. By employing selective preparation of Dicke states, the initialization of a permutation-invariant quantum error-correction code was explored in Ref. \cite{2019Chunfeng}. This method facilitated the rapid generation of highly symmetric codes with exceptional fidelity. To realize the permutation-invariant quantum error-correction code, each logical qubit is encoded on several physical qubits. However, previous studies have paid little attention to methods for entangling states of two or several quantum systems in a controlled manner. A natural question arises regarding how to entangle logical qubits with selective interaction. This motivates us to delve into entanglement of multiple systems.

In this paper, we present a theoretical proposal for generating entangled Dicke states in the ultrastrong coupling regime. A series of unitary transformations have been applied to the Hamiltonian to derive the effective interaction. Additionally, we investigate how to utilize selective resonant interactions to obtain entangled Dicke states in a finite system \cite{PR2004Chu,PRA2009Son,PRL2013Luo,RMP1990H,NJP2015Eckardt}. Furthermore, by utilizing the soft temporal quantum control, we can effectively mitigate unwanted off-resonant contributions. This approach allows for an accurate description of high-fidelity quantum states even in scenarios involving strong perturbations and long evolution times \cite{PhysRevLett.121.050402}. The collective operators can be mapped to Bosonic operators by means of Holstein-Primakoff transformation \cite{1940Holstein}, thereby opening up exciting possibility for employing the same approach to generate magnon NOON states. 

This paper is structured as follows. In Sec. \ref{SecII}, we derive the effective Hamiltonian by choosing appropriate unitary transformations. Sec. \ref{SecIII} presents a detailed scheme for generating entangled Dicke states with high-fidelity quantum state using Gaussian soft control techniques. In Sec. \ref{SecIV}, the Holstein-Primakoff transformation is utilized to establish selective interactions between two distinct magnons, thereby providing the potential for generating magnon NOON states in the thermodynamic limit. Furthermore, in Sec. \ref{SecV}, we analyze the influence of decoherence on the dynamical evolution and evaluate the experimental feasibility of our proposed scheme. The results reveal that our approach can still achieve entangled target states with high fidelity.
Finally, we draw our conclusions in Sec. \ref{SecVI}. 
\section{The effective Hamiltonian for finite qubits}
\label{SecII}
We consider two qubit ensembles, each of which consists of two-level systems, are situated with a common quantum resonator. The $j$-th ensemble comprising $N_{j}$ qubits is coupled to the corresponding periodically driving field. The Hamiltonian is given by ($\hbar = 1$ throughout this work)
\begin{equation}\label{eq_01}
\hat{H}= \hat{H}_{0} + \hat{H}_{\rm int}+ \hat{H}_{d}(t),
\end{equation}
where
\begin{equation}\label{eq_02}
\begin{split}
    \hat{H}_{0}=&\ \omega_{c}\hat{a}^{\dag}\hat{a}-\sum_{j=1}^{2}\varepsilon_{j}{\hat{S}^{z}_{j}},\\
    \hat{H}_{\rm int} =& \sum_{j=1}^{2}g_{j}(\hat{a}+\hat{a}^{\dag})\hat{S}^{x}_{j},\\
    \hat{H}_{d}(t) =& \sum_{j=1}^{2}\left[\omega_{q_{j}}+A_{j}\cos{({\omega}t)}\right]\hat{S}^{x}_{j},
    \end{split}
\end{equation}
The notation $\hat{a}^{\dag}$ ($\hat{a}$) denotes the creation (annihilation) operator of the cavity field with frequency $\omega_{c}$, the notation $\varepsilon_{j}$ is the atomic frequency in the $j$-th ensemble, $g_{j}$ denotes the coupling strength between resonator and qubits in the $j$-th ensemble. The $j$-th ensemble is subjected to a periodic driving field with amplitude $A_{j}$ and frequency $\omega$. The notation $\omega_{q_{j}}$ represents the static energy split of the biased term for the $j$-th driving field. $\hat{S}^{\alpha}_{j}=\sum_{r_{j}=1}^{N_{j}}\hat{s}_{r_{j}}^{\alpha}$ ($\alpha=x,y,z$) are the collective operators with $\hat{s}_{r_{j}}^{\alpha}$ being the atomic operators acting on the $r_{j}$-th site of $j$-th ensemble. We can verify each ensemble operators satisfy the $\mathfrak{su}(2)$ algebra relations: $[\hat{S}_{i}^{\alpha}, \hat{S}_{j}^{\beta}]= i\delta_{ij}\epsilon_{\alpha\beta\gamma}\hat{S}^{\gamma}_{j}$ ($\alpha, \beta, \gamma \in\{x, y, z\}$) with $\delta_{ij}$ and $\epsilon_{\alpha\beta\gamma}$ being the Kronecker delta and Levi-Civita symbols. To achieve a better understanding of the system, we apply the following unitary transformation:
\begin{equation*}
\hat{U}=\exp{\left[\sum_{j=1}^{2}\frac{g_j}{\omega_c}(\hat{a}-{\hat{a}^\dag)}\hat{S}^{x}_{j}\right]}\times \exp{\left[i\frac{\pi}{\sqrt{2}}\sum_{j=1}^{2}(\hat{S}^{x}_{j}+\hat{S}^{z}_{j})\right]}.
\end{equation*}
The transformed Hamiltonian can be expressed as 
\begin{equation}\label{eq_03}
\hat{H}^{\prime}(t) = \hat{U}^{\dag}\hat{H}\hat{U}=\hat{H}_{0}^{\prime}(t)+\hat{H}_{\rm int}^{\prime},
\end{equation}
where
\begin{eqnarray*}
\hat{H}_{0}^{\prime}(t)&=&\omega_{c}\hat{a}^{\dag}\hat{a}+\hat{F}(\hat{S}^{z}_{1},\hat{S}^{z}_{2})
+\sum_{j=1}^{2} A_{j}\cos{({\omega}t)}\hat{S}^{z}_{j},\\
\hat{H}_{\rm int}^{\prime}&=&-\sum_{j=1}^{2}\frac{\varepsilon_{j}}{2}\left[\hat{D}(\beta_{j})\hat{S}^{+}_{j}+\hat{D}^{\dag}(\beta_{j})\hat{S}^{-}_{j}\right],
\end{eqnarray*}
where $\hat{D} (\beta_{j}) = \exp \left( \beta_{j} \hat{a}^{\dagger} - \beta_{j}^{*}\hat{a} \right)$ is the displacement operator with $\beta_{j} = g_{j}/\omega_{c}$, and 
\begin{equation*}
\hat{F}(\hat{S}^{z}_{1},\hat{S}^{z}_{2})=\sum_{j=1}^{2}\omega_{q_{j}}\hat{S}^{z}_{j}-\omega_{c}\sum_{j=1}^{2}\left(\beta_{j}\hat{S}^{z}_{j}\right)^{2}
-2\frac{g_{1}g_{2}}{\omega_{c}}\hat{S}^{z}_{1}\hat{S}^{z}_{2}.
\end{equation*}
Moving to the rotating frame defined by
\begin{equation}\label{eq_04}
  \begin{split}
     \hat{U}_{0}(t) &=\exp\left[-i\int_{0}^{t}d\tau\hat{H}_{0}^{\prime}(\tau)\right],
  \end{split}
\end{equation}
we further obtain the following transformed Hamiltonian \cite{PhysRevA.47.5138}
\begin{small}
\begin{equation}\label{Hint}
      \hat{H}_{\rm I}^{\prime}(t) =-\sum_{j=1}^{2}\frac{\varepsilon_{j}}{2}\left[\hat{D}(\beta_{j}e^{i\omega_{c}t})\hat{S}^{+}_{j} e^{i\eta_{j}\sin{({\omega}t)}}e^{i\hat{f} _{j}(\hat{S}^{z}_{1},\hat{S}^{z}_{2})t}+\rm{H.c.}\right],
\end{equation}
\end{small}
where 
\begin{equation*}
\begin{split}
  \hat{f}_{1}(\hat{S}^{z}_{1},\hat{S}^{z}_{2})&=\hat{F}(\hat{S}^{z}_{1}+1,\hat{S}^{z}_{2})-\hat{F}(\hat{S}^{z}_{1},\hat{S}^{z}_{2})\\
  &=\omega_{q_{1}}-\frac{2g_{1}g_{2}}{\omega_{c}}\hat{S}^{z}_{2}
  -\frac{g_{1}^{2}}{\omega_{c}}(2\hat{S}^{z}_{1}+1),\\
  \hat{f}_{2}(\hat{S}^{z}_{1},\hat{S}^{z}_{2})&=\hat{F}(\hat{S}^{z}_{1},\hat{S}^{z}_{2}+1)-\hat{F}(\hat{S}^{z}_{1},\hat{S}^{z}_{2})\\
  &=\omega_{q_{2}}-\frac{2g_{1}g_{2}}{\omega_{c}}\hat{S}^{z}_{1}
  -\frac{g_{2}^{2}}{\omega_{c}}(2\hat{S}^{z}_{2}+1).
\end{split}
\end{equation*}
Utilizing the Jacobi-Anger identity \cite{PhysRevA.96.043849}
\begin{equation}\label{J}
  e^{i\eta_{j}\sin{({\omega}t)}}=\sum_{q=-\infty}^{\infty}\mathcal{J}_{q}(\eta_{j})e^{iq{\omega}t},
\end{equation}
with $\mathcal{J}_{q}(\eta_{j})$ being the $q$-th order Bessel function of the first kind with $\eta_{j}=A_{j}/{\omega}$, the Hamiltonian in Eq.~(\ref{Hint}) can be rewritten as
\begin{small}
\begin{equation}\label{H_int(q)}
     \hat{H}^{\prime}_{\rm I} (t) = -\sum_{q=-\infty}^{\infty}\sum_{j=1}^{2}\left[\frac{\varepsilon_{j}}{2}\mathcal{J}_{q}(\eta_{j})\hat{D}[\beta_{j}(t)]\hat{S}^{+}_{j}
e^{i\left[q\omega+\hat{f}_{j}(\hat{S}^{z}_{1},\hat{S}^{z}_{2})\right]t}
   +\rm{H.c.}\right ],
\end{equation}
\end{small}
where $\beta_{j}(t)=\beta_{j}e^{i\omega_{c}t}$. Since the Hamiltonian is permutation symmetric for each qubit ensemble, we introduce the following normalized ${N_{j}}$-qubit Dicke states with $k_{j}$ atomic excitations for convenience \cite{PhysRevA.84.042324,PhysRevA.80.052302}
\begin{equation*}
    \ket{W_{N_{j}}^{k_{j}}}=(C_{k_{j}}^{N_{j}})^{-1/2}\sum_{m}P_{m}
\ket{e_{1},e_{2},\cdots,e_{k_{j}},g_{k_{j}+1}\cdots,g_{N_{j}}},
\end{equation*}
where $\sum_{m}P_{m}(\bullet)$ indicates the sum over all particle permutations for $j$-th qubit ensembles and $C_{k_{j}}^{N_{j}}={N_{j}!}/{k_{j}!(N_{j}-k_{j})!}$ is the binomial coefficient. In the Dicke states basis, the collective operators can be reduced to the following $(N_{j}+1)$-dimensional permutation symmetric subspace 
\begin{subequations}\label{S}
  \begin{align}
        \hat{S}^{z}_{j}&=\sum_{k_{j}=0}^{N_{j}}(k_{j}-\frac{N_{j}}{2})\ket{W_{N_{j}}^{k_{j}}}\bra{W_{N_{j}}^{k_{j}}},\\
        \hat{S}^{+}_{j}&=\sum_{k_{j}=0}^{N_{j}}h_{j}(N_{j},k_{j})\ket{W_{N_{j}}^{k_{j}+1}}\bra{W_{N_{j}}^{k_{j}}},\\
        \hat{S}^{-}_{j}&=\sum_{k_{j}=0}^{N_{j}}h_{j}(N_{j},k_{j})\ket{W_{N_{j}}^{k_{j}}}\bra{W_{N_{j}}^{k_{j}+1}},
  \end{align}
\end{subequations}
where $h_{j}(N_{j},k_{j})=\sqrt{(k_{j}+1)(N_{j}-k_{j})}$. In terms of the Fock state basis \cite{2022Yu}, the displacement operator can be rewritten as 
\begin{equation}\label{Dmn}
     \hat{D}\left[\beta_{j}(t)\right]=\sum_{m,n=0}^{\infty} \bra{m}\hat{D}\left[\beta_{j}(t)\right]\ket{n}\hat{A}_{mn},
\end{equation}
where $\hat{A}_{m, n} = \ket{m} \bra{n}$ and matrix elements ${D}_{mn}\left[\beta_{j}(t)\right]=\bra{m}\hat{D}\left[\beta_{j}(t)\right]\ket{n}$ can be written as 
{\small
\begin{equation*}
D_{mn}\left[\beta_{j}(t)\right]=\left\{\begin{array}{ll}
e^{-\frac{1}{2}\beta^{2}_{j}}  L_{n}^{(0)}\left(\beta^{2}_{j}\right), & m = n; \\
e^{-\frac{1}{2}\beta^{2}_{j}} \beta^{s}_{j} e^{is\omega_c t}\sqrt{\frac{n !}{m !}} L_{n}^{(s)}\left(\beta^{2}_{j}\right), & m > n;\\
e^{-\frac{1}{2}\beta^{2}_{j}} (-\beta_{j})^{s}e^{-is\omega_c t} \sqrt{\frac{m !}{n !}} L_{m}^{(s)}\left(\beta^{2}_{j}\right), & m < n.
\end{array}\right.
\end{equation*}}
Here $L^{(s)}_{n}(\beta^{2}_{j})$ is an associated Laguerre polynomial with $s=|m-n|$. The cases $s = 0$ and $s > 0$ correspond to carrier transition and the $s$-th sidebands transition, respectively. In terms of Dicke states and Fock states, the Hamiltonian in Eq.~(\ref{H_int(q)}) can be reduced to the following Dicke state subspace
\begin{equation}
\label{H_dicke}
\hat{H}^{\prime}_{\rm I}(t)=\hat{H}_{\rm car}(t)+\hat{H}_{\rm red}(t)+\hat{H}_{\rm blue}(t),
\end{equation}
where
\begin{widetext}
\begin{equation}\label{H_I}
\begin{split}
\hat{H}_{\rm car}(t)=\sum_{n=0}^{\infty}\sum_{k_{1},k_{2}}&\left[G_{1}^{(0)}(k_{1},\eta_{1})e^{i\delta^{(0)}_{1}t}\hat{W}_{N_{1}}^{k_{1}+1,k_{1}}\otimes
\hat{W}_{N_{2}}^{k_{2},k_{2}}\otimes\hat{A}_{n,n}\right.\\
&+\left.G^{(0)}_{2}(k_{2}, \eta_{2})e^{i\delta^{(0)}_{2}t}\hat{W}_{N_{1}}^{k_{1},k_{1}}\otimes
  \hat{W}_{N_{2}}^{k_{2}+1,k_{2}}\otimes\hat{A}_{n,n}\right]+\rm{H.c.},\\
 \hat{H}_{\rm red}(t)= \sum_{n=0}^{\infty}\sum_{s=1}^{\infty}\sum_{k_{1},k_{2}}&\left[G^{(s)}_{1}(k_{1},\eta_{1})e^{i\delta^{(s)}_{1}t}\hat{W}_{N_{1}}^{k_{1}+1,k_{1}}\otimes
  \hat{W}_{N_{2}}^{k_{2},k_{2}}\otimes\hat{A}_{n+s,n}\right.\\
  &+\left.G^{(s)}_{2}(k_{2},\eta_{2})e^{i\delta^{(s)}_{2}t}\hat{W}_{N_{1}}^{k_{1},k_{1}}\otimes\hat{W}_{N_{2}}^{k_{2}+1,k_{2}}\otimes\hat{A}_{n+s,n}\right]+\rm{H.c.},\\
  \hat{H}_{\rm blue}(t)=\sum_{n=0}^{\infty}\sum_{s=1}^{\infty}\sum_{k_{1},k_{2}}&\left[(-1)^{s}G^{(s)}_{1}(k_{1},\eta_{1})e^{i\delta^{(-s)}_{1}t}\hat{W}_{N_{1}}^{k_{1}+1,k_{1}}\otimes
  \hat{W}_{N_{2}}^{k_{2},k_{2}}\otimes\hat{A}_{n,n+s}\right.\\&
  +\left. (-1)^{s}G^{(s)}_{2}(k_{2},\eta_{2})e^{i\delta^{(-s)}_{2}t}\hat{W}_{N_{1}}^{k_{1},k_{1}}\otimes\hat{W}_{N_{2}}^{k_{2},k_{2}+1}\otimes\hat{A}_{n,n+s}\right]+\rm{H.c.}.
\end{split} 
\end{equation}
\end{widetext}
Here $ \hat{W}_{N_{j}}^{k_{j}, k_{j}^{\prime}} = \ket{W_{N_{j}}^{k_{j}}}\bra{W_{N_{j}}^{k_{j}^{\prime}}}$ and
$G_{j}^{(s)}(k_{j}, \eta _{j})=-\frac{1}{2}{\varepsilon_{j}}\mathcal{J}_{q}(\eta_{j})h_{j}(N_{j}, k_{j})D_{mn}(\beta_{j})$
with $\delta^{(\pm s)}_{j}=q\omega+\Delta_{j}\pm s\omega_{c}$
and 
\begin{eqnarray*}
\Delta_{1}(k_{1},k_{2})&=&\omega_{q_{1}}+\frac{g_{1}^2}{\omega_{c}}(N_{1}-2k_{1}-1)+\frac{g_{1}g_{2}}{\omega_{c}}(N_{2}-2k_{2}),\\
\Delta_{2}(k_{1},k_{2})&=&\omega_{q_{2}}+\frac{g_{2}^2}{\omega_{c}}(N_{2}-2k_{2}-1)+\frac{g_{1}g_{2}}{\omega_{c}}(N_{1}-2k_{1}).
\end{eqnarray*}
\begin{figure}
\centering
\includegraphics[width=0.48\textwidth]{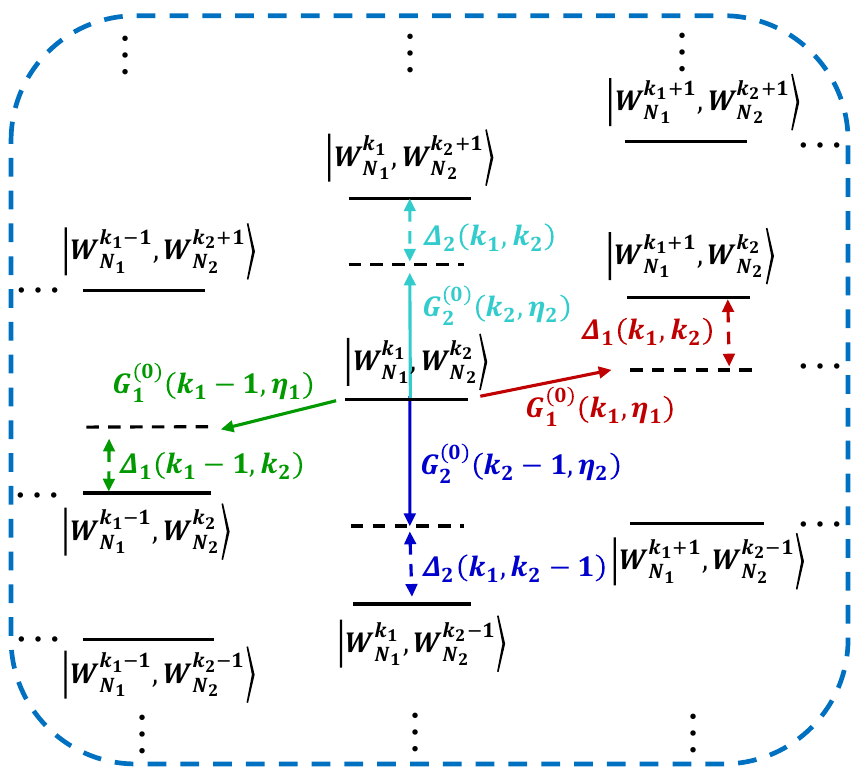}
\caption{Schematic diagram of transition with taking quantum state $\ket{W_{N_{1}}^{k_{1}},W_{N_{2}}^{k_{2}}}$ as an example. Different driving Rabi frequencies $G^{(0)}_{1}(k_{1},\eta_{1})$ (solid magenta line), $G^{(0)}_{1}(k_{1}-1,\eta_{1})$ (solid green line), $G^{(0)}_{2}(k_{2},\eta_{2})$ (solid blue line) and $G^{(0)}_{2}(k_{2}-1,\eta_{2})$ (solid cyan line) that close to resonant transition result in the selective transitions.}\label{fig_01}
\end{figure}
Obviously, we can tune the driving parameters to obtain the desired effective Hamiltonians. If the driving frequency
$\omega=-\Delta_{1}(k_{1}, k_{2})/{q_{0}}$, $-\Delta_{1}(k_{1}-1,k_{2})/{q_{0}}$, $-\Delta_{2} (k_{1},k_{2})/{q_{0}}$, or $-\Delta_{2}(k_{1},k_{2}-1)/{q_{0}}$, the on-resonance terms only appear in the carrier terms [i.e., $\hat{H}_{\rm car}(t)$ in Eq.~(\ref{H_dicke})]. Neglecting the off-resonance terms in the expansion of the Hamiltonian $\hat{H}^{\prime}_{\rm I}(t)$, we can obtain effective Hamiltonian based on the secular approximation \cite{PhysRevA.49.89}. Then we can investigate the desired Dicke state transitions based on the selective interactions. Furthermore, assuming the initial state of the system is $\ket{W_{N_{1}}^{k_{1}}}\otimes\ket{W_{N_{2}}^{k_{2}}}\otimes\ket{0}$ under the new frame. We can verify that such initial state corresponds to the following state in the original frame 
\begin{equation}\label{quantum state}
\ket{W_{N_{1}}^{k_{1}}}_{x}\otimes\ket{W_{N_{2}}^{k_{2}}}_{x}\otimes\ket{\xi_{k_{1},k_{2}}}=\hat{U}\ket{W_{N_{1}}^{k_{1}}}\otimes\ket{W_{N_{2}}^{k_{2}}}\otimes\ket{0},
\end{equation}
where $\ket{\xi_{k_{1},k_{2}}}=e^{-|\xi_{k_{1},k_{2}}|^{2}/2}\sum_{n=0}^{\infty}(\xi_{k_{1},k_{2}}^{n}/\sqrt{n!})\ket{n}$ with $\xi_{k_{1},k_{2}}= (N_1/2-k_1)\beta_{1}+(N_2/2-k_2)\beta_{2}$ and $\hat{U}$ is given by Eq. (\ref{eq_03}) . In fact, $\ket{W_{N_{j}}^{k_{j}}}_{x}$ is also a Dicke state, such that $\hat{\textbf{S}}^{2}_{j}\ket{W_{N_{j}}^{k_{j}}}_{x} = N_{j}/2(N_{j}/2+1)\ket{W_{N_{j}}^{k_{j}}}_{x}$ and $\hat{S}_{j}^{x}\ket{W_{N_{j}}^{k_{j}}}_{x}=(k_{j}-N_{j}/2)\ket{W_{N_{j}}^{k_{j}}}_{x}$. It is worth mentioning that if we consider the conditions $\epsilon_{j} \rightarrow 0$, $\omega_{q_{j}}>0$ and without driving, the ground state in the frame associated with unitary transformation $\hat{U}$ is $\ket{W_{N_{1}}^{0}}\otimes\ket{W_{N_{2}}^{0}}\otimes\ket{0}$, which corresponds to the state $\ket{W_{N_{1}}^{0}}_{x}\otimes\ket{W_{N_{2}}^{0}}_{x}\otimes\ket{\xi_{0,0}}$ with $\xi_{0,0} = (N_{1}\beta_{1}+N_{2}\beta_{2})/2$ in the original frame. By tuning the driving frequency to satisfy on-resonant condition and choosing proper initial state, we can obtain a subspace spanned by involved states. For instance, if we tune the driving frequency $\omega=-\Delta_{1}(k_{1}, k_{2})/{q_{0}}$ and choose $\ket{W_{N_{1}}^{k_{1}}}\otimes\ket{W_{N_{2}}^{k_{2}}}\otimes\ket{0}$ as initial state, we can obtain the subspace spanned by $\left\{\ket{W_{N_{1}}^{k_{1}}}\otimes\ket{W_{N_{2}}^{k_{2}}}\otimes\ket{0},\ket{W_{N_{1}}^{k_{1}+1}}\otimes\ket{W_{N_{2}}^{k_{2}}}\otimes\ket{0}\right\}$. For our choice, the carrier transition is activated and the field part of state remains on the ``vacuum" state. Considering such special case, we will omit it in the following discussions and hereafter study the generation of Dicke states in the new frame for simplicity, unless specified otherwise. The transition diagram for two Dicke states is displayed in Fig. \ref{fig_01}. In this figure, we choose quantum state $\ket{W_{N_{1}}^{k_{1}},W_{N_{2}}^{k_{2}}}$ ($\ket{W_{N_{1}}^{k_{1}},W_{N_{2}}^{k_{2}}}\equiv\ket{W_{N_{1}}^{k_{1}}}\otimes\ket{W_{N_{2}}^{k_{2}}}$) be the initial state as an example. Driving a carrier transition close to resonance results in different effective interactions. Based on these properties, we can realize the so-called selective interactions, which allows to entangle two Dicke states by tuning the desired driving parameters.

The terms of $\hat{W}_{N_{1}}^{k^{\prime}_{1}+1,k^{\prime}_{1}}\otimes\hat{W}_{N_{2}}^{k^{\prime}_{2},k^{\prime}_{2}}$ and its Hermitian are time-independent only when the driving frequency $\omega=-\Delta_{1}(k^{\prime}_{1},k^{\prime}_{2})/{q_{0}}$ is tuned. To safely disregard the impact of non-resonant terms within the effective Hamiltonian, it is necessary to satisfy the rotating wave approximation (RWA) condition, which effectively suppresses undesired transitions. Therefore, the conditions ${|q\omega+\Delta_{2}(k_{1},k_{2})}|\gg{|G^{(0)}_{2}(k_{2},\eta_{2})|}$, and ${|q\omega+\Delta_{1}(k_{1},k_{2})|}\gg{|G^{(0)}_{1}(k_{1},\eta_{1})|}$ for $k_{1}\neq{k^{\prime}_{1}}$ or $k_{2}\neq{k^{\prime}_{2}}$ or $q\neq{q_{0}}$ need to be fulfilled within the ultrastrong coupling regime. According to the constraint, only the transition $\ket{W_{N_{1}}^{k^{\prime}_{1}},W_{N_{2}}^{k^{\prime}_{2}}}\leftrightarrow \ket{W_{N_{1}}^{k^{\prime}_{1}+1},W_{N_{2}}^{k^{\prime}_{2}}}$ is possible, and the other transitions are not permitted. In this case, the effective Hamiltonian is given as
\begin{equation}\label{eq_eff1}
  \hat{H}^{(1)}_{\rm eff}=G^{(0)}_{1}(k^{\prime}_{1},\eta_{1})
\left[\hat{W}_{N_{1}}^{k^{\prime}_{1}+1,k^{\prime}_{1}}\otimes\hat{W}_{N_{2}}^{k^{\prime}_{2},k^{\prime}_{2}}+\rm{H.c.}\right].
\end{equation}
The terms of $\hat{W}_{N_{1}}^{k^{\prime}_{1},k^{\prime}_{1}}\otimes\hat{W}_{N_{2}}^{k^{\prime}_{2}+1,k^{\prime}_{2}}$ and its Hermitian are time-independent only when the driving frequency reaches the resonance condition $\omega=-{\Delta_{2}(k^{\prime}_{1},k^{\prime}_{2})}/{q_{0}}$.
Similarly, the conditions ${|q\omega+\Delta_{1}(k_{1},k_{2})|}\gg{|G^{(0)}_{1}(k_{1},\eta_{1})|}$, and ${|q\omega+\Delta_{2}(k_{1},k_{2})|}\gg{|G^{(0)}_{2}(k_{2},\eta_{2})|}$ for $k_{1}\neq{k^{\prime}_{1}}$ or $k_{2}\neq{k^{\prime}_{2}}$ or $q\neq{q_{0}}$ need to be satisfied in the ultrastrong coupling regime. Under these circumstances, only the transition $\ket{W_{N_{1}}^{k^{\prime}_{1}},W_{N_{2}}^{k^{\prime}_{2}}} \leftrightarrow\ket{W_{N_{1}}^{k^{\prime}_{1}},W_{N_{2}}^{k^{\prime}_{2}+1}}$ is permitted, and other transitions are forbidden.Accordingly, the effective Hamiltonian reads
\begin{equation}\label{eq_eff2}
  \hat{H}^{(2)}_{\rm eff}=G^{(0)}_{2}(k^{\prime}_{2},\eta_{2})
\left[\hat{W}_{N_{1}}^{k^{\prime}_{1},k^{\prime}_{1}}\otimes\hat{W}_{N_{2}}^{k^{\prime}_{2}+1,k^{\prime}_{2}}+\rm{H.c.}\right].
\end{equation}
Based on these conditions described above, we can achieve the desired transition between the initial and final states. In the following section, we will explore the applications of the so-called selective interactions.
\begin{figure}
\centering
\includegraphics[width=0.48\textwidth]{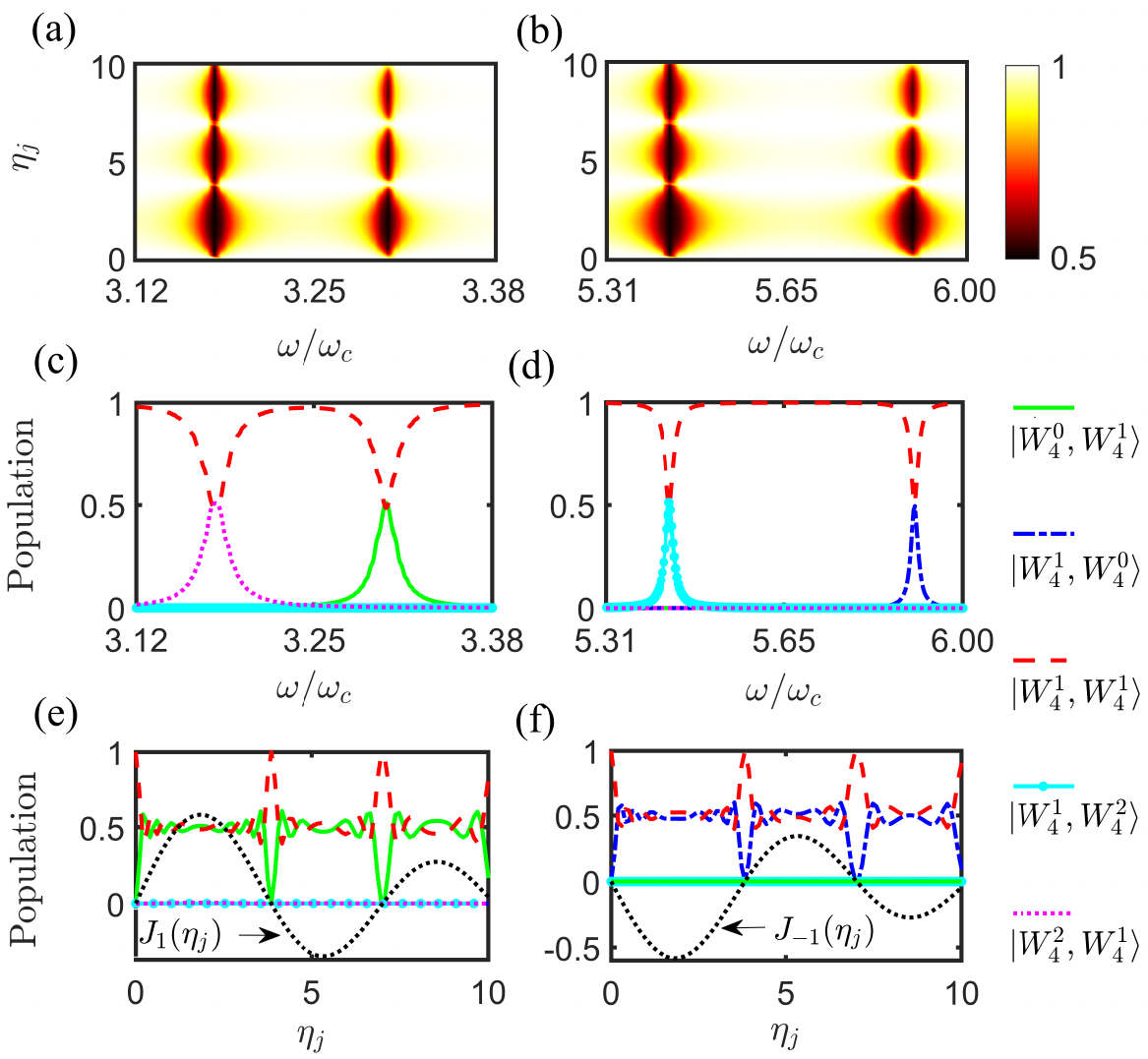}
\caption{Numerical results of the entangled Dicke states based on the selective resonant interaction. Panels (a) and (b) show the average population of initial state $\ket{W_{4}^{1},W_{4}^{1}}$ as the function of driving frequency $\omega$ and parameter $\eta_{j}$. Panels (c) and (d) show the population transferred from the initial state $\ket{W_{4}^{1},W_{4}^{1}}$ to the Dicke states $\ket{W_{4}^{1},W_{4}^{0}}$, $\ket{W_{4}^{0},W_{4}^{1}}$, $\ket{W_{4}^{2},W_{4}^{1}}$, and $\ket{W_{4}^{1},W_{4}^{2}}$ as the function of $\omega/\omega_{c}$ at the fixed parameter $\eta_{j}=1.09$. Panels (e) and (f) show the population from the initial state $\ket{W_{4}^{1},W_{4}^{1}}$ to the Dicke states $\ket{W_{4}^{0},W_{4}^{1}}$, and $\ket{W_{4}^{1},W_{4}^{0}}$ at the fixed driving parameter $\omega$=3.3025 or 5.9053. The dynamics is governed by the Hamiltonian in Eq. (\ref{eq_02}). The other parameters are chosen as $\omega_{c}/2\pi = 2.2 $ GHz, $g_ {1}=0.25 \omega_{c}$, $g_{2}=0.49 \omega_{c}$, $\varepsilon_{1}=\varepsilon_{2}= 0.01 \omega_{c}$, $\omega_{q_{1}} = 2.87 \omega_{c}$, $\omega_{q_{2}} = 4.94 \omega_{c}$.}\label{fig_02}
\end{figure}

\section{The applications of selective transition}
\label{SecIII}
\subsection{Selective interaction dynamics in two qubit ensembles}
\label{SecIIIA}
Through the implementation of controllable interactions, we possess the capability to generate a wide range of entangled Dicke states. To illustrate selective transitions, we will explain how to achieve the desired state in a specific example. Considering that each ensemble constitutes four qubits (i.e., $N_{1}=N_{2}=4$, $q_{0}=-1$) with the initial state $\ket{W_{4}^{1},W_{4}^{1}}$, one can selectively obtain various final states such as $\ket{W_{4}^{1},W_{4}^{0}}$, $\ket{W_{4}^{0},W_{4}^{1}}$, $\ket{W_{4}^{2},W_{4}^{1}}$, and $\ket{W_{4}^{1},W_{4}^{2}}$. With the mechanism of the dominated role of the selective interaction in the dynamics of the system at hand, we have shown the calculated time-averaged population: $\overline{P}=\frac{1}{T}\int_{0}^{T}|\inner{\beta}{\psi(t)}|^{2}dt$ with $\ket{\beta} = \ket{W_{4}^{1},W_{4}^{1}}$ as a function of the modulation frequency and modulation index in Fig. \ref{fig_02}. In the regime of high modulation frequency ($\omega \gg G^{(0)}_{j}$), the dynamics is relatively robust and dictated by the Bessel functions. It can be easily seen that the pearl-stripes align along the $\eta_j$ axis at resonances, while the local minima along the chains indicate the values of $\eta_{j}$ at which population trapping takes place. Between the stripes (along the $\omega/\omega_c$ axis), the average population of $\ket{W_{4}^{1},W_{4}^{1}}$ remains due to the atoms being driven far off-resonance. In Figs. \ref{fig_02} (c) and (d), we further show how the population of each quantum state changes with the driving frequency, while assuming a fixed value for the parameter $\eta_{j}$. The resonances can be identified at the intersection of peaks and dips. Once the resonance condition is satisfied, the population dynamics exhibits coherent Rabi oscillations at those peaks. Far away from the resonances, the system has a small probability of evolution due to the off-resonance. We also note that the effective Rabi coupling is directly proportional to $\mathcal{J}_{q}(\eta_j)$. Consequently, at the zeros of Bessel function [$\mathcal{J}_{q}(\eta_{j})$ = 0], the dynamics freezes and leads to population trapping \cite{PhysRevB.67.165301}. In Figs. \ref{fig_02} (e) and (f), we show the results for the case of resonances, and the peaks appear at the zeros of $\mathcal{J}_{q} (\eta_{j})$ = 0. We find that the dynamical evolution is switched off and the corresponding transition is strongly suppressed. Through an appropriate choice of $\eta_{j}$, we observe either population trapping or coherent dynamics between the initial state and final states. In short, an invariant population at the driving-induced resonance indicates the population trapping.

\begin{figure}
\centering
\includegraphics[width=0.48\textwidth]{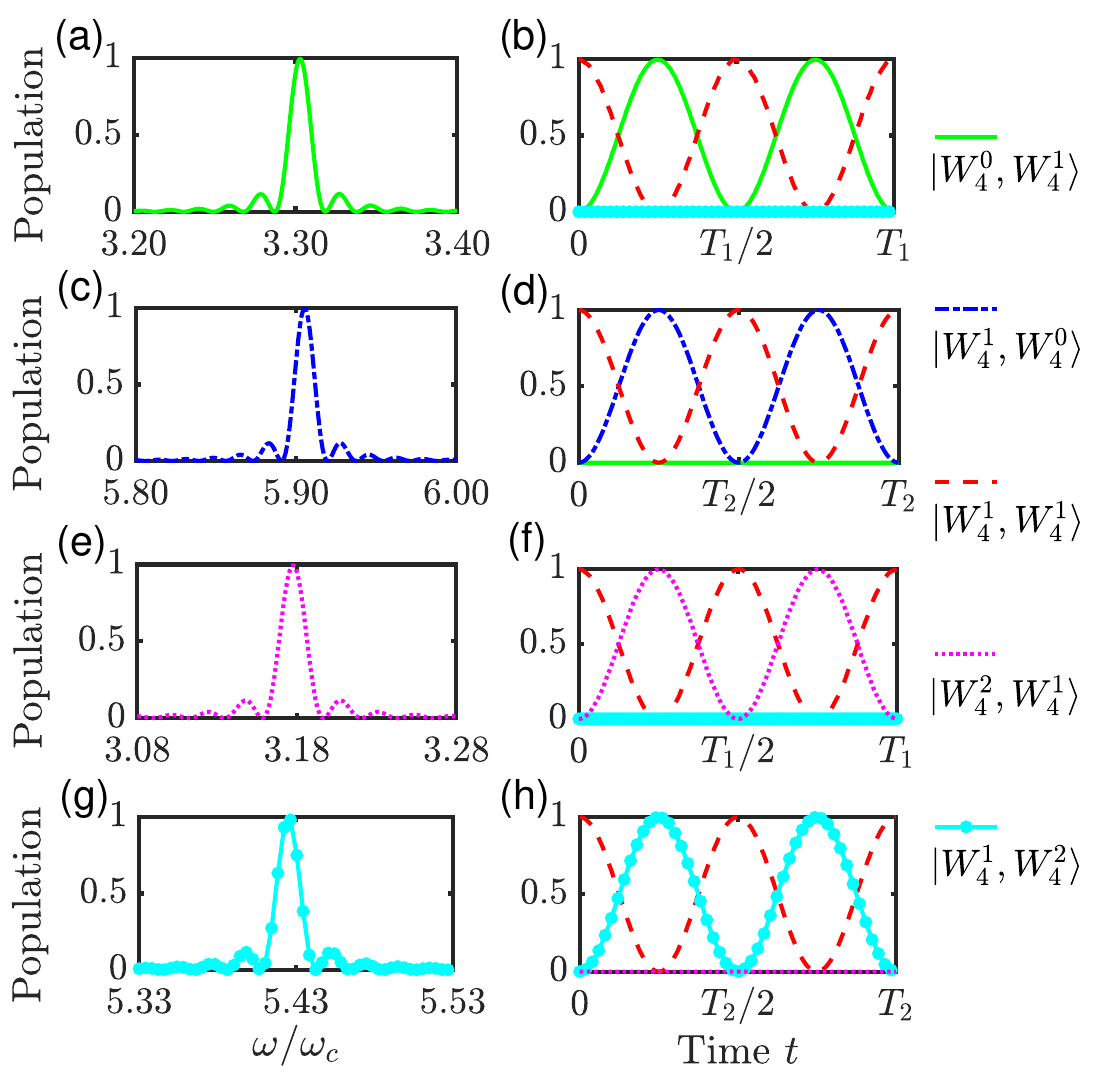}
\caption{Numerical results of the dynamics evolution based on the selective resonant interaction. Panels (a), (c), (e), and (g) indicate the population of the states $\ket{W_{4}^{0},W_{4}^{1}}$, $\ket{W_{4}^{1},W_{4}^{0}}$, $\ket{W_{4}^{2},W_{4}^{1}}$, and $\ket{W_{4}^{1},W_{4}^{2}}$ generated from $\ket{W_{4}^{1},W_{4}^{1}}$ versus $\omega/\omega_{c}$. Panels (b), (d), (f) and (h) show the dynamics evolution of the Dicke states $\ket{W_{4}^{0},W_{4}^{1}}$, $\ket{W_{4}^{1},W_{4}^{0}}$, $\ket{W_{4}^{2},W_{4}^{1}}$, and $\ket{W_{4}^{1},W_{4}^{2}}$ generated from $\ket{W_{4}^{1},W_{4}^{1}}$ versus time $t$. The dynamics is governed by the Hamiltonian in Eq. (\ref{eq_02}). The other parameters are chosen as $\eta_{1}=\eta_{2}=1.09$, $\omega_{c}/2 \pi = 2.2 $ GHz, $ g_ {1} = 0.25 \omega_{c}$, $g_{2}=0.49 \omega_{c}$, $\varepsilon_{1}=\varepsilon_{2}= 0.01 \omega_{c}$, $\omega_{q_{1}} = 2.87 \omega_{c}$, $\omega_{q_{2}} = 4.94 \omega_{c}$.}\label{fig_03}
\end{figure}

For highlighting the nonlinearity of energy level and characteristic selective transitions, a comprehensive analysis of both Fig. \ref{fig_02} and Fig. \ref{fig_03} reveals that assuming the driving frequency $\omega=\Delta_{1}({0,1})=3.3025\omega_{c}$, the Rabi oscillation between $\ket{W_{4}^{1},W_{4}^{1}}$ and $\ket{W_{4}^{0},W_{4}^{1}}$ can be obtained with a period of approximately $T_{1} = 2\pi/[G^{(0)}_{1}(0,1.09)] \approx 0.1002{\rm \mu s}$. Similarly, we can manipulate the driving frequency $\omega=\Delta_{2}({1,0})=5.9053\omega_{c}$ to achieve the evolution state from $\ket{W_{4}^{1},W_{4}^{1}}$ to $\ket{W_{4}^{1},W_{4}^{0}}$ with the time period of $T_{2} = 2\pi/[G^{(0)}_{2}(0,1.09)]  \approx 0.1095 {\rm \mu s}$. Furthermore, when the driving frequency $\omega=\Delta_{1}({1,1})=3.1775\omega_{c}$, the evolution of the system leads to the population transfer from $\ket{W_{4}^{1},W_{4}^{1}}$ to $\ket{W_{4}^{2},W_{4}^{1}}$ with the period of $T_{1} = 2\pi/[G^{(0)}_{1}(1,1.09)] \approx 0.0818 {\rm \mu s}$. While adjusting the driving frequency $\omega=\Delta_{2}({1,1})=5.4251\omega_{c}$, the Hamiltonian in Eq. (\ref{eq_02}) results in the creation of the quantum state $\ket{W_{4}^{1}} \otimes \ket{W_{4}^{2}}$ with the time period of $T_{2} = 2 \pi/[G^{(0)}_{2}(1,1.09)] \approx 0.0894 {\rm \mu s}$.
\subsection{Generation of entangled Dicke states}
\label{SecIIIB}
In terms of implementing the controllable interaction, it is also possible to achieve the desired entangled Dicke states through a multi-step process. Here, we present an example of entangling two Dicke states by desired interaction. The target state reads
\begin{equation}
 \ket{\psi(t_{f})}= \frac{1}{\sqrt{2}}(\ket{W_{4}^{4},W_{4}^{0}}+\ket{W_{4}^{0},W_{4}^{4}}).
\end{equation}
The corresponding dynamic evolution process is shown in Fig. \ref{fig_04}. We will provide a detailed, step-by-step explanation of how to achieve the target state.

\begin{itemize}
\item Step 1. Initially, we prepare the entire system in the state $\ket{\psi(t_0)}=\ket{W_{4}^{0},W_{4}^{0}}$. Applying the resonance Hamiltonian with the driving frequency $\omega=\Delta_{1}({0,0})$ to ensure selective resonance between $\ket{W_{4}^{0},W_{4}^{0}}$ and $\ket{W_{4}^{1},W_{4}^{0}}$, we obtain the evolution state $\ket{\psi(t_{1})}=(\ket{W_{4}^{0},W_{4}^{0}}-i\ket{W_{4}^{1},W_{4}^{0}})/{\sqrt{2}}$ at time $t_{1}={\pi}/{4\Omega_{1}^{0}} \approx 0.0654 {\rm \mu s}$ with $\Omega_{1}^{0}\equiv G^{(0)}_{1}(0,0.18)$.
\item Step 2. Specially, we set $\omega=\Delta_{1}({1,0})$ to achieve the selective resonance between $\ket{W_{4}^{1},W_{4}^{0}}$ and $\ket{W_{4}^{2},W_{4}^{0}}$ as illustrated in Fig. \ref{fig_04}. Then the evolution state $\ket{\psi(t_{2})}=(\ket{W_{4}^{0},W_{4}^{0}}-\ket{W_{4}^{2},W_{4}^{0}})/{\sqrt{2}}$ after a period of time $t_{2}=t_{1}+{\pi}/{2\Omega_{2}^{0}} \approx 0.1722{\rm \mu s}$ with $\Omega_{2}^{0}\equiv G^{(0)}_{1}(1,0.18)$. 
\begin{table*}
\centering
\caption{The generation of entangled Dicke state $ \left(\ket{W_{4}^{4},W_{4}^{0}}+\ket{W_{4}^{0},W_{4}^{4}}\right)/\sqrt{2}$ with the initial state $\ket{W_{4}^{0},W_{4}^{0}}$. The dynamics is governed by the Hamiltonian in Eq. (\ref{eq_02}).}
\label{table_I}
\setlength{\tabcolsep}{2.2mm}
\begin{tabular}{cccccc}
\hline
\hline
{\rm Step~$l$} & Driving frequency $\omega/\omega_{c}$ & Time duration ($\mu s$) & Final state for step~$l$ & Fidelity (abrupt) & Fidelity (soft) \\
\hline
1 & 3.5475 & ${\pi}/{4G^{(0)}_{1}(0,0.18)} \approx 0.0654 $ & $ 
\left(\ket{W_{4}^{0},W_{4}^{0}}-i\ket{W_{4}^{1},W_{4}^{0}}\right)/{\sqrt{2}}$ & 0.9999 & 0.9999\\
2 & 3.4225 & ${\pi}/{2G^{(0)}_{1}(1,0.18)} \approx 0.1068 $ & $ \left(\ket{W_{4}^{0},W_{4}^{0}}-\ket{W_{4}^{2},W_{4}^{0}}\right)/{\sqrt{2}}$ & 0.9989 & 0.9992\\
3 & 3.2975 & ${\pi}/{2G^{(0)}_{1}(2,0.18)} \approx 0.1068 $ & $ \left(\ket{W_{4}^{0},W_{4}^{0}}+i\ket{W_{4}^{3},W_{4}^{0}}\right)/{\sqrt{2}}$ & 0.9954 & 0.9962\\
4 & 3.1725 & ${\pi}/{2G^{(0)}_{1}(3,0.18)} \approx 0.1308 $ & $ \left(\ket{W_{4}^{0},W_{4}^{0}}+\ket{W_{4}^{4},W_{4}^{0}}\right)/{\sqrt{2}}$ & 0.9864 & 0.9918\\
5 & 6.1503 & ${\pi}/{2G^{(0)}_{2}(0,0.18)} \approx 0.0994 $ & $ \left(\ket{W_{4}^{4},W_{4}^{0}}-i\ket{W_{4}^{0},W_{4}^{1}}\right)/{\sqrt{2}}$ & 0.9782 & 0.9877\\
6 & 5.6701 & ${\pi}/{2G^{(0)}_{2}(1,0.18)} \approx 0.0812 $ & $ \left(\ket{W_{4}^{4},W_{4}^{0}}-\ket{W_{4}^{0},W_{4}^{2}}\right)/{\sqrt{2}}$ & 0.9700 & 0.9827\\
7 & 5.1899 & ${\pi}/{2G^{(0)}_{2}(2,0.18)} \approx 0.0812 $ & $ \left(\ket{W_{4}^{4},W_{4}^{0}}+i\ket{W_{4}^{0},W_{4}^{3}}\right)/{\sqrt{2}} $ & 0.9409 & 0.9781\\
8 & 4.7097 & ${\pi}/{2G^{(0)}_{2}(3,0.18)}\approx 0.0994 $ & $ \left(\ket{W_{4}^{4},W_{4}^{0}}+\ket{W_{4}^{0},W_{4}^{4}}\right)/{\sqrt{2}}$ & 0.9322 & 0.9624\\
\hline
\hline
\end{tabular}
\end{table*}
\begin{figure}
\centering
\includegraphics[width=0.48\textwidth]{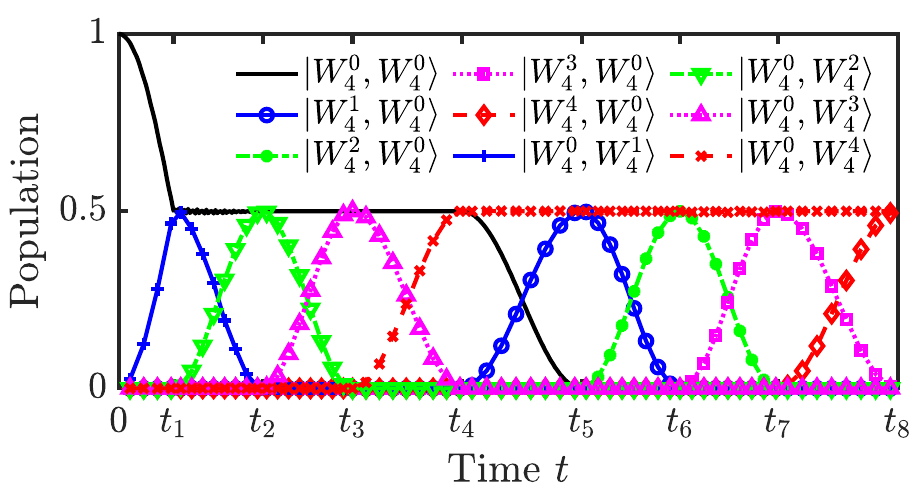}
\caption{The figure shows the population of entangled Dicke state for $N_{1} = N_{2} = 4$ from $(\ket{W_{4}^{4},W_{4}^{0}}+\ket{W_{4}^{0},W_{4}^{4}})/{\sqrt{2}}$ versus time $t$. The dynamics is governed by the Hamiltonian in Eq. (\ref{eq_02}). The other parameters are chosen as $\eta_{1}=\eta_{2}=0.18$, $\omega_{c}/2\pi = 2.2 $ GHz, $g_{1}=0.25 \omega_{c}$, $g_{2}=0.49 \omega_{c}$, $\varepsilon_{1}=\varepsilon_{2}= 0.01 \omega_{c}$, $\omega_{q_{1}} = 2.87 \omega_{c}$, $\omega_{q_{2}} = 4.94 \omega_{c}$.}\label{fig_04}
\end{figure}
\item Step 3. Certainly, we can adjust $\omega=\Delta_{1}({2,0})$ to hold perfect oscillations transition between $\ket{W_{4}^{2},W_{4}^{0}}$ and $\ket{W_{4}^{3},W_{4}^{0}}$. The resulting state is $\ket{\psi(t_{3})}=(\ket{W_{4}^{0},W_{4}^{0}}+i\ket{W_{4}^{3},W_{4}^{0}})/{\sqrt{2}}$ at time $t_{3}=t_{2}+\pi/{2\Omega_{3}^{0}} \approx 0.2790 {\rm \mu s}$ with $\Omega_{3}^{0}= G^{(0)}_{1}(2,0.18)$.

\item Step 4. We tune $\omega=\Delta_{1}({3,0})$ to attain the selective resonance between $\ket{W_{4}^{4},W_{4}^{0}}$ and $\ket{W_{4}^{3},W_{4}^{0}}$. The quantum state evolves into $\ket{\psi(t_{4})}=(\ket{W_{4}^{0},W_{4}^{0}}+\ket{W_{4}^{4},W_{4}^{0}})/{\sqrt{2}}$ at the moment $t_{4}=t_{3}+{\pi}/{2\Omega_{4}^{0}} \approx 0.4098 {\rm \mu s}$ with $\Omega_{4}^{0}\equiv G^{(0)}_{1}(3,0.18)$.

\item Step 5. When the frequency $\omega=\Delta_{2}({0,0})$ satisfies the condition of resonance, and the evolved state changes into $\ket{\psi(t_{5})}=(\ket{W_{4}^{4},W_{4}^{0}}-i\ket{W_{4}^{0},W_{4}^{1}})/{\sqrt{2}}$
in the time $t_{5}=t_{4}+{\pi}/{2\Omega_{5}^{0}} \approx 0.5092 {\rm \mu s}$ with $\Omega_{5}^{0}= G^{(0)}_{2}(0,0.18)$.

\item Step 6. Subsequently, we set $\omega=\Delta_{2}({0,1})$ to maintain selective resonance almost strictly between the state $\ket{W_{4}^{0},W_{4}^{1}}$ and $\ket{W_{4}^{0},W_{4}^{2}}$. After $t_{6}=t_{5}+{\pi}/{2\Omega_{6}^{0}} \approx 0.5904 {\rm \mu s}$ with $\Omega_{6}^{0}= G^{(0)}_{2}(1,0.18)$, the quantum state $\ket{\psi(t_{6})}=(\ket{W_{4}^{4},W_{4}^{0}}-\ket{W_{4}^{0},W_{4}^{2}})/{\sqrt{2}}$ can be realized.

\item Step 7. After applying the effective Hamiltonian  $\hat{H}^{(2)}_{\rm eff}=G^{(0)}_{2}(2,0.28)\left[\hat{W}_{4}^{0,0}\otimes\hat{W}_{4}^{3,2}+\rm{H.c.}\right]$ on the state $\ket{\psi(t_{6})}$ and setting the driving frequency $\omega=\Delta_{2}({0,2})$,
we can reach the state $\ket{\psi(t_{7})}=(\ket{W_{4}^{4},W_{4}^{0}}+i\ket{W_{4}^{0},W_{4}^{3}})/{\sqrt{2}}$ at $t_{7}=t_{6}+{\pi}/{2\Omega_{7}^{0}}\approx 0.6716 {\rm \mu s}$ with $\Omega_{7}^{0}= G^{(0)}_{2}(2,0.18)$.

\item Step 8. Finally, we tune the driving frequency $\omega=\Delta_{2}({0,3})$ to derive the transition between $\ket{W_{4}^{0},W_{4}^{3}}$ and $\ket{W_{4}^{0},W_{4}^{4}}$. In the serious of operations, the quantum state is excited to $\ket{\psi(t_{8})}=(\ket{W_{4}^{4},W_{4}^{0}}+\ket{W_{4}^{0},W_{4}^{4}})/{\sqrt{2}}$ at time $t_{8}=t_{7}+{\pi}/{2\Omega_{8}^{0}}\approx 0.7709 {\rm \mu s}$ with $\Omega_{8}^{0}= G^{(0)}_{2}(3,0.18)$. 
\end{itemize}
If all terms in the Eq. (\ref{H_int(q)}) are fast-time varying, higher order effective Hamiltonian will dominate the dynamics \cite{PhysRevA.106.022431}. Based on the higher order Hamiltonian, the number of steps can be reduced. Tuning the driving frequency to obtain the two-atomic excitation process, we can reduce the eight-step operations to four-step operations \cite{Mu:20}. The numerical results of above eight-step protocol are listed in Table \ref{table_I}. The numerical simulation of the fidelity is defined as $F(\ket{\psi_{T}},\ket{\psi(t)})=|\bra{\psi_{T}}\psi(t)\rangle|^{2}$ with $\ket{\psi_{T}}$ and $\ket{\psi(t)}$ being the target state and evolution state, respectively. Through the aforementioned procedure, we have successfully achieved the desired target state. The final state for the eighth step protocol reads
\begin{equation}
\ket{\psi(t_{8})}=\frac{1}{\sqrt{2}}(\ket{W_{4}^{4},W_{4}^{0}}+\ket{W_{4}^{0},W_{4}^{4}})\otimes \ket{0}.
\end{equation}
In the original frame, such state can be recast as follows
\begin{equation}
\ket{\psi'(t_{8})}=\frac{1}{\sqrt{2}}(\ket{W_{4}^{4},W_{4}^{0}}_{x}\otimes\ket{\alpha}+\ket{W_{4}^{0},W_{4}^{4}}_{x}\otimes\ket{-\alpha}),
\end{equation}
where $\alpha = 2(\beta_{2}-\beta_{1})$. Hence, we obtain the tripartite entangled states. Then, we will show how to obtain entangled two Dicke states based on projective measurement. The orthogonal even and odd Shcr\"{o}dinger cat states take as following form
\begin{equation}
\ket{C_{\alpha}^{\pm}}=\mathcal{N}_{\pm}^{-1}(\ket{\alpha}\pm \ket{-\alpha}),
\end{equation}
where $\mathcal{N}_{\pm}=\sqrt{2\left[1\pm \exp(-2|\alpha|^{2})\right]}$ are the respective normalization factors. Here, even or odd refers to the photon number of the cat states in Fock space, that is, a
state containing only even or odd photon numbers. In terms of cat state basis, the final states can be rewritten as
\begin{small}
\begin{equation}
    \ket{\psi'(t_{8})}=\frac{1}{2}\left[\mathcal{N}_{+}\ket{D^{+}_{x}}\otimes\ket{C_{\alpha}^{+}}+\mathcal{N}_{-}\ket{D^{-}_{x}}\otimes\ket{C_{\alpha}^{-}}\right],
\end{equation}
\end{small}
where $\ket{D^{\pm}_{x}}=(\ket{W_{4}^{4},W_{4}^{0}}_{x}\pm\ket{W_{4}^{0},W_{4}^{4}}_{x})/\sqrt{2}$ are entangled atomic Dicke states. Performing projective measurement on the system in the cat states basis \cite{Izumi2018}, the atomic Dicke states can be derived. If the projective measurement result is even (odd) cat state $\ket{C_{\alpha}^{\pm}}$, we can obtain the atomic final states $\ket{D^{\pm}_{x}}$ with probability $[1\pm \exp(-2|\alpha|^{2})]/2$. Hence, we obtain the entangled state for two qubit ensembles. 
\subsection{Fidelity enhancement via Gaussian soft quantum control}
\label{SecIIIC}
In the subsection \ref{SecIIIB}, we assume that the drive is abruptly turned on at an initial time $t_{l-1}$ and lasts till a final time $t_{l}$ for $l$-th step. To improve the fidelity of the entangled Dicke states, we can employ Gaussian time-dependent control techniques, and achieve precise coupling to resonant terms through modulation of the coupling constant over time. The Gaussian soft control \cite{PhysRevLett.121.050402} offers the advantage of maintaining higher fidelity of the corresponding target quantum states within a broader parameter range, which accurately satisfy the conditions of RWA and avoid unnecessary transitions. Such temporal modulation significantly suppresses the non-resonant effects of the interaction while remaining insensitive to operation time, which can extremely improve the robustness of the scheme. Next, we change the time-independent Rabi frequency $\Omega_{l}^{0}$ to a time-dependent Gaussian form \cite{PhysRevApplied.20.014014}
\begin{figure}
\centering
\includegraphics[width=0.48\textwidth]{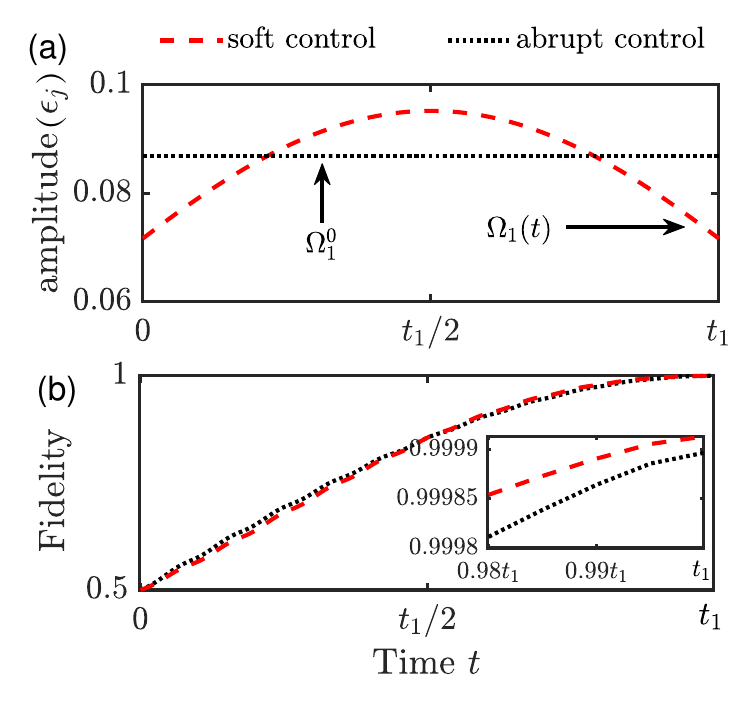}
\caption{Panel (a) shows the abrupt driving and soft driving as the function of time $t$ for ${\rm Step~1}$. Panel (b) shows the fidelity of evolution state $\left(\ket{W_{4}^{0},W_{4}^{0}}-i\ket{W_{4}^{1},W_{4}^{0}}\right)/{\sqrt{2}}$ in ${\rm Step~1}$. Insets illustrates the fidelity of final state under both driving conditions. The dynamics is governed by the Hamiltonian in Eq. (\ref{eq_02}). The other parameters are chosen consistent with Fig. \ref{fig_04}.}\label{fig_05}
\end{figure}
\begin{equation}
  \Omega_{l}(t)=\mu \Omega_{l}^{0}\exp\Big(-\frac{[t-(t_{l-1}+t_{l})/2]^2}{2\sigma^{2}_{l}}\Big),
\end{equation}
where $\Omega_{l}^{0}$ represents the corresponding Rabi frequency. $t_{l-1}$ ($t_{l}$) is the minimum (maximum) of each segment evolution time in the eight-step operation. $\sqrt{2} \sigma_{l}$ is the width of the Gaussian pulse applied during each step of evolution. In order to maintain the necessary coherent process for the preparation of entangled states, the Gaussian pulses need to be satisfied
\begin{equation}\label{eq_15}
    A_{l}=\int_{t_{l-1}}^{t_{l}} \Omega_{l}(t) \, dt =
\begin{cases}
    \frac{\pi}{4}, & {\rm For~Step}~1 \\
    \frac{\pi}{2}, & {\rm For~other~Steps}
\end{cases}
\end{equation}
According to Eq. (\ref{eq_15}), we have imposed a condition for calculating the pulse area within a finite time interval, ranging from $t_{l-1}$ to $t_{l}$. This condition ensures that Gaussian pulses can be reliably substituted for the previous Rabi frequency $\Omega_{l}^{0}$, i.e., $\mu >1$. Furthermore, parameter $\sigma_{l}$ is determined by the following relation
\begin{equation}
\sigma_{l}=
{\frac{\sqrt{\pi}}{2\sqrt2 n\mu \Omega_{l}(0) {\rm erf}[(t_{l}-t_{l-1})/(2\sqrt{2}\sigma_{l})]}}.
\end{equation}
\begin{figure}
\centering
\includegraphics[width=0.48\textwidth]{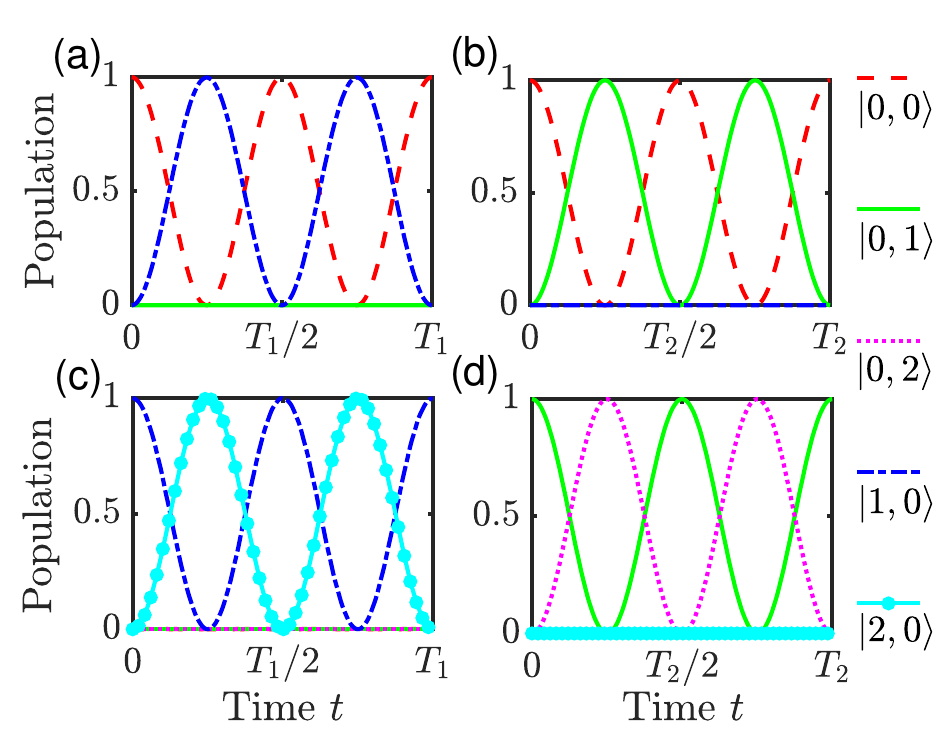}
\caption{Panels (a)-(d) show the populations with different magnon Fock states as the initial states. The dynamics is governed by the Hamiltonian in Eq. (\ref{{H}^{m}_{int}}). The other parameters are chosen as $\eta_{1}=\eta_{2}=0.12$, $\omega_{c}/2\pi = 2.2 $ GHz, $g_ {1} = 0.30 \omega_{c}$, $g_{2} = 0.49 \omega_{c}$, $\varepsilon_{1}=\varepsilon_{2}= 0.01 \omega_{c}$, $\omega_{q_{1}} = 2.78 \omega_{c}$, $\omega_{q_{2}} = 4.94 \omega_{c}$.}
\label{fig_06}
\end{figure}
Here the function ${\rm erf}(x)=\frac{2}{\sqrt \pi}\int^{x}_{0}dze^{-z^{2}}$, and $n$=2 for $l$=1 and $n$=1 for all other cases. We are able to achieve soft control by applying different Gaussian pulses to each step and then obtain high-fidelity target quantum states. In Fig. \ref{fig_05}, we show the temporal evolution of both drive amplitudes in the initial state. Correspondingly, we present the fidelity of the desired target state throughout the evolution, with insets indicating that soft control yields a higher fidelity compared to abrupt control upon completion of the process. The situation can be explained as follows: when the absence of a modulation for $\Omega_{l}(t)$, i.e., abrupt driving $\Omega_{l}^{0}$, unwanted terms in the effective Hamiltonian can be neglected by the RWA provided that the $\Delta_{j}$ is sufficiently large compared with $\Omega_{l}^{0}$. However, after applying Gaussian soft control, we can calculate the Hamiltonian by using Mangus expansion for a time interval of [$t_{l-1}$, $t_{l}$], where the mean coefficient $g_{M}\varpropto\mu\Omega_{l}^{0}\exp(-\frac{1}{2}\sigma_{l}^2\Delta_{j})$. A simple inspection of $g_{M}$ reveals that the effective couplings decay exponentially with $\Delta_{j}$ at the fixed $\mu$, and the contribution to the average Hamiltonian decreases accordingly. In other words, the high-frequency oscillation term can be greatly suppressed when the detuning amount is relatively large. In addition, we find that $g_{M}$ is a monotonically increasing function with $\mu$ and larger $g_{M}$ can have negative consequences on the coupling term of the resonance as the result of selecting a high value for $\mu$. Thus, combined with practical implementations, it is essential to rationally choose $\mu$ to achieve high-fidelity quantum states. For the above consideration, we choose $\mu=\sqrt{6/5}$ to achieve soft control. The corresponding results, presented in the last column of Table \ref{table_I}, show improvement in fidelity compared to abrupt control. So our soft control scheme enables highly selective coupling between different resonance constituents of composite systems. 
\begin{table*}
\centering
\caption{The generation of magnon NOON state $(\ket{2,0}+\ket{0,2})/\sqrt{2}$ with the initial state $\ket{\psi(0)}=\ket{0,0}$. The dynamics is governed by the Hamiltonian in Eq. (\ref{{H}^{m}_{int}}).}
\label{table_II}
\setlength{\tabcolsep}{3mm}
\begin{tabular}{cccccc}
\hline
\hline
Step $i$ & Driving frequency $\omega/\omega_{c}$ & Time duration ($\mu s$) & Evolution state $\ket{\psi(t_{i})}$ & Fidelity (abrupt) & Fidelity (soft)\\
\hline
1 & 50.0900 & ${\pi}/{4G^{(0)}_{m1}(0,0.12)} \approx {0.0140}$ & $(\ket{0,0}-i\ket{1,0})/\sqrt{2}$ & 0.9994 & 0.9999\\
2 & 49.9100 & ${\pi}/{2G^{(0)}_{m1}(0,0.12)} \approx 0.0198 $ & $(\ket{0,0}-\ket{2,0})/\sqrt{2}$ & 0.9947 & 0.9998\\
3 & 82.1199 & ${\pi}/{2G^{(0)}_{m2}(0,0.12)} \approx 0.0302 $ & $(\ket{2,0}-i\ket{0,1})/\sqrt{2}$ & 0.9846 & 0.9997\\
4 & 81.6397 & ${\pi}/{2G^{(0)}_{m2}(1,0.12)} \approx 0.0214 $ & $(\ket{2,0}+\ket{0,2})/\sqrt{2}$ & 0.9715 & 0.9995\\
\hline
\hline
\end{tabular}
\end{table*}
\section{The thermodynamic limit case}
\label{SecIV}
Now, we consider the effect of nonlinear energy levels in the thermodynamic limit ($N_{1}\gg 1$ and $N_{2}\gg 1$). With the Holstein-Primakoff transformation, the collective operators can be written as \cite{PhysRev.58.1098}
\begin{subequations}
\begin{align}
   \hat{S}^{z}_{j}&=\hat{m}_{j}^{\dag}\hat{m}_{j}-s_{j},\\
   \hat{S}^{+}_{j}&=\hat{m}_{j}^{\dag}{(2s_{j}-\hat{m}_{j}^{\dag}\hat{m}_{j})^{\frac{1}{2}}},\\
   \hat{S}^{-}_{j}&={(2s_{j}-\hat{m}_{j}^{\dag}\hat{m}_{j})^{\frac{1}{2}}}\hat{m}_{j},
  \end{align}
\end{subequations}
where $s_{j} = N_{j}/2$ is pseudospin quantum number for the $j$-th qubit ensemble, and $\hat{m}_{j}^{\dag}$ ($\hat{m}_{j})$ denotes the creation (annihilation) operator of the two different magnons. For the low-lying magnon excitations, i.e., $\langle\hat{m}_{j}^{\dag}\hat{m}_{j}\rangle/2s_{j}\ll1$, we can safely approximate the collective operators as the following relations: $\hat{S}^{+}_{j}\approx{\sqrt{2s_{j}}\hat{m}_{j}^{\dag}},\hat{S}^{-}_{j}\approx{\sqrt{2s_{j}}\hat{m}_{j}}$. Then, the Eq. (\ref{H_int(q)}) reads
\begin{widetext}
\begin{equation}\label{{H}^{m}_{int}}
\begin{split}
  \hat{H}^{m}_{\rm I}(t)=-\sum_{q=-\infty}^{\infty}&\mathcal{J}_{q}(\eta_{j})\left[
\varepsilon_{1}\sqrt{\frac{s_{1}}{2}}\hat{D}(\beta_{1}e^{i\omega_{c}t})\hat{m}_{1}^{\dag}e^{i[q{\omega}+\Delta_{1}+2\frac{g_{1}g_{2}}{\omega_{c}}(s_{2}-\hat{m}_{2}^{\dag}\hat{m}_{2})+\frac{g_{1}^2}{\omega_{c}}(2s_{1}-2\hat{m}_{1}^{\dag}\hat{m}_{1}-1)]t}\right.\\
&\left.+\varepsilon_{2}\sqrt{\frac{s_{2}}{2}}\hat{D}(\beta_{2}e^{i\omega_{c}t})\hat{m}_{2}^{\dag}e^{i[q\omega+\Delta_{2}+2\frac{g_{1}g_{2}}{\omega_{c}}(s_{1}-\hat{m}_{1}^{\dag}\hat{m}_{1})+\frac{g_{2}^2}{\omega_{c}}(2s_{2}-2\hat{m}_{2}^{\dag}\hat{m}_{2}-1)]t}+\rm{H.c.}\right].
\end{split}
\end{equation}
In the subspace of two different magnon Fock states, the effective Hamiltonian (\ref{H_int(q)}) can be rewritten as
\begin{equation}\label{Heff(m)}
\begin{split}
  \hat{H}^{m}_{\rm I}(t)=
\sum_{q=-\infty}^{\infty}\sum_{m_{1},m_{2}}&\left[G^{(0)}_{m1}(m_{1},\eta_{1})
e^{iq{\omega}t}e^{i\Delta_{1}({m}_{1},{m}_{2})t}
\hat{M}_{m_{1}+1,m_{1}}\otimes\hat{M}_{m_{2},m_{2}}\right.\\
&\left.+G^{(0)}_{m2}(m_{2},\eta_{2})e^{iq{\omega}t}e^{i\Delta_{2}({m}_{1},{m}_{2})t}
\hat{M}_{m_{1},m_{1}}\otimes\hat{M}_{m_{2}+1,m_{2}}+\rm{H.c.}\right],
\end{split}
\end{equation}
\end{widetext}
where $m_{j}$ is the eigenvalue of the magnon Fock operator $\hat{m}^{\dag}_{j}\hat{m}_{j}$, $\hat{M}_{m_{j},m_{j}^{\prime}}=\ket{m_{j}}\bra{m_{j}^{\prime}}$,
$G^{(0)}_{mj}(m_{j},\eta_{j})=-\varepsilon_{j}\sqrt{{s_{j}(m_{j}+1)}}\mathcal{J}_{q}(\eta_{j})e^{{-\frac{1}{2}\beta_{j}^{2}}}/\sqrt{2}$,
and
{\small
\begin{eqnarray*}\label{Delta^{m}}
\Delta_{1}({m}_{1},{m}_{2})&=&\Delta_{1}+\frac{g_{1}^2}{\omega_{c}}(2s_{1}-2m_{1}-1)+2\frac{g_{1}g_{2}}{\omega_{c}}(s_{2}-m_{2}),\\
\Delta_{2}({m}_{1},{m}_{2})&=&\Delta_{2}+\frac{g_{2}^2}{\omega_{c}}(2s_{2}-2m_{2}-1)+2\frac{g_{1}g_{2}}{\omega_{c}}(s_{1}-m_{1}).
\end{eqnarray*}}
Similarly, in order to maintain the same approach as that mentioned above, we can tune the driving frequency $\omega$ to obtain the desired selective interactions based on the RWA.
The terms of $\hat{M}_{m^{\prime}_{1}+1,m^{\prime}_{1}}\otimes\hat{M}_{m^{\prime}_{2},m^{\prime}_{2}}$ and its Hermitian are time-independent only when the driving frequency $\omega=-{\Delta_{1}(m^{\prime}_{1},m^{\prime}_{2})}/{q_{0}}$.
If one can verify that the condition ${|q\omega+\Delta_{2}(m_{1},m_{2})}|\gg{|G^{(0)}_{m2}(m_{2},\eta_{2})|}$ holds to maintain for any $q$, $m_{1}$ and $m_{2}$.
And ${|q\omega+\Delta_{1}(m_{1},m_{2})|}\gg{|G^{(0)}_{m1}(m_{1},\eta_{1})|}$ for $m_{1}\neq{m^{\prime}_{1}}$ or $m_{2}\neq{m^{\prime}_{2}}$ or $q\neq{q_{0}}$ remains valid within the ultrastrong coupling regime.
In this case, the transition occurs between $\ket{m^{\prime}_{1},m^{\prime}_{2}}(\ket{m^{\prime}_{1},m^{\prime}_{2}}\equiv \ket{m^{\prime}_{1}}\otimes\ket{m^{\prime}_{2}})$ and
$\ket{m^{\prime}_{1}+1,m^{\prime}_{2}}$, the effective Hamiltonian reads
\begin{equation}\label{eq_eff1(m)}
  \hat{H}^{(1)}_{\rm eff}=G^{(0)}_{m1}(m^{\prime}_{1},\eta_{1})
[\hat{M}_{m^{\prime}_{1}+1,m^{\prime}_{1}}\otimes\hat{M}_{m^{\prime}_{2},m^{\prime}_{2}}+\rm{H.c.}].
\end{equation}
The terms of $\hat{M}_{m^{\prime}_{1},m^{\prime}_{1}}\otimes\hat{M}_{m^{\prime}_{2}+1,m^{\prime}_{2}}$ and its Hermitian are time-independent only when the driving frequency is tuned $\omega=-{\Delta_{2}(m^{\prime}_{1},m^{\prime}_{2})}/{q_{0}}$.
If we can ensure that the conditions ${|q\omega+\Delta_{1}(m_{1},m_{2})|}\gg{|G^{(0)}_{m1}(m_{1},\eta_{1})|}$ and ${|q\omega+\Delta_{2}(m^{\prime}_{1},m^{\prime}_{2})|}\gg{|G^{(0)}_{m2}(m_{2},\eta_{2})|}$ for $m_{1}\neq{m^{\prime}_{1}}$ or $m_{2}\neq{m^{\prime}_{2}}$ or $q\neq{q_{0}}$ are realized in the ultrastrong coupling regime. In such case, the transition occurs between $\ket{m^{\prime}_{1},m^{\prime}_{2}}$ and
$\ket{m^{\prime}_{1},m^{\prime}_{2}+1}$, the effective Hamiltonian reads
\begin{equation}\label{eq_eff2(m)}
  \hat{H}^{(2)}_{\rm eff}=G^{(0)}_{m2}(m^{\prime}_{2},\eta_{2})
[\hat{M}_{m^{\prime}_{1},m^{\prime}_{1}}\otimes\hat{M}_{m^{\prime}_{2}+1,m^{\prime}_{2}}+\rm{H.c.}].
\end{equation}
\begin{figure}[tbp]
\centering
\includegraphics[angle=0,width=0.48\textwidth]{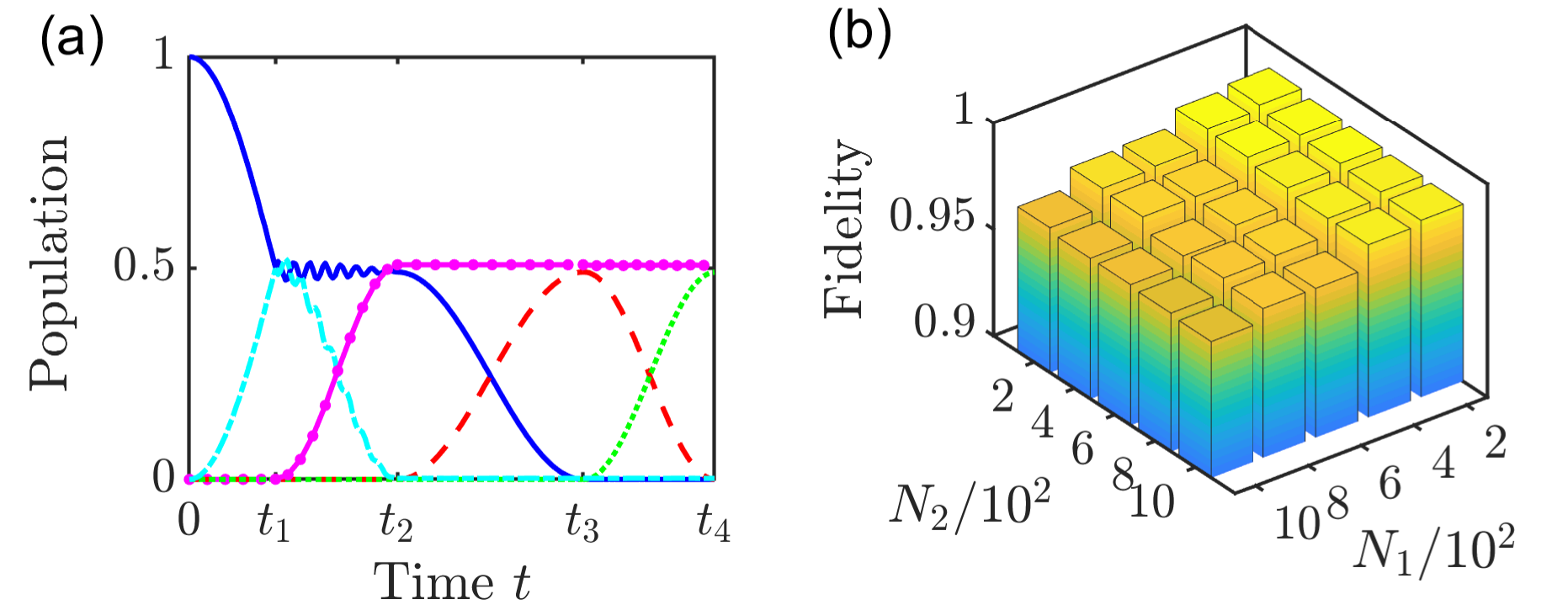}
\caption{Panel (a) shows the populations of the magnon NOON states from
$\ket{0,0}$ versus time $t$. Evolution states $\ket{0,0}$ (solid cyan line), $(\ket{0,0}-i\ket{1,0})/\sqrt{2}$ (dotted blue line), $(\ket{0,0}-\ket{2,0})/\sqrt{2}$ (dotted magenta line), $(\ket{2,0}-i\ket{0,1})/\sqrt{2}$ (dotted red line), and $(\ket{2,0}+\ket{0,2})/\sqrt{2}$ (dotted green line).
Pnael (b) shows the fidelity with choosing different total qubit number $N_{1}$ and $N_{2}$ for generating magnon NOON states. The dynamics is governed by the Hamiltonian in Eq. (\ref{{H}^{m}_{int}}). The parameters are chosen as $\eta_{1}=\eta_{2}=0.12$, $\omega_{c}/2\pi = 2.2 $ GHz, $g_ {1}=0.30 \omega_{c}$, $g_{2}=0.49 \omega_{c}$, $\varepsilon_{1}=\varepsilon_{2}= 0.01 \omega_{c}$, $\Delta_{1} = 2.78 \omega_{c}$, $\Delta_2 = 4.94 \omega_{c}$.}\label{fig_07}
\end{figure} 
So far, the model Hamiltonian in the thermodynamic limit has been simplified as a controlled selective interaction under frequency modulation. Therefore, the applications of selective interactions can also be actualized under the thermodynamic limit.

For the sake of brevity, we focus on a system with large number of qubits shown in Fig. \ref{fig_06}, where the initial state $\ket{\psi(0)}=\ket{0, 0}$ with the total qubit number $N_{1} = N_{2} = 200$. Tuning the driving frequency $\omega=\Delta_{1}({0,0})$, the transition of the quantum states between $\ket{0,0}$ and $\ket{1,0}$ can be achieved with a period of time $T_{1} = 2\pi/[G^{(0)}_{m1}(0,0.12)] \approx 0.1123 {\rm \mu s}$. When the driving frequency is set to $\omega=\Delta_{2}({0,0})$, we can derive the evolution state $\ket{\psi(t)}=\cos{\left[G^{(0)}_{m2}(0,0.12)t\right]}\ket{0,0}
-i\sin{\left[G^{(0)}_{m2}(0,0,12)t\right]}\ket{0,1}$ with a period of time $T_{2} = 2\pi/[G^{(0)}_{m2}(0,0.12)] \approx 0.1210 {\rm \mu s}$. Similarly, supposing that the initial state is prepared as $\ket{\psi(0)}=\ket{1,0}$, we obtain the transition of the quantum states between $\ket{1,0}$ and $\ket{2,0}$ by tuning the driving frequency $\omega=\Delta_{1}({1,0})$ with a period of time $T_{1} = 2\pi/[G^{(0)}_{m1}(1,0.12)] \approx 0.0794 {\rm \mu s}$. Taking into account $\ket{\psi(0)}=\ket{0,1}$ as the initial state, the transition of the quantum state between $\ket{0,1}$ and $\ket{0,2}$ can be obtained with a period of time $T_{2} = 2\pi/[G^{(0)}_{m2}(1,0.12)]  \approx 0.0856 {\rm \mu s}$ by adjusting the driving frequency $\omega=\Delta_{2}({0,1})$.

In the following contents, we will show how to prepare the magnon NOON states with selective resonance interaction. This is clearly illustrated in Fig. \ref{fig_07} and detailed data in Table \ref{table_II}.
If the magnons are not initially distributed throughout the system, the initial state can be expressed as $\ket{\psi(0)}=\ket{0,0}$. By tuning the driving frequency $\omega=\Delta_{1}({0,0})$, the evolved state is $\ket{\psi(t_{1})}=(\ket{0,0}-i\ket{1,0})/\sqrt{2}$ after a duration of $t_{1}={\pi}/{4G^{(0)}_{m1}} (0, 0.12)\approx 0.0140{\mu s}$.
Again adjusting the driving frequency $\omega=\Delta_{1}({1,0})$, the evolved state can be expressed as $\ket{\psi(t_{2})}=(\ket{0,0}-\ket{2,0})/\sqrt{2}$ at the time $t_{2}=t_{1}+{\pi}/{2G^{(0)}_{m1}(0,0.12)}\approx 0.0338{\mu s}$.
The next step involves tuning the driving frequency to $\omega=\Delta_{2}({0,0})$, which results in the state $\ket{\psi(t_{3})}=(\ket{2,0}-i\ket{0,1})/\sqrt{2}$ after the time $t_{3}=t_{2}+{\pi}/{2G^{(0)}_{m2}(0,0.12)}\approx 0.0640{\mu s}$.
Ultimately, when the driving frequency $\omega=\Delta_{2}({0,1})$ is satisfied, the quantum state becomes
$\ket{\psi(t_{4})}=(\ket{2,0}+\ket{0,2})/\sqrt{2}$ at the moment $t_{4}=t_{3}+{\pi}/{2G^{(0)}_{m2}(1,0.12)}\approx 0.0854{\mu s}$. In addition, we have applied the Gaussian soft control to the system, and the fidelity of the target quantum state can be found through Table \ref{table_II}. To further investigate the generation of the magnon NOON state \cite{doi:10.1080/00107510802091298,PhysRevA.101.033809}, we conduct simulations to assess the fidelity for larger total qubit numbers $N_{1}$ and $N_{2}$. On the right panel of Fig. \ref{fig_07}, it is evident to hold high fidelity as the two total qubit numbers increase. Notably, when $N_{1} = N_{2}= 1 \times 10^{3}$, the fidelity value can reach $F=0.9644$.
\begin{figure}[tbp]
\centering
\includegraphics[angle=0,width=0.48\textwidth]{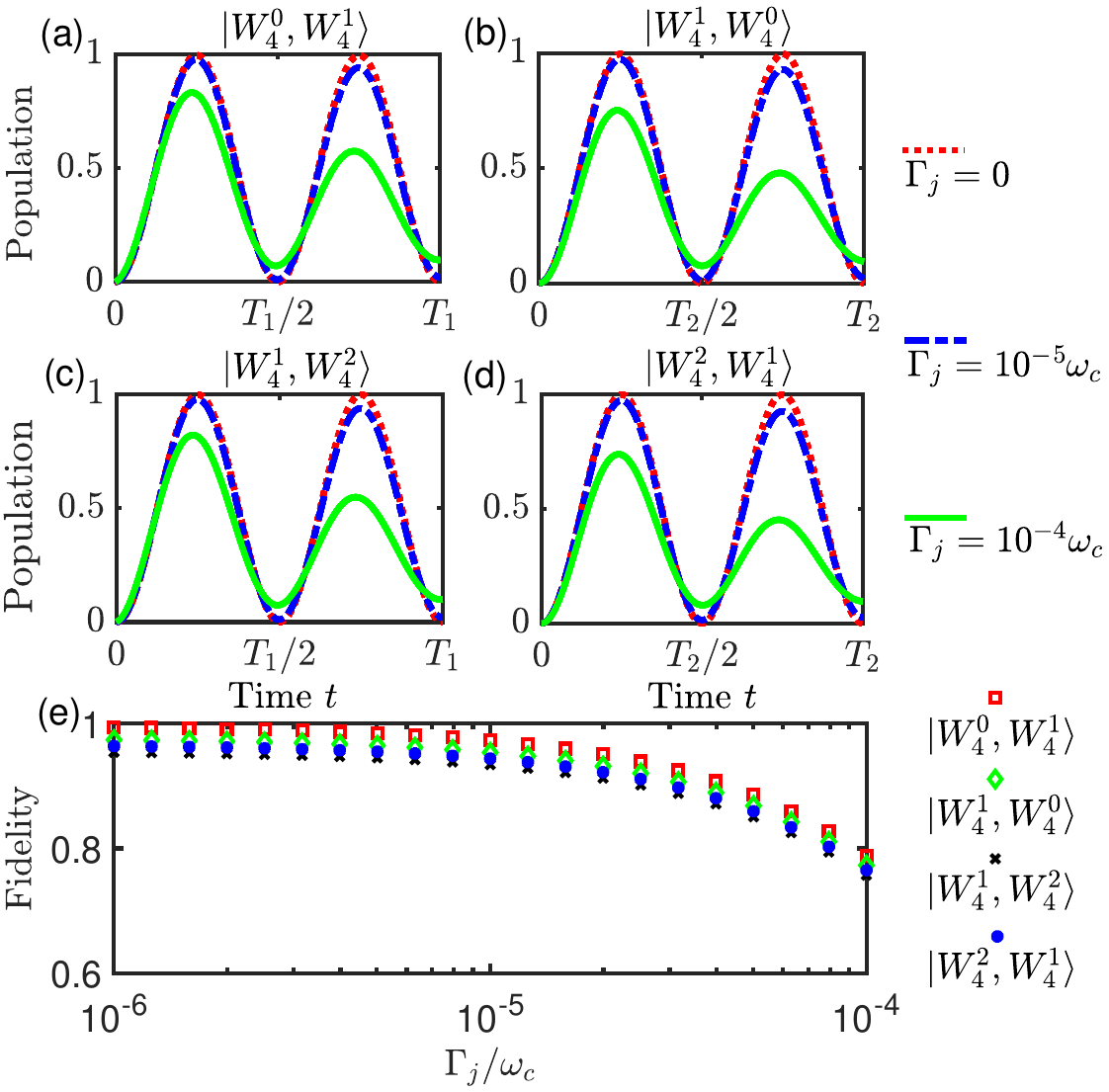}
\caption{Panels (a)-(d) show the dynamic evolution of the target state $\ket{W_{4}^{0},W_{4}^{1}}$, $\ket{W_{4}^{1},W_{4}^{0}}$, $\ket{W_{4}^{2},W_{4}^{1}}$, and $\ket{W_{4}^{1},W_{4}^{2}}$ starting from the initial state $\ket{W_{4}^{1},W_{4}^{1}}$ under various dissipative processes. Panel (e) shows the fidelity of target state evolving half a period under different dissipation conditions. The dynamics is governed by the Hamiltonian in Eq. (\ref{H_Gamma}). The parameter $\kappa=10^{-5}\omega_{c}$ and other parameters are consistent as shown in Fig. \ref{fig_02}.}
\label{fig_08}
\end{figure}
The feasibility of the NOON state has been established by us. Strictly speaking, the effective $G^{(0)}_{mj} \propto \epsilon_{j}\sqrt{N_{j}}$, as indicated by the effective Hamiltonian in Eq. (\ref{Heff(m)}). When the number of atoms becomes sizable, a significant coupling strength emerges, and validity of the effective Hamiltonian should be examined. To avoid undesired transitions, appropriate number of qubits $N_{j}$ and $\epsilon_{j}$ should be considered, ensuring that the high-frequency vibration term can be safely disregarded. By doing so, we can achieve a series of desired entangled states with high fidelity.
\section{Decoherence and experimental feasibility}
\label{SecV}
In this section, we first provide a detailed analysis when the system suffers from decoherence, considering scenarios with both a few and a large number of qubits. Subsequently, we present experimentally feasible parameters for implementing our proposed scheme.
\subsection{Decoherence of the dynamic system}
In order to better discuss the scheme we have proposed, we now describe the system dynamics by considering the dissipation channels taken into account. Here, we employ a dressed-state master equation by considering the standard master equation breaks down in the ultra-strong coupling regime \cite{PhysRevA.84.043832,PhysRevA.98.053834}. With the Born-Markov approximation and assuming each qubit ensemble suffering collective dissipation \cite{PhysRevB.108.054302,PhysRevA.101.053818,PhysRevA.95.063824}, the dressed-state master equation at zero temperature reads
\begin{small}
\begin{equation}\label{H_Gamma}
    \frac{d\hat{\rho}}{dt}=-i[\hat{H},\hat{\rho}]+\sum_{m,n>m}\left(\Gamma_{q_{1}}^{nm}+\Gamma_{q_{2}}^{nm}+\Gamma_{c}^{nm}\right)\mathcal{D}[\ket{m}\bra{n}]\hat{\rho},
\end{equation}
\end{small}
where the Lindblad operator $\mathcal{D}[\hat{\mathcal{O}}]\hat{\rho}=2\hat{\mathcal{O}}\hat{\rho}\hat{\mathcal{O}^{\dag}}-\hat{\rho}\hat{\mathcal{O}^{\dag}}\hat{\mathcal{O}}-\hat{\mathcal{O}^{\dag}}\hat{\mathcal{O}}\hat{\rho}$. $\ket{m}$ and $\ket{n}$ are eigenstates of the undriven (i.e., $A_{j}=0$) Hamiltonian given by Eq. (\ref{eq_01}) with eigenvalues $\omega_{m}$ and $\omega_{n}$. Here the eigenstates are labeled by an increasing order (i.e., $\omega_{n}>\omega_{m}$ for $n>m$ ). The qubit ensembles and resonator decay rates can be written as 
\begin{eqnarray*}\label{dressed_master_eq}
&&\Gamma_{q_{j}}^{nm}=\frac{\Gamma_{j}}{N_{j}}\frac{\Delta_{nm}}{\omega_{q_{j}}}|\bra{m}\hat{S}_{j}^{-}+\hat{S}_{j}^{+}\ket{n}|^{2},\\
&&\Gamma_{c}^{nm}=\kappa\frac{\Delta_{nm}}{\omega_{c}}|\bra{m}\hat{a}+\hat{a}^{\dag}\ket{n}|^{2},
\end{eqnarray*}
where $\kappa$ and $\Gamma_{j}$ are the bare loss rates for the resonator and the $j$-th qubit ensemble, and $\Delta_{nm}=\omega_{n}-\omega_{m}$.
\begin{table}[tbp]
\caption{The fidelity of the corresponding target states is obtained with the initial state $\ket{W_{4}^{1},W_{4}^{1}}$ under different dissipation rate. The dynamics is governed by the Hamiltonian in Eq. (\ref{H_Gamma}).}
\label{table_III}
\setlength{\tabcolsep}{3.5mm}
\begin{tabular}{cccc}
\hline
\hline
 & $\Gamma_{j}=0$ & $\Gamma_{j}=10^{-5}\omega_c$ & $\Gamma_{j}=10^{-4}\omega_c$ \\ \hline
$\ket{W_{4}^{0},W_{4}^{1}}$ & 0.9963   & 0.9778  & 0.8262      \\ 
$\ket{W_{4}^{1},W_{4}^{0}}$ & 0.9997   & 0.9749  & 0.7774      \\ 
$\ket{W_{4}^{2},W_{4}^{1}}$ & 0.9959   & 0.9735  & 0.7891      \\ 
$\ket{W_{4}^{1},W_{4}^{2}}$ & 0.9997   & 0.9794  & 0.8143      \\ 
\hline
\hline
\end{tabular}
\end{table}
To evaluate the robustness of the proposal, we introduce the definition of the fidelity as: $F\left[\ket{\psi_{T}},\rho(t)\right]=\bra{\psi_{T}}\rho(t)\ket{\psi_{T}}$ with $\rho(t)$ being the evolution density matrix. In Figs. \ref{fig_08} (a)-(d), we depict the temporal dynamics of the selective transition under the dissipation. We consider an initial state $\ket{W_{4}^{1}, W_{4}^{1}}$ and assume $\Gamma_{1} =\Gamma_{2}$ for convenience. In the ideal case when $\Gamma_{j} = 0$, coherent oscillations of the initial state and target state occur, indicating that photon loss has minimal impact on the system. However, as the rate of dissipation increases, the coherence oscillations are destroyed by disspation, and the fidelity of the target states is decreased. Fig. \ref{fig_08} (e) further shows the fidelity of target state evolving half a period under different dissipation rates. Table \ref{table_III} shows that with an increase in the dissipation, the accuracy of the target state diminishes. Nonetheless, the results show that the selective transition can be achieved under dissipative processes, and the Dicke states can be created within the coherent time.

\begin{figure}[tbp]
\centering
\includegraphics[angle=0,width=0.48\textwidth]{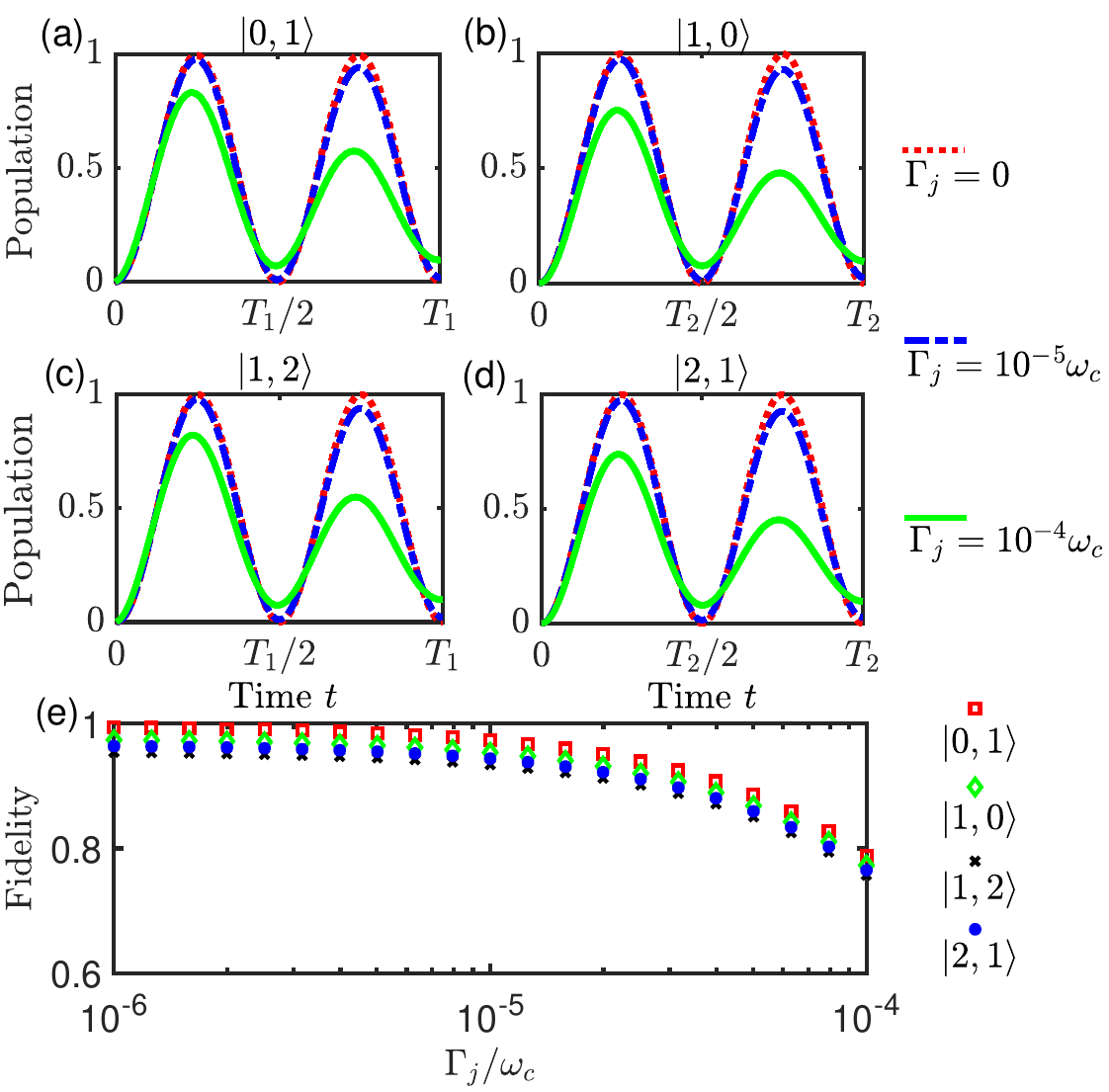}
\caption{Panels (a)-(d) show the dynamic evolution process of the target state $\ket{0, 1}$, $\ket{1, 0}$, $\ket{1, 2}$, $\ket{2, 1}$ starting from the initial state $\ket{1, 1}$ under dissipative processes. Panel (e) shows the fidelity of the final state and target state. The dynamics is governed by the Hamiltonian in Eq. (\ref{H_Gamma2}). The parameter $\kappa=10^{-5}\omega_c$ and other selected parameters are consistent as shown in Fig. \ref{fig_06}.}
\label{fig_09}
\end{figure}

In the thermodynamic limit $N_{j}\gg 1$, we show the resulting coupling scales as $\epsilon_{j}\sqrt{N_{j}}$. However, we emphasize that this previous approximation can be highly effective even in the more practical scenario of a finite number of atoms, specifically when the number of excitations in the system is significantly less than the total number $N_{j}$ of atoms in the ensemble. Hence, when $\Gamma_{1}=\Gamma_{2}$ and $N_1=N_2$, the dressed-state master equation can be obtained by Holstein-Primakoff transformation \cite{PhysRevA.101.053818}. Consequently, the dressed-state master equation after Holstein-Primakoff transformation can be recast as
\begin{small}
\begin{equation}\label{H_Gamma2}
    \frac{d\hat{\tilde{\rho}}}{dt}=-i[\hat{\tilde{H}},\hat{\tilde{\rho}}]+\sum_{m,n>m}\left(\tilde{\Gamma}_{q_{1}}^{nm}+\tilde{\Gamma}_{q_{2}}^{nm}+\tilde{\Gamma}_{c}^{nm}\right)\mathcal{D}[\ket{\tilde{m}}\bra{\tilde{n}}]\hat{\tilde{\rho}}.
\end{equation}
\end{small}
Here the tilde indicates symbols after Holstein-Primakoff transformation, $\ket{\tilde{m}}$ and $\ket{\tilde{n}}$ are eigenstates of the undriven Hamiltonian given by Eq. (\ref{eq_01}) after Holstein-Primakoff transformation with eigenvalues $\tilde{\omega}_{m}$ and $\tilde{\omega}_{n}$. The resonator and qubit ensembles decay rates associated with given system can be written as 
\begin{eqnarray*}\label{dressed_master_eq_{2}}
&&\tilde{\Gamma}_{q_{j}}^{nm}=\Gamma_{j}\frac{\tilde{\Delta}_{nm}}{\omega_{q_{j}}}|\bra{\tilde{m}}\hat{m}_{j}+\hat{m}_{j}^{\dag}\ket{\tilde{n}}|^{2},\\
&&\tilde{\Gamma}_{c}^{nm}=\kappa\frac{\Delta_{nm}}{\omega_{c}}|\bra{\tilde{m}}\hat{a}+\hat{a}^{\dag}\ket{\tilde{n}}|^{2},
\end{eqnarray*}
where $\tilde{\Delta}_{nm}=\tilde{\omega}_{n}-\tilde{\omega}_{m}$.
In Fig. \ref{fig_09}, we show the evolution of the target state under the different dissipative rates, revealing a similar trend to the previous scenario with a noticeable decrease in the fidelity of the target state as the dissipation rate increases. Additionally, we provide the corresponding fidelity values in the presence of decoherence, which aligns perfectly with the aforementioned results. These results highlight the high robustness of the entire system under lower levels of dissipation. Furthermore, Table \ref{table_IV} illustrates that as dissipation increases, the accuracy of the target state diminishes, and coherence gradually decreases over time.

\subsection{Experimental consideration and feasibility}
To assess the feasibility of our entangled Dicke states generation scheme in a realistic experiment, we now discuss the relevant achievable experimental parameters. As outlined in the main text, our protocol works in the ultrastrong light-matter coupling regime. For few number of atoms case, our proposal can be realized in circuit QED systems. In such superconducting circuit experiments, qubit and resonator frequencies are usually in the range $\omega_{c, (q_{j})}/2\pi\sim 1-10$ GHz \cite{PhysRevA.95.053824,Yoshihara2017,PhysRevResearch.3.033275}. The coupling strengths range from $g/\omega_{c}\simeq 0.12$ to theoretical limits $g/\omega_{c}\sim 1-3$ for flux qubits \cite{Niemczyk2010,PhysRevLett.108.120501}. In this work, we choose $\omega_{c}/2\pi = 2.2$ GHz, $\omega_{q_{1}}/2\pi = 2.87\omega_{c}\simeq 6.31$ GHz, and $\omega_{q_{2}}/2\pi = 2.87\omega_{c}\simeq 10.87$ GHz. We choose $g_{1}/\omega_{c} = 0.25$ and $g_{2}/\omega_{c} = 0.49$. Thus the physical parameters of scheme are accessible in current technologies. Recent experimental works has also demonstrated that dissipation and dephasing rates in a flux qubit are of the order of $2\pi\times 10$ kHz \cite{RevModPhys.91.025005,PhysRevLett.113.123601,PhysRevB.93.104518}. For transmission-line resonators, quality factors on the order of $10^{6}$ have been realized \cite{PhysRevLett.107.240501}, which indicates that quantum coherence of systems is of the order of $1 - 10$ ms within current experimental capabilities \cite{10.1063/1.3693409}. For the large number of two-level systems, our proposal can also be studied with a hybrid system consisting of qubit ensembles and superconducting resonator \cite{PhysRevA.89.042321}. Each qubit ensemble can be realized using the NV centers in a diamond crystal, and each qubit ensemble is coupled to a superconducting resonator placed in between these two qubit ensembles. The coupling strengths between the resonator and qubit ensembles can be tuned into ultrastrong coupling regimes by either engineering the hybrid structure in advance or tuning the excitation frequencies of qubit ensembles via the external magnetic fields.
\begin{table}[tbp]
\caption{The fidelity of the corresponding target states is obtained with the initial state $\ket{1, 1}$ under different dissipation conditions. The dynamics is governed by the Hamiltonian in Eq. (\ref{H_Gamma2}).}
\label{table_IV}
\setlength{\tabcolsep}{3.5mm}
\begin{tabular}{cccc}
\hline
\hline
 & $\Gamma_{j}=0$ & $\Gamma_{j}=10^{-5}\omega_c$ & $\Gamma_{j}=10^{-4}\omega_c$ \\ \hline
$\ket{0, 1}$ & 0.9953   & 0.9775  & 0.7961       \\ 
$\ket{1, 0}$ & 0.9986   & 0.9745  & 0.7448       \\ 
$\ket{2, 1}$ & 0.9951   & 0.9729  & 0.7525       \\ 
$\ket{1, 2}$ & 0.9984   & 0.9783  & 0.7838       \\ 
\hline
\hline
\end{tabular}
\end{table}

\section{Conclusion}
\label{SecVI}
In conclusion, we have presented a scheme for entangling two Dicke states in a periodic modulated system. The effectiveness of this method can be verified through numerical simulations. For some special cases, the ensemble-ensemble entangled state can be obtained by means of an even-odd cat states projective measurement. Additionally, the Gaussian soft control can be utilized to enhance the fidelity of the target states. In the thermodynamics limit, our protocol can also be applied to attain the selective interactions between two distinct magnon systems by means of the Holstein-Primakoff transformation. Furthermore, it has successfully shown the creation of high fidelity magnon states in a large number of qubits, which establishes a reliable approach for generating magnon NOON state in the thermodynamic limit. Finally, in the presence of decoherence during dynamical evolution, our scheme can achieve highly reliable target entangled states. This provides further evidence for the experimental feasibility of our approach in preparing entangled states with high fidelity.

\section*{Acknowledgments}
The work is supported by the scientific research Project of the Education Department of Jilin Province (Grant Nos. JJKH20241408KJ and JJKH20231293KJ). C.W. is supported by the National Research Foundation, Singapore and A*STAR under its Quantum Engineering Programme (NRF2021-QEP2-02-P03).

\bibliography{ref}

\begin{thebibliography}{88}%
\makeatletter
\providecommand \@ifxundefined [1]{%
 \@ifx{#1\undefined}
}%
\providecommand \@ifnum [1]{%
 \ifnum #1\expandafter \@firstoftwo
 \else \expandafter \@secondoftwo
 \fi
}%
\providecommand \@ifx [1]{%
 \ifx #1\expandafter \@firstoftwo
 \else \expandafter \@secondoftwo
 \fi
}%
\providecommand \natexlab [1]{#1}%
\providecommand \enquote  [1]{``#1''}%
\providecommand \bibnamefont  [1]{#1}%
\providecommand \bibfnamefont [1]{#1}%
\providecommand \citenamefont [1]{#1}%
\providecommand \href@noop [0]{\@secondoftwo}%
\providecommand \href [0]{\begingroup \@sanitize@url \@href}%
\providecommand \@href[1]{\@@startlink{#1}\@@href}%
\providecommand \@@href[1]{\endgroup#1\@@endlink}%
\providecommand \@sanitize@url [0]{\catcode `\\12\catcode `\$12\catcode
  `\&12\catcode `\#12\catcode `\^12\catcode `\_12\catcode `\%12\relax}%
\providecommand \@@startlink[1]{}%
\providecommand \@@endlink[0]{}%
\providecommand \url  [0]{\begingroup\@sanitize@url \@url }%
\providecommand \@url [1]{\endgroup\@href {#1}{\urlprefix }}%
\providecommand \urlprefix  [0]{URL }%
\providecommand \Eprint [0]{\href }%
\providecommand \doibase [0]{https://doi.org/}%
\providecommand \selectlanguage [0]{\@gobble}%
\providecommand \bibinfo  [0]{\@secondoftwo}%
\providecommand \bibfield  [0]{\@secondoftwo}%
\providecommand \translation [1]{[#1]}%
\providecommand \BibitemOpen [0]{}%
\providecommand \bibitemStop [0]{}%
\providecommand \bibitemNoStop [0]{.\EOS\space}%
\providecommand \EOS [0]{\spacefactor3000\relax}%
\providecommand \BibitemShut  [1]{\csname bibitem#1\endcsname}%
\let\auto@bib@innerbib\@empty
\bibitem [{\citenamefont {Silveira}\ and\ \citenamefont
  {Angelo}(2017)}]{2017Silveira}%
  \BibitemOpen
  \bibfield  {author} {\bibinfo {author} {\bibfnamefont {L.~S.}\ \bibnamefont
  {Silveira}}\ and\ \bibinfo {author} {\bibfnamefont {R.~M.}\ \bibnamefont
  {Angelo}},\ }\bibfield  {title} {\bibinfo {title} {Classical-hidden-variable
  description for entanglement dynamics of two-qubit pure states},\ }\href
  {https://doi.org/10.1103/PhysRevA.95.062105} {\bibfield  {journal} {\bibinfo
  {journal} {Phys. Rev. A}\ }\textbf {\bibinfo {volume} {95}},\ \bibinfo
  {pages} {062105} (\bibinfo {year} {2017})}\BibitemShut {NoStop}%
\bibitem [{\citenamefont {Zhu}\ and\ \citenamefont {Hayashi}(2019)}]{2019Zhu}%
  \BibitemOpen
  \bibfield  {author} {\bibinfo {author} {\bibfnamefont {H.}~\bibnamefont
  {Zhu}}\ and\ \bibinfo {author} {\bibfnamefont {M.}~\bibnamefont {Hayashi}},\
  }\bibfield  {title} {\bibinfo {title} {General framework for verifying pure
  quantum states in the adversarial scenario},\ }\href
  {https://doi.org/10.1103/PhysRevA.100.062335} {\bibfield  {journal} {\bibinfo
   {journal} {Phys. Rev. A}\ }\textbf {\bibinfo {volume} {100}},\ \bibinfo
  {pages} {062335} (\bibinfo {year} {2019})}\BibitemShut {NoStop}%
\bibitem [{\citenamefont {Qiao}\ \emph {et~al.}(2020)\citenamefont {Qiao},
  \citenamefont {Li}, \citenamefont {Dong}, \citenamefont {Chen}, \citenamefont
  {Zhou},\ and\ \citenamefont {Li}}]{2020Qiao}%
  \BibitemOpen
  \bibfield  {author} {\bibinfo {author} {\bibfnamefont {Y.-F.}\ \bibnamefont
  {Qiao}}, \bibinfo {author} {\bibfnamefont {H.-Z.}\ \bibnamefont {Li}},
  \bibinfo {author} {\bibfnamefont {X.-L.}\ \bibnamefont {Dong}}, \bibinfo
  {author} {\bibfnamefont {J.-Q.}\ \bibnamefont {Chen}}, \bibinfo {author}
  {\bibfnamefont {Y.}~\bibnamefont {Zhou}},\ and\ \bibinfo {author}
  {\bibfnamefont {P.-B.}\ \bibnamefont {Li}},\ }\bibfield  {title} {\bibinfo
  {title} {Phononic-waveguide-assisted steady-state entanglement of
  silicon-vacancy centers},\ }\href
  {https://doi.org/10.1103/PhysRevA.101.042313} {\bibfield  {journal} {\bibinfo
   {journal} {Phys. Rev. A}\ }\textbf {\bibinfo {volume} {101}},\ \bibinfo
  {pages} {042313} (\bibinfo {year} {2020})}\BibitemShut {NoStop}%
\bibitem [{\citenamefont {Contreras-Tejada}\ \emph {et~al.}(2022)\citenamefont
  {Contreras-Tejada}, \citenamefont {Palazuelos},\ and\ \citenamefont
  {Vicente}}]{2022Contreras}%
  \BibitemOpen
  \bibfield  {author} {\bibinfo {author} {\bibfnamefont {P.}~\bibnamefont
  {Contreras-Tejada}}, \bibinfo {author} {\bibfnamefont {C.}~\bibnamefont
  {Palazuelos}},\ and\ \bibinfo {author} {\bibfnamefont {J.~I.~d.}\
  \bibnamefont {Vicente}},\ }\bibfield  {title} {\bibinfo {title} {Asymptotic
  survival of genuine multipartite entanglement in noisy quantum networks
  depends on the topology},\ }\href
  {https://doi.org/10.1103/PhysRevLett.128.220501} {\bibfield  {journal}
  {\bibinfo  {journal} {Phys. Rev. Lett.}\ }\textbf {\bibinfo {volume} {128}},\
  \bibinfo {pages} {220501} (\bibinfo {year} {2022})}\BibitemShut {NoStop}%
\bibitem [{\citenamefont {Morvan}\ \emph {et~al.}(2022)\citenamefont {Morvan},
  \citenamefont {Chen}, \citenamefont {Larson}, \citenamefont {Santiago},\ and\
  \citenamefont {Siddiqi}}]{2022Morvan}%
  \BibitemOpen
  \bibfield  {author} {\bibinfo {author} {\bibfnamefont {A.}~\bibnamefont
  {Morvan}}, \bibinfo {author} {\bibfnamefont {L.}~\bibnamefont {Chen}},
  \bibinfo {author} {\bibfnamefont {J.~M.}\ \bibnamefont {Larson}}, \bibinfo
  {author} {\bibfnamefont {D.~I.}\ \bibnamefont {Santiago}},\ and\ \bibinfo
  {author} {\bibfnamefont {I.}~\bibnamefont {Siddiqi}},\ }\bibfield  {title}
  {\bibinfo {title} {Optimizing frequency allocation for fixed-frequency
  superconducting quantum processors},\ }\href
  {https://doi.org/10.1103/PhysRevResearch.4.023079} {\bibfield  {journal}
  {\bibinfo  {journal} {Phys. Rev. Res.}\ }\textbf {\bibinfo {volume} {4}},\
  \bibinfo {pages} {023079} (\bibinfo {year} {2022})}\BibitemShut {NoStop}%
\bibitem [{\citenamefont {Eddins}\ \emph {et~al.}(2022)\citenamefont {Eddins},
  \citenamefont {Motta}, \citenamefont {Gujarati}, \citenamefont {Bravyi},
  \citenamefont {Mezzacapo}, \citenamefont {Hadfield},\ and\ \citenamefont
  {Sheldon}}]{2022Eddins}%
  \BibitemOpen
  \bibfield  {author} {\bibinfo {author} {\bibfnamefont {A.}~\bibnamefont
  {Eddins}}, \bibinfo {author} {\bibfnamefont {M.}~\bibnamefont {Motta}},
  \bibinfo {author} {\bibfnamefont {T.~P.}\ \bibnamefont {Gujarati}}, \bibinfo
  {author} {\bibfnamefont {S.}~\bibnamefont {Bravyi}}, \bibinfo {author}
  {\bibfnamefont {A.}~\bibnamefont {Mezzacapo}}, \bibinfo {author}
  {\bibfnamefont {C.}~\bibnamefont {Hadfield}},\ and\ \bibinfo {author}
  {\bibfnamefont {S.}~\bibnamefont {Sheldon}},\ }\bibfield  {title} {\bibinfo
  {title} {Doubling the size of quantum simulators by entanglement forging},\
  }\href {https://doi.org/10.1103/PRXQuantum.3.010309} {\bibfield  {journal}
  {\bibinfo  {journal} {PRX Quantum}\ }\textbf {\bibinfo {volume} {3}},\
  \bibinfo {pages} {010309} (\bibinfo {year} {2022})}\BibitemShut {NoStop}%
\bibitem [{\citenamefont {Alsina}\ and\ \citenamefont
  {Latorre}(2016)}]{2016Alsina}%
  \BibitemOpen
  \bibfield  {author} {\bibinfo {author} {\bibfnamefont {D.}~\bibnamefont
  {Alsina}}\ and\ \bibinfo {author} {\bibfnamefont {J.~I.}\ \bibnamefont
  {Latorre}},\ }\bibfield  {title} {\bibinfo {title} {Experimental test of
  mermin inequalities on a five-qubit quantum computer},\ }\href
  {https://doi.org/10.1103/PhysRevA.94.012314} {\bibfield  {journal} {\bibinfo
  {journal} {Phys. Rev. A}\ }\textbf {\bibinfo {volume} {94}},\ \bibinfo
  {pages} {012314} (\bibinfo {year} {2016})}\BibitemShut {NoStop}%
\bibitem [{\citenamefont {Su}\ \emph {et~al.}(2017)\citenamefont {Su},
  \citenamefont {Tang}, \citenamefont {Wu}, \citenamefont {Cai}, \citenamefont
  {Yang}, \citenamefont {Li}, \citenamefont {Liu}, \citenamefont {Lu},
  \citenamefont {\.{Z}ukowski},\ and\ \citenamefont {Pan}}]{2017Su}%
  \BibitemOpen
  \bibfield  {author} {\bibinfo {author} {\bibfnamefont {Z.-E.}\ \bibnamefont
  {Su}}, \bibinfo {author} {\bibfnamefont {W.-D.}\ \bibnamefont {Tang}},
  \bibinfo {author} {\bibfnamefont {D.}~\bibnamefont {Wu}}, \bibinfo {author}
  {\bibfnamefont {X.-D.}\ \bibnamefont {Cai}}, \bibinfo {author} {\bibfnamefont
  {T.}~\bibnamefont {Yang}}, \bibinfo {author} {\bibfnamefont {L.}~\bibnamefont
  {Li}}, \bibinfo {author} {\bibfnamefont {N.-L.}\ \bibnamefont {Liu}},
  \bibinfo {author} {\bibfnamefont {C.-Y.}\ \bibnamefont {Lu}}, \bibinfo
  {author} {\bibfnamefont {M.}~\bibnamefont {\.{Z}ukowski}},\ and\ \bibinfo
  {author} {\bibfnamefont {J.-W.}\ \bibnamefont {Pan}},\ }\bibfield  {title}
  {\bibinfo {title} {Experimental test of the irreducible four-qubit
  greenberger-horne-zeilinger paradox},\ }\href
  {https://doi.org/10.1103/PhysRevA.95.030103} {\bibfield  {journal} {\bibinfo
  {journal} {Phys. Rev. A}\ }\textbf {\bibinfo {volume} {95}},\ \bibinfo
  {pages} {030103} (\bibinfo {year} {2017})}\BibitemShut {NoStop}%
\bibitem [{\citenamefont {Luo}\ and\ \citenamefont {Fei}(2021)}]{2021Alsina}%
  \BibitemOpen
  \bibfield  {author} {\bibinfo {author} {\bibfnamefont {M.-X.}\ \bibnamefont
  {Luo}}\ and\ \bibinfo {author} {\bibfnamefont {S.-M.}\ \bibnamefont {Fei}},\
  }\bibfield  {title} {\bibinfo {title} {Robust multipartite entanglement
  without entanglement breaking},\ }\href
  {https://doi.org/10.1103/PhysRevResearch.3.043120} {\bibfield  {journal}
  {\bibinfo  {journal} {Phys. Rev. Res.}\ }\textbf {\bibinfo {volume} {3}},\
  \bibinfo {pages} {043120} (\bibinfo {year} {2021})}\BibitemShut {NoStop}%
\bibitem [{\citenamefont {Orito}\ \emph {et~al.}(2022)\citenamefont {Orito},
  \citenamefont {Kuno},\ and\ \citenamefont {Ichinose}}]{2022Orito}%
  \BibitemOpen
  \bibfield  {author} {\bibinfo {author} {\bibfnamefont {T.}~\bibnamefont
  {Orito}}, \bibinfo {author} {\bibfnamefont {Y.}~\bibnamefont {Kuno}},\ and\
  \bibinfo {author} {\bibfnamefont {I.}~\bibnamefont {Ichinose}},\ }\bibfield
  {title} {\bibinfo {title} {Quantum information spreading in random spin
  chains with topological order},\ }\href
  {https://doi.org/10.1103/PhysRevB.106.104204} {\bibfield  {journal} {\bibinfo
   {journal} {Phys. Rev. B}\ }\textbf {\bibinfo {volume} {106}},\ \bibinfo
  {pages} {104204} (\bibinfo {year} {2022})}\BibitemShut {NoStop}%
\bibitem [{\citenamefont {Gonz\'{a}lez-Guti\'{e}rrez}\ and\ \citenamefont
  {Torres}(2019)}]{2019Gonz}%
  \BibitemOpen
  \bibfield  {author} {\bibinfo {author} {\bibfnamefont {C.~A.}\ \bibnamefont
  {Gonz\'{a}lez-Guti\'{e}rrez}}\ and\ \bibinfo {author} {\bibfnamefont {J.~M.}\
  \bibnamefont {Torres}},\ }\bibfield  {title} {\bibinfo {title} {Atomic bell
  measurement via two-photon interactions},\ }\href
  {https://doi.org/10.1103/PhysRevA.99.023854} {\bibfield  {journal} {\bibinfo
  {journal} {Phys. Rev. A}\ }\textbf {\bibinfo {volume} {99}},\ \bibinfo
  {pages} {023854} (\bibinfo {year} {2019})}\BibitemShut {NoStop}%
\bibitem [{\citenamefont {Wong-Campos}\ \emph {et~al.}(2017)\citenamefont
  {Wong-Campos}, \citenamefont {Moses}, \citenamefont {Johnson},\ and\
  \citenamefont {Monroe}}]{2017Wong}%
  \BibitemOpen
  \bibfield  {author} {\bibinfo {author} {\bibfnamefont {J.~D.}\ \bibnamefont
  {Wong-Campos}}, \bibinfo {author} {\bibfnamefont {S.~A.}\ \bibnamefont
  {Moses}}, \bibinfo {author} {\bibfnamefont {K.~G.}\ \bibnamefont {Johnson}},\
  and\ \bibinfo {author} {\bibfnamefont {C.}~\bibnamefont {Monroe}},\
  }\bibfield  {title} {\bibinfo {title} {Demonstration of two-atom entanglement
  with ultrafast optical pulses},\ }\href
  {https://doi.org/10.1103/PhysRevLett.119.230501} {\bibfield  {journal}
  {\bibinfo  {journal} {Phys. Rev. Lett.}\ }\textbf {\bibinfo {volume} {119}},\
  \bibinfo {pages} {230501} (\bibinfo {year} {2017})}\BibitemShut {NoStop}%
\bibitem [{\citenamefont {Seidelmann}\ \emph {et~al.}(2022)\citenamefont
  {Seidelmann}, \citenamefont {Schimpf}, \citenamefont {Bracht}, \citenamefont
  {Cosacchi}, \citenamefont {Vagov}, \citenamefont {Rastelli}, \citenamefont
  {Reiter},\ and\ \citenamefont {Axt}}]{2022Seidelmann}%
  \BibitemOpen
  \bibfield  {author} {\bibinfo {author} {\bibfnamefont {T.}~\bibnamefont
  {Seidelmann}}, \bibinfo {author} {\bibfnamefont {C.}~\bibnamefont {Schimpf}},
  \bibinfo {author} {\bibfnamefont {T.~K.}\ \bibnamefont {Bracht}}, \bibinfo
  {author} {\bibfnamefont {M.}~\bibnamefont {Cosacchi}}, \bibinfo {author}
  {\bibfnamefont {A.}~\bibnamefont {Vagov}}, \bibinfo {author} {\bibfnamefont
  {A.}~\bibnamefont {Rastelli}}, \bibinfo {author} {\bibfnamefont {D.~E.}\
  \bibnamefont {Reiter}},\ and\ \bibinfo {author} {\bibfnamefont {V.~M.}\
  \bibnamefont {Axt}},\ }\bibfield  {title} {\bibinfo {title} {Two-photon
  excitation sets limit to entangled photon pair generation from quantum
  emitters},\ }\href {https://doi.org/10.1103/PhysRevLett.129.193604}
  {\bibfield  {journal} {\bibinfo  {journal} {Phys. Rev. Lett.}\ }\textbf
  {\bibinfo {volume} {129}},\ \bibinfo {pages} {193604} (\bibinfo {year}
  {2022})}\BibitemShut {NoStop}%
\bibitem [{\citenamefont {Chaisson}\ \emph {et~al.}(2022)\citenamefont
  {Chaisson}, \citenamefont {Poitras}, \citenamefont {Richard}, \citenamefont
  {Castonguay-Page}, \citenamefont {Glinel}, \citenamefont {Landry},\ and\
  \citenamefont {Hamel}}]{2022Chaisson}%
  \BibitemOpen
  \bibfield  {author} {\bibinfo {author} {\bibfnamefont {Z.~M.~E.}\
  \bibnamefont {Chaisson}}, \bibinfo {author} {\bibfnamefont {P.~F.}\
  \bibnamefont {Poitras}}, \bibinfo {author} {\bibfnamefont {M.}~\bibnamefont
  {Richard}}, \bibinfo {author} {\bibfnamefont {Y.}~\bibnamefont
  {Castonguay-Page}}, \bibinfo {author} {\bibfnamefont {P.-H.}\ \bibnamefont
  {Glinel}}, \bibinfo {author} {\bibfnamefont {V.}~\bibnamefont {Landry}},\
  and\ \bibinfo {author} {\bibfnamefont {D.~R.}\ \bibnamefont {Hamel}},\
  }\bibfield  {title} {\bibinfo {title} {Phase-stable source of high-quality
  three-photon polarization entanglement by cascaded down-conversion},\ }\href
  {https://doi.org/10.1103/PhysRevA.105.063705} {\bibfield  {journal} {\bibinfo
   {journal} {Phys. Rev. A}\ }\textbf {\bibinfo {volume} {105}},\ \bibinfo
  {pages} {063705} (\bibinfo {year} {2022})}\BibitemShut {NoStop}%
\bibitem [{\citenamefont {Wang}\ \emph {et~al.}(2016)\citenamefont {Wang},
  \citenamefont {Chen}, \citenamefont {Li}, \citenamefont {Huang},
  \citenamefont {Liu}, \citenamefont {Chen}, \citenamefont {Luo}, \citenamefont
  {Su}, \citenamefont {Wu}, \citenamefont {Li}, \citenamefont {Lu},
  \citenamefont {Hu}, \citenamefont {Jiang}, \citenamefont {Peng},
  \citenamefont {Li}, \citenamefont {Liu}, \citenamefont {Chen}, \citenamefont
  {Lu},\ and\ \citenamefont {Pan}}]{2016Wangxilin}%
  \BibitemOpen
  \bibfield  {author} {\bibinfo {author} {\bibfnamefont {X.-L.}\ \bibnamefont
  {Wang}}, \bibinfo {author} {\bibfnamefont {L.-K.}\ \bibnamefont {Chen}},
  \bibinfo {author} {\bibfnamefont {W.}~\bibnamefont {Li}}, \bibinfo {author}
  {\bibfnamefont {H.-L.}\ \bibnamefont {Huang}}, \bibinfo {author}
  {\bibfnamefont {C.}~\bibnamefont {Liu}}, \bibinfo {author} {\bibfnamefont
  {C.}~\bibnamefont {Chen}}, \bibinfo {author} {\bibfnamefont {Y.-H.}\
  \bibnamefont {Luo}}, \bibinfo {author} {\bibfnamefont {Z.-E.}\ \bibnamefont
  {Su}}, \bibinfo {author} {\bibfnamefont {D.}~\bibnamefont {Wu}}, \bibinfo
  {author} {\bibfnamefont {Z.-D.}\ \bibnamefont {Li}}, \bibinfo {author}
  {\bibfnamefont {H.}~\bibnamefont {Lu}}, \bibinfo {author} {\bibfnamefont
  {Y.}~\bibnamefont {Hu}}, \bibinfo {author} {\bibfnamefont {X.}~\bibnamefont
  {Jiang}}, \bibinfo {author} {\bibfnamefont {C.-Z.}\ \bibnamefont {Peng}},
  \bibinfo {author} {\bibfnamefont {L.}~\bibnamefont {Li}}, \bibinfo {author}
  {\bibfnamefont {N.-L.}\ \bibnamefont {Liu}}, \bibinfo {author} {\bibfnamefont
  {Y.-A.}\ \bibnamefont {Chen}}, \bibinfo {author} {\bibfnamefont {C.-Y.}\
  \bibnamefont {Lu}},\ and\ \bibinfo {author} {\bibfnamefont {J.-W.}\
  \bibnamefont {Pan}},\ }\bibfield  {title} {\bibinfo {title} {Experimental
  ten-photon entanglement},\ }\href
  {https://doi.org/10.1103/PhysRevLett.117.210502} {\bibfield  {journal}
  {\bibinfo  {journal} {Phys. Rev. Lett.}\ }\textbf {\bibinfo {volume} {117}},\
  \bibinfo {pages} {210502} (\bibinfo {year} {2016})}\BibitemShut {NoStop}%
\bibitem [{\citenamefont {Tzemos}\ and\ \citenamefont
  {Contopoulos}(2021)}]{2021Tzemos}%
  \BibitemOpen
  \bibfield  {author} {\bibinfo {author} {\bibfnamefont {A.~C.}\ \bibnamefont
  {Tzemos}}\ and\ \bibinfo {author} {\bibfnamefont {G.}~\bibnamefont
  {Contopoulos}},\ }\bibfield  {title} {\bibinfo {title} {Ergodicity and
  born’s rule in an entangled three-qubit bohmian system},\ }\href
  {https://doi.org/10.1103/PhysRevE.104.054211} {\bibfield  {journal} {\bibinfo
   {journal} {Phys. Rev. E}\ }\textbf {\bibinfo {volume} {104}},\ \bibinfo
  {pages} {054211} (\bibinfo {year} {2021})}\BibitemShut {NoStop}%
\bibitem [{\citenamefont {Kounalakis}\ \emph {et~al.}(2022)\citenamefont
  {Kounalakis}, \citenamefont {Bauer},\ and\ \citenamefont
  {Blanter}}]{2022Kounalakis}%
  \BibitemOpen
  \bibfield  {author} {\bibinfo {author} {\bibfnamefont {M.}~\bibnamefont
  {Kounalakis}}, \bibinfo {author} {\bibfnamefont {G.~E.~W.}\ \bibnamefont
  {Bauer}},\ and\ \bibinfo {author} {\bibfnamefont {Y.~M.}\ \bibnamefont
  {Blanter}},\ }\bibfield  {title} {\bibinfo {title} {Analog quantum control of
  magnonic cat states on a chip by a superconducting qubit},\ }\href
  {https://doi.org/10.1103/PhysRevLett.129.037205} {\bibfield  {journal}
  {\bibinfo  {journal} {Phys. Rev. Lett.}\ }\textbf {\bibinfo {volume} {129}},\
  \bibinfo {pages} {037205} (\bibinfo {year} {2022})}\BibitemShut {NoStop}%
\bibitem [{\citenamefont {Wang}\ \emph {et~al.}(2018)\citenamefont {Wang},
  \citenamefont {Li}, \citenamefont {Yin},\ and\ \citenamefont
  {Zeng}}]{Wang2018}%
  \BibitemOpen
  \bibfield  {author} {\bibinfo {author} {\bibfnamefont {Y.}~\bibnamefont
  {Wang}}, \bibinfo {author} {\bibfnamefont {Y.}~\bibnamefont {Li}}, \bibinfo
  {author} {\bibfnamefont {Z.-q.}\ \bibnamefont {Yin}},\ and\ \bibinfo {author}
  {\bibfnamefont {B.}~\bibnamefont {Zeng}},\ }\bibfield  {title} {\bibinfo
  {title} {16-qubit ibm universal quantum computer can be fully entangled},\
  }\href {https://doi.org/10.1038/s41534-018-0095-x} {\bibfield  {journal}
  {\bibinfo  {journal} {npj Quantum Information}\ }\textbf {\bibinfo {volume}
  {4}},\ \bibinfo {pages} {46} (\bibinfo {year} {2018})}\BibitemShut {NoStop}%
\bibitem [{\citenamefont {Hou}\ \emph {et~al.}(2016)\citenamefont {Hou},
  \citenamefont {Zhong}, \citenamefont {Tian}, \citenamefont {Dong},
  \citenamefont {Qi}, \citenamefont {Li}, \citenamefont {Wang}, \citenamefont
  {Nori}, \citenamefont {Xiang}, \citenamefont {Li},\ and\ \citenamefont
  {Guo}}]{2016Houzhibo}%
  \BibitemOpen
  \bibfield  {author} {\bibinfo {author} {\bibfnamefont {Z.}~\bibnamefont
  {Hou}}, \bibinfo {author} {\bibfnamefont {H.-S.}\ \bibnamefont {Zhong}},
  \bibinfo {author} {\bibfnamefont {Y.}~\bibnamefont {Tian}}, \bibinfo {author}
  {\bibfnamefont {D.}~\bibnamefont {Dong}}, \bibinfo {author} {\bibfnamefont
  {B.}~\bibnamefont {Qi}}, \bibinfo {author} {\bibfnamefont {L.}~\bibnamefont
  {Li}}, \bibinfo {author} {\bibfnamefont {Y.}~\bibnamefont {Wang}}, \bibinfo
  {author} {\bibfnamefont {F.}~\bibnamefont {Nori}}, \bibinfo {author}
  {\bibfnamefont {G.-Y.}\ \bibnamefont {Xiang}}, \bibinfo {author}
  {\bibfnamefont {C.-F.}\ \bibnamefont {Li}},\ and\ \bibinfo {author}
  {\bibfnamefont {G.-C.}\ \bibnamefont {Guo}},\ }\bibfield  {title} {\bibinfo
  {title} {Full reconstruction of a 14-qubit state within four hours},\ }\href
  {https://doi.org/10.1088/1367-2630/18/8/083036} {\bibfield  {journal}
  {\bibinfo  {journal} {New J. Phys.}\ }\textbf {\bibinfo {volume} {18}},\
  \bibinfo {pages} {083036} (\bibinfo {year} {2016})}\BibitemShut {NoStop}%
\bibitem [{\citenamefont {Monz}\ \emph {et~al.}(2011)\citenamefont {Monz},
  \citenamefont {Schindler}, \citenamefont {Barreiro}, \citenamefont {Chwalla},
  \citenamefont {Nigg}, \citenamefont {Coish}, \citenamefont {Harlander},
  \citenamefont {H\"ansel}, \citenamefont {Hennrich},\ and\ \citenamefont
  {Blatt}}]{2011Monz}%
  \BibitemOpen
  \bibfield  {author} {\bibinfo {author} {\bibfnamefont {T.}~\bibnamefont
  {Monz}}, \bibinfo {author} {\bibfnamefont {P.}~\bibnamefont {Schindler}},
  \bibinfo {author} {\bibfnamefont {J.~T.}\ \bibnamefont {Barreiro}}, \bibinfo
  {author} {\bibfnamefont {M.}~\bibnamefont {Chwalla}}, \bibinfo {author}
  {\bibfnamefont {D.}~\bibnamefont {Nigg}}, \bibinfo {author} {\bibfnamefont
  {W.~A.}\ \bibnamefont {Coish}}, \bibinfo {author} {\bibfnamefont
  {M.}~\bibnamefont {Harlander}}, \bibinfo {author} {\bibfnamefont
  {W.}~\bibnamefont {H\"ansel}}, \bibinfo {author} {\bibfnamefont
  {M.}~\bibnamefont {Hennrich}},\ and\ \bibinfo {author} {\bibfnamefont
  {R.}~\bibnamefont {Blatt}},\ }\bibfield  {title} {\bibinfo {title} {14-qubit
  entanglement: Creation and coherence},\ }\href
  {https://doi.org/10.1103/PhysRevLett.106.130506} {\bibfield  {journal}
  {\bibinfo  {journal} {Phys. Rev. Lett.}\ }\textbf {\bibinfo {volume} {106}},\
  \bibinfo {pages} {130506} (\bibinfo {year} {2011})}\BibitemShut {NoStop}%
\bibitem [{\citenamefont {Song}\ \emph {et~al.}(2017)\citenamefont {Song},
  \citenamefont {Xu}, \citenamefont {Liu}, \citenamefont {Yang}, \citenamefont
  {Zheng}, \citenamefont {Deng}, \citenamefont {Xie}, \citenamefont {Huang},
  \citenamefont {Guo}, \citenamefont {Zhang}, \citenamefont {Zhang},
  \citenamefont {Xu}, \citenamefont {Zheng}, \citenamefont {Zhu}, \citenamefont
  {Wang}, \citenamefont {Chen}, \citenamefont {Lu}, \citenamefont {Han},\ and\
  \citenamefont {Pan}}]{2017Song}%
  \BibitemOpen
  \bibfield  {author} {\bibinfo {author} {\bibfnamefont {C.}~\bibnamefont
  {Song}}, \bibinfo {author} {\bibfnamefont {K.}~\bibnamefont {Xu}}, \bibinfo
  {author} {\bibfnamefont {W.}~\bibnamefont {Liu}}, \bibinfo {author}
  {\bibfnamefont {C.-p.}\ \bibnamefont {Yang}}, \bibinfo {author}
  {\bibfnamefont {S.-B.}\ \bibnamefont {Zheng}}, \bibinfo {author}
  {\bibfnamefont {H.}~\bibnamefont {Deng}}, \bibinfo {author} {\bibfnamefont
  {Q.}~\bibnamefont {Xie}}, \bibinfo {author} {\bibfnamefont {K.}~\bibnamefont
  {Huang}}, \bibinfo {author} {\bibfnamefont {Q.}~\bibnamefont {Guo}}, \bibinfo
  {author} {\bibfnamefont {L.}~\bibnamefont {Zhang}}, \bibinfo {author}
  {\bibfnamefont {P.}~\bibnamefont {Zhang}}, \bibinfo {author} {\bibfnamefont
  {D.}~\bibnamefont {Xu}}, \bibinfo {author} {\bibfnamefont {D.}~\bibnamefont
  {Zheng}}, \bibinfo {author} {\bibfnamefont {X.}~\bibnamefont {Zhu}}, \bibinfo
  {author} {\bibfnamefont {H.}~\bibnamefont {Wang}}, \bibinfo {author}
  {\bibfnamefont {Y.-A.}\ \bibnamefont {Chen}}, \bibinfo {author}
  {\bibfnamefont {C.-Y.}\ \bibnamefont {Lu}}, \bibinfo {author} {\bibfnamefont
  {S.}~\bibnamefont {Han}},\ and\ \bibinfo {author} {\bibfnamefont {J.-W.}\
  \bibnamefont {Pan}},\ }\bibfield  {title} {\bibinfo {title} {10-qubit
  entanglement and parallel logic operations with a superconducting circuit},\
  }\href {https://doi.org/10.1103/PhysRevLett.119.180511} {\bibfield  {journal}
  {\bibinfo  {journal} {Phys. Rev. Lett.}\ }\textbf {\bibinfo {volume} {119}},\
  \bibinfo {pages} {180511} (\bibinfo {year} {2017})}\BibitemShut {NoStop}%
\bibitem [{\citenamefont {Kranzl}\ \emph {et~al.}(2022)\citenamefont {Kranzl},
  \citenamefont {Joshi}, \citenamefont {Maier}, \citenamefont {Brydges},
  \citenamefont {Franke}, \citenamefont {Blatt},\ and\ \citenamefont
  {Roos}}]{2022Kranzl}%
  \BibitemOpen
  \bibfield  {author} {\bibinfo {author} {\bibfnamefont {F.}~\bibnamefont
  {Kranzl}}, \bibinfo {author} {\bibfnamefont {M.~K.}\ \bibnamefont {Joshi}},
  \bibinfo {author} {\bibfnamefont {C.}~\bibnamefont {Maier}}, \bibinfo
  {author} {\bibfnamefont {T.}~\bibnamefont {Brydges}}, \bibinfo {author}
  {\bibfnamefont {J.}~\bibnamefont {Franke}}, \bibinfo {author} {\bibfnamefont
  {R.}~\bibnamefont {Blatt}},\ and\ \bibinfo {author} {\bibfnamefont {C.~F.}\
  \bibnamefont {Roos}},\ }\bibfield  {title} {\bibinfo {title} {Controlling
  long ion strings for quantum simulation and precision measurements},\ }\href
  {https://doi.org/10.1103/PhysRevA.105.052426} {\bibfield  {journal} {\bibinfo
   {journal} {Phys. Rev. A}\ }\textbf {\bibinfo {volume} {105}},\ \bibinfo
  {pages} {052426} (\bibinfo {year} {2022})}\BibitemShut {NoStop}%
\bibitem [{\citenamefont {Guo}\ \emph {et~al.}(2022)\citenamefont {Guo},
  \citenamefont {Wu}, \citenamefont {Huang}, \citenamefont {Feng},
  \citenamefont {Huang}, \citenamefont {Yang}, \citenamefont {Ma},
  \citenamefont {Yao}, \citenamefont {Zhou},\ and\ \citenamefont
  {Duan}}]{2022Guo}%
  \BibitemOpen
  \bibfield  {author} {\bibinfo {author} {\bibfnamefont {W.-X.}\ \bibnamefont
  {Guo}}, \bibinfo {author} {\bibfnamefont {Y.-K.}\ \bibnamefont {Wu}},
  \bibinfo {author} {\bibfnamefont {Y.-Y.}\ \bibnamefont {Huang}}, \bibinfo
  {author} {\bibfnamefont {L.}~\bibnamefont {Feng}}, \bibinfo {author}
  {\bibfnamefont {C.-X.}\ \bibnamefont {Huang}}, \bibinfo {author}
  {\bibfnamefont {H.-X.}\ \bibnamefont {Yang}}, \bibinfo {author}
  {\bibfnamefont {J.-Y.}\ \bibnamefont {Ma}}, \bibinfo {author} {\bibfnamefont
  {L.}~\bibnamefont {Yao}}, \bibinfo {author} {\bibfnamefont {Z.-C.}\
  \bibnamefont {Zhou}},\ and\ \bibinfo {author} {\bibfnamefont {L.-M.}\
  \bibnamefont {Duan}},\ }\bibfield  {title} {\bibinfo {title} {Picosecond
  ion-qubit manipulation and spin-phonon entanglement with resonant laser
  pulses},\ }\href {https://doi.org/10.1103/PhysRevA.106.022608} {\bibfield
  {journal} {\bibinfo  {journal} {Phys. Rev. A}\ }\textbf {\bibinfo {volume}
  {106}},\ \bibinfo {pages} {022608} (\bibinfo {year} {2022})}\BibitemShut
  {NoStop}%
\bibitem [{\citenamefont {Song}\ \emph {et~al.}(2022)\citenamefont {Song},
  \citenamefont {Tao}, \citenamefont {Qiu},\ and\ \citenamefont
  {Wei}}]{2022Song}%
  \BibitemOpen
  \bibfield  {author} {\bibinfo {author} {\bibfnamefont {G.-Z.}\ \bibnamefont
  {Song}}, \bibinfo {author} {\bibfnamefont {M.-J.}\ \bibnamefont {Tao}},
  \bibinfo {author} {\bibfnamefont {J.}~\bibnamefont {Qiu}},\ and\ \bibinfo
  {author} {\bibfnamefont {H.-R.}\ \bibnamefont {Wei}},\ }\bibfield  {title}
  {\bibinfo {title} {Quantum entanglement creation based on quantum scattering
  in one-dimensional waveguides},\ }\href
  {https://doi.org/10.1103/PhysRevA.106.032416} {\bibfield  {journal} {\bibinfo
   {journal} {Phys. Rev. A}\ }\textbf {\bibinfo {volume} {106}},\ \bibinfo
  {pages} {032416} (\bibinfo {year} {2022})}\BibitemShut {NoStop}%
\bibitem [{\citenamefont {Masson}\ and\ \citenamefont
  {Parkins}(2019)}]{PhysRevA.99.023822}%
  \BibitemOpen
  \bibfield  {author} {\bibinfo {author} {\bibfnamefont {S.~J.}\ \bibnamefont
  {Masson}}\ and\ \bibinfo {author} {\bibfnamefont {S.}~\bibnamefont
  {Parkins}},\ }\bibfield  {title} {\bibinfo {title} {Extreme spin squeezing in
  the steady state of a generalized dicke model},\ }\href
  {https://doi.org/10.1103/PhysRevA.99.023822} {\bibfield  {journal} {\bibinfo
  {journal} {Phys. Rev. A}\ }\textbf {\bibinfo {volume} {99}},\ \bibinfo
  {pages} {023822} (\bibinfo {year} {2019})}\BibitemShut {NoStop}%
\bibitem [{\citenamefont {Stockton}\ \emph {et~al.}(2004)\citenamefont
  {Stockton}, \citenamefont {van Handel},\ and\ \citenamefont
  {Mabuchi}}]{PhysRevA.70.022106}%
  \BibitemOpen
  \bibfield  {author} {\bibinfo {author} {\bibfnamefont {J.~K.}\ \bibnamefont
  {Stockton}}, \bibinfo {author} {\bibfnamefont {R.}~\bibnamefont {van
  Handel}},\ and\ \bibinfo {author} {\bibfnamefont {H.}~\bibnamefont
  {Mabuchi}},\ }\bibfield  {title} {\bibinfo {title} {Deterministic dicke-state
  preparation with continuous measurement and control},\ }\href
  {https://doi.org/10.1103/PhysRevA.70.022106} {\bibfield  {journal} {\bibinfo
  {journal} {Phys. Rev. A}\ }\textbf {\bibinfo {volume} {70}},\ \bibinfo
  {pages} {022106} (\bibinfo {year} {2004})}\BibitemShut {NoStop}%
\bibitem [{\citenamefont {Johnsson}\ \emph {et~al.}(2020)\citenamefont
  {Johnsson}, \citenamefont {Mukty}, \citenamefont {Burgarth}, \citenamefont
  {Volz},\ and\ \citenamefont {Brennen}}]{PhysRevLett.125.190403}%
  \BibitemOpen
  \bibfield  {author} {\bibinfo {author} {\bibfnamefont {M.~T.}\ \bibnamefont
  {Johnsson}}, \bibinfo {author} {\bibfnamefont {N.~R.}\ \bibnamefont {Mukty}},
  \bibinfo {author} {\bibfnamefont {D.}~\bibnamefont {Burgarth}}, \bibinfo
  {author} {\bibfnamefont {T.}~\bibnamefont {Volz}},\ and\ \bibinfo {author}
  {\bibfnamefont {G.~K.}\ \bibnamefont {Brennen}},\ }\bibfield  {title}
  {\bibinfo {title} {Geometric pathway to scalable quantum sensing},\ }\href
  {https://doi.org/10.1103/PhysRevLett.125.190403} {\bibfield  {journal}
  {\bibinfo  {journal} {Phys. Rev. Lett.}\ }\textbf {\bibinfo {volume} {125}},\
  \bibinfo {pages} {190403} (\bibinfo {year} {2020})}\BibitemShut {NoStop}%
\bibitem [{\citenamefont {Hakoshima}\ and\ \citenamefont
  {Matsuzaki}(2020)}]{PhysRevA.102.042610}%
  \BibitemOpen
  \bibfield  {author} {\bibinfo {author} {\bibfnamefont {H.}~\bibnamefont
  {Hakoshima}}\ and\ \bibinfo {author} {\bibfnamefont {Y.}~\bibnamefont
  {Matsuzaki}},\ }\bibfield  {title} {\bibinfo {title} {Efficient detection of
  inhomogeneous magnetic fields from a single spin with dicke states},\ }\href
  {https://doi.org/10.1103/PhysRevA.102.042610} {\bibfield  {journal} {\bibinfo
   {journal} {Phys. Rev. A}\ }\textbf {\bibinfo {volume} {102}},\ \bibinfo
  {pages} {042610} (\bibinfo {year} {2020})}\BibitemShut {NoStop}%
\bibitem [{\citenamefont {Liu}\ \emph {et~al.}(2019)\citenamefont {Liu},
  \citenamefont {Yu}, \citenamefont {Shang}, \citenamefont {Zhu},\ and\
  \citenamefont {Zhang}}]{2019Liu}%
  \BibitemOpen
  \bibfield  {author} {\bibinfo {author} {\bibfnamefont {Y.-C.}\ \bibnamefont
  {Liu}}, \bibinfo {author} {\bibfnamefont {X.-D.}\ \bibnamefont {Yu}},
  \bibinfo {author} {\bibfnamefont {J.}~\bibnamefont {Shang}}, \bibinfo
  {author} {\bibfnamefont {H.}~\bibnamefont {Zhu}},\ and\ \bibinfo {author}
  {\bibfnamefont {X.}~\bibnamefont {Zhang}},\ }\bibfield  {title} {\bibinfo
  {title} {Efficient verification of dicke states},\ }\href
  {https://doi.org/10.1103/PhysRevApplied.12.044020} {\bibfield  {journal}
  {\bibinfo  {journal} {Phys. Rev. Appl.}\ }\textbf {\bibinfo {volume} {12}},\
  \bibinfo {pages} {044020} (\bibinfo {year} {2019})}\BibitemShut {NoStop}%
\bibitem [{\citenamefont {T\'oth}(2012)}]{PhysRevA.85.022322}%
  \BibitemOpen
  \bibfield  {author} {\bibinfo {author} {\bibfnamefont {G.}~\bibnamefont
  {T\'oth}},\ }\bibfield  {title} {\bibinfo {title} {Multipartite entanglement
  and high-precision metrology},\ }\href
  {https://doi.org/10.1103/PhysRevA.85.022322} {\bibfield  {journal} {\bibinfo
  {journal} {Phys. Rev. A}\ }\textbf {\bibinfo {volume} {85}},\ \bibinfo
  {pages} {022322} (\bibinfo {year} {2012})}\BibitemShut {NoStop}%
\bibitem [{\citenamefont {Pezz\`e}\ \emph {et~al.}(2018)\citenamefont
  {Pezz\`e}, \citenamefont {Smerzi}, \citenamefont {Oberthaler}, \citenamefont
  {Schmied},\ and\ \citenamefont {Treutlein}}]{RevModPhys.90.035005}%
  \BibitemOpen
  \bibfield  {author} {\bibinfo {author} {\bibfnamefont {L.}~\bibnamefont
  {Pezz\`e}}, \bibinfo {author} {\bibfnamefont {A.}~\bibnamefont {Smerzi}},
  \bibinfo {author} {\bibfnamefont {M.~K.}\ \bibnamefont {Oberthaler}},
  \bibinfo {author} {\bibfnamefont {R.}~\bibnamefont {Schmied}},\ and\ \bibinfo
  {author} {\bibfnamefont {P.}~\bibnamefont {Treutlein}},\ }\bibfield  {title}
  {\bibinfo {title} {Quantum metrology with nonclassical states of atomic
  ensembles},\ }\href {https://doi.org/10.1103/RevModPhys.90.035005} {\bibfield
   {journal} {\bibinfo  {journal} {Rev. Mod. Phys.}\ }\textbf {\bibinfo
  {volume} {90}},\ \bibinfo {pages} {035005} (\bibinfo {year}
  {2018})}\BibitemShut {NoStop}%
\bibitem [{\citenamefont {Dicke}(1954)}]{PhysRev.93.99}%
  \BibitemOpen
  \bibfield  {author} {\bibinfo {author} {\bibfnamefont {R.~H.}\ \bibnamefont
  {Dicke}},\ }\bibfield  {title} {\bibinfo {title} {Coherence in spontaneous
  radiation processes},\ }\href {https://doi.org/10.1103/PhysRev.93.99}
  {\bibfield  {journal} {\bibinfo  {journal} {Phys. Rev.}\ }\textbf {\bibinfo
  {volume} {93}},\ \bibinfo {pages} {99} (\bibinfo {year} {1954})}\BibitemShut
  {NoStop}%
\bibitem [{\citenamefont {Huang}\ and\ \citenamefont
  {Tian}(2023)}]{PhysRevA.107.063713}%
  \BibitemOpen
  \bibfield  {author} {\bibinfo {author} {\bibfnamefont {J.-F.}\ \bibnamefont
  {Huang}}\ and\ \bibinfo {author} {\bibfnamefont {L.}~\bibnamefont {Tian}},\
  }\bibfield  {title} {\bibinfo {title} {Modulation-based superradiant phase
  transition in the strong-coupling regime},\ }\href
  {https://doi.org/10.1103/PhysRevA.107.063713} {\bibfield  {journal} {\bibinfo
   {journal} {Phys. Rev. A}\ }\textbf {\bibinfo {volume} {107}},\ \bibinfo
  {pages} {063713} (\bibinfo {year} {2023})}\BibitemShut {NoStop}%
\bibitem [{\citenamefont {Xiao}\ \emph {et~al.}(2007)\citenamefont {Xiao},
  \citenamefont {Zou},\ and\ \citenamefont {Guo}}]{2007Xiao}%
  \BibitemOpen
  \bibfield  {author} {\bibinfo {author} {\bibfnamefont {Y.-F.}\ \bibnamefont
  {Xiao}}, \bibinfo {author} {\bibfnamefont {X.-B.}\ \bibnamefont {Zou}},\ and\
  \bibinfo {author} {\bibfnamefont {G.-C.}\ \bibnamefont {Guo}},\ }\bibfield
  {title} {\bibinfo {title} {Generation of atomic entangled states with
  selective resonant interaction in cavity quantum electrodynamics},\ }\href
  {https://doi.org/10.1103/PhysRevA.75.012310} {\bibfield  {journal} {\bibinfo
  {journal} {Phys. Rev. A}\ }\textbf {\bibinfo {volume} {75}},\ \bibinfo
  {pages} {012310} (\bibinfo {year} {2007})}\BibitemShut {NoStop}%
\bibitem [{\citenamefont {Prevedel}\ \emph {et~al.}(2009)\citenamefont
  {Prevedel}, \citenamefont {Cronenberg}, \citenamefont {Tame}, \citenamefont
  {Paternostro}, \citenamefont {Walther}, \citenamefont {Kim},\ and\
  \citenamefont {Zeilinger}}]{2009Prevedel}%
  \BibitemOpen
  \bibfield  {author} {\bibinfo {author} {\bibfnamefont {R.}~\bibnamefont
  {Prevedel}}, \bibinfo {author} {\bibfnamefont {G.}~\bibnamefont
  {Cronenberg}}, \bibinfo {author} {\bibfnamefont {M.~S.}\ \bibnamefont
  {Tame}}, \bibinfo {author} {\bibfnamefont {M.}~\bibnamefont {Paternostro}},
  \bibinfo {author} {\bibfnamefont {P.}~\bibnamefont {Walther}}, \bibinfo
  {author} {\bibfnamefont {M.~S.}\ \bibnamefont {Kim}},\ and\ \bibinfo {author}
  {\bibfnamefont {A.}~\bibnamefont {Zeilinger}},\ }\bibfield  {title} {\bibinfo
  {title} {Experimental realization of dicke states of up to six qubits for
  multiparty quantum networking},\ }\href
  {https://doi.org/10.1103/PhysRevLett.103.020503} {\bibfield  {journal}
  {\bibinfo  {journal} {Phys. Rev. Lett.}\ }\textbf {\bibinfo {volume} {103}},\
  \bibinfo {pages} {020503} (\bibinfo {year} {2009})}\BibitemShut {NoStop}%
\bibitem [{\citenamefont {Shao}\ \emph {et~al.}(2010)\citenamefont {Shao},
  \citenamefont {Chen}, \citenamefont {Zhang}, \citenamefont {Zhao},\ and\
  \citenamefont {Yeon}}]{2010Shao}%
  \BibitemOpen
  \bibfield  {author} {\bibinfo {author} {\bibfnamefont {X.-Q.}\ \bibnamefont
  {Shao}}, \bibinfo {author} {\bibfnamefont {L.}~\bibnamefont {Chen}}, \bibinfo
  {author} {\bibfnamefont {S.}~\bibnamefont {Zhang}}, \bibinfo {author}
  {\bibfnamefont {Y.-F.}\ \bibnamefont {Zhao}},\ and\ \bibinfo {author}
  {\bibfnamefont {K.-H.}\ \bibnamefont {Yeon}},\ }\bibfield  {title} {\bibinfo
  {title} {Deterministic generation of arbitrary multi-atom symmetric dicke
  states by a combination of quantum zeno dynamics and adiabatic passage},\
  }\href {https://doi.org/10.1209/0295-5075/90/50003} {\bibfield  {journal}
  {\bibinfo  {journal} {Europhysics Letters}\ }\textbf {\bibinfo {volume}
  {90}},\ \bibinfo {pages} {50003} (\bibinfo {year} {2010})}\BibitemShut
  {NoStop}%
\bibitem [{\citenamefont {Ekert}(1991)}]{PhysRevLett.67.661}%
  \BibitemOpen
  \bibfield  {author} {\bibinfo {author} {\bibfnamefont {A.~K.}\ \bibnamefont
  {Ekert}},\ }\bibfield  {title} {\bibinfo {title} {Quantum cryptography based
  on bell's theorem},\ }\href {https://doi.org/10.1103/PhysRevLett.67.661}
  {\bibfield  {journal} {\bibinfo  {journal} {Phys. Rev. Lett.}\ }\textbf
  {\bibinfo {volume} {67}},\ \bibinfo {pages} {661} (\bibinfo {year}
  {1991})}\BibitemShut {NoStop}%
\bibitem [{\citenamefont {Wang}\ \emph {et~al.}(2021)\citenamefont {Wang},
  \citenamefont {Noh}, \citenamefont {Lebreuilly}, \citenamefont {Girvin},\
  and\ \citenamefont {Jiang}}]{PhysRevApplied.15.044026}%
  \BibitemOpen
  \bibfield  {author} {\bibinfo {author} {\bibfnamefont {C.-H.}\ \bibnamefont
  {Wang}}, \bibinfo {author} {\bibfnamefont {K.}~\bibnamefont {Noh}}, \bibinfo
  {author} {\bibfnamefont {J.}~\bibnamefont {Lebreuilly}}, \bibinfo {author}
  {\bibfnamefont {S.}~\bibnamefont {Girvin}},\ and\ \bibinfo {author}
  {\bibfnamefont {L.}~\bibnamefont {Jiang}},\ }\bibfield  {title} {\bibinfo
  {title} {Photon-number-dependent hamiltonian engineering for cavities},\
  }\href {https://doi.org/10.1103/PhysRevApplied.15.044026} {\bibfield
  {journal} {\bibinfo  {journal} {Phys. Rev. Appl.}\ }\textbf {\bibinfo
  {volume} {15}},\ \bibinfo {pages} {044026} (\bibinfo {year}
  {2021})}\BibitemShut {NoStop}%
\bibitem [{\citenamefont {Ji}\ \emph {et~al.}(2019)\citenamefont {Ji},
  \citenamefont {Liu}, \citenamefont {Zhou}, \citenamefont {Xiu}, \citenamefont
  {Dong}, \citenamefont {Dong}, \citenamefont {Gao},\ and\ \citenamefont
  {Yi}}]{PhysRevA.99.023808}%
  \BibitemOpen
  \bibfield  {author} {\bibinfo {author} {\bibfnamefont {Y.~Q.}\ \bibnamefont
  {Ji}}, \bibinfo {author} {\bibfnamefont {Y.~L.}\ \bibnamefont {Liu}},
  \bibinfo {author} {\bibfnamefont {S.~J.}\ \bibnamefont {Zhou}}, \bibinfo
  {author} {\bibfnamefont {X.~M.}\ \bibnamefont {Xiu}}, \bibinfo {author}
  {\bibfnamefont {L.}~\bibnamefont {Dong}}, \bibinfo {author} {\bibfnamefont
  {H.~K.}\ \bibnamefont {Dong}}, \bibinfo {author} {\bibfnamefont {Y.~J.}\
  \bibnamefont {Gao}},\ and\ \bibinfo {author} {\bibfnamefont {X.~X.}\
  \bibnamefont {Yi}},\ }\bibfield  {title} {\bibinfo {title} {Fast conversion
  of dicke states $|{D}_{n}^{(2)}\ensuremath{\rangle}$ to
  $|{D}_{n+1}^{(2)}\ensuremath{\rangle}$ by transitionless quantum driving},\
  }\href {https://doi.org/10.1103/PhysRevA.99.023808} {\bibfield  {journal}
  {\bibinfo  {journal} {Phys. Rev. A}\ }\textbf {\bibinfo {volume} {99}},\
  \bibinfo {pages} {023808} (\bibinfo {year} {2019})}\BibitemShut {NoStop}%
\bibitem [{\citenamefont {Liu}\ \emph {et~al.}(2023)\citenamefont {Liu},
  \citenamefont {Ji}, \citenamefont {Han}, \citenamefont {Cui}, \citenamefont
  {Zhang},\ and\ \citenamefont
  {Wang}}]{https://doi.org/10.1002/qute.202200173}%
  \BibitemOpen
  \bibfield  {author} {\bibinfo {author} {\bibfnamefont {Y.-L.}\ \bibnamefont
  {Liu}}, \bibinfo {author} {\bibfnamefont {Y.-Q.}\ \bibnamefont {Ji}},
  \bibinfo {author} {\bibfnamefont {X.}~\bibnamefont {Han}}, \bibinfo {author}
  {\bibfnamefont {W.-X.}\ \bibnamefont {Cui}}, \bibinfo {author} {\bibfnamefont
  {S.}~\bibnamefont {Zhang}},\ and\ \bibinfo {author} {\bibfnamefont {H.-F.}\
  \bibnamefont {Wang}},\ }\bibfield  {title} {\bibinfo {title} {Fast conversion
  of three-particle dicke states to four-particle dicke states with rydberg
  superatoms},\ }\href {https://doi.org/https://doi.org/10.1002/qute.202200173}
  {\bibfield  {journal} {\bibinfo  {journal} {Advanced Quantum Technologies}\
  }\textbf {\bibinfo {volume} {6}},\ \bibinfo {pages} {2200173} (\bibinfo
  {year} {2023})}\BibitemShut {NoStop}%
\bibitem [{\citenamefont {Li}\ \emph {et~al.}(2022)\citenamefont {Li},
  \citenamefont {Braverman}, \citenamefont {Colombo}, \citenamefont {Shu},
  \citenamefont {Kawasaki}, \citenamefont {Adiyatullin}, \citenamefont
  {Pedrozo-Pe\~nafiel}, \citenamefont {Mendez},\ and\ \citenamefont
  {Vuleti\ifmmode~\acute{c}\else \'{c}\fi{}}}]{PRXQuantum.3.020308}%
  \BibitemOpen
  \bibfield  {author} {\bibinfo {author} {\bibfnamefont {Z.}~\bibnamefont
  {Li}}, \bibinfo {author} {\bibfnamefont {B.}~\bibnamefont {Braverman}},
  \bibinfo {author} {\bibfnamefont {S.}~\bibnamefont {Colombo}}, \bibinfo
  {author} {\bibfnamefont {C.}~\bibnamefont {Shu}}, \bibinfo {author}
  {\bibfnamefont {A.}~\bibnamefont {Kawasaki}}, \bibinfo {author}
  {\bibfnamefont {A.~F.}\ \bibnamefont {Adiyatullin}}, \bibinfo {author}
  {\bibfnamefont {E.}~\bibnamefont {Pedrozo-Pe\~nafiel}}, \bibinfo {author}
  {\bibfnamefont {E.}~\bibnamefont {Mendez}},\ and\ \bibinfo {author}
  {\bibfnamefont {V.}~\bibnamefont {Vuleti\ifmmode~\acute{c}\else
  \'{c}\fi{}}},\ }\bibfield  {title} {\bibinfo {title} {Collective spin-light
  and light-mediated spin-spin interactions in an optical cavity},\ }\href
  {https://doi.org/10.1103/PRXQuantum.3.020308} {\bibfield  {journal} {\bibinfo
   {journal} {PRX Quantum}\ }\textbf {\bibinfo {volume} {3}},\ \bibinfo {pages}
  {020308} (\bibinfo {year} {2022})}\BibitemShut {NoStop}%
\bibitem [{\citenamefont {Zhang}\ and\ \citenamefont
  {Duan}(2014)}]{Zhang_2014}%
  \BibitemOpen
  \bibfield  {author} {\bibinfo {author} {\bibfnamefont {Z.}~\bibnamefont
  {Zhang}}\ and\ \bibinfo {author} {\bibfnamefont {L.~M.}\ \bibnamefont
  {Duan}},\ }\bibfield  {title} {\bibinfo {title} {Quantum metrology with dicke
  squeezed states},\ }\href {https://doi.org/10.1088/1367-2630/16/10/103037}
  {\bibfield  {journal} {\bibinfo  {journal} {New J. Phys.}\ }\textbf {\bibinfo
  {volume} {16}},\ \bibinfo {pages} {103037} (\bibinfo {year}
  {2014})}\BibitemShut {NoStop}%
\bibitem [{\citenamefont {Apellaniz}\ \emph {et~al.}(2015)\citenamefont
  {Apellaniz}, \citenamefont {L\"ucke}, \citenamefont {Peise}, \citenamefont
  {Klempt},\ and\ \citenamefont {T\'oth}}]{2015Apellaniz}%
  \BibitemOpen
  \bibfield  {author} {\bibinfo {author} {\bibfnamefont {I.}~\bibnamefont
  {Apellaniz}}, \bibinfo {author} {\bibfnamefont {B.}~\bibnamefont {L\"ucke}},
  \bibinfo {author} {\bibfnamefont {J.}~\bibnamefont {Peise}}, \bibinfo
  {author} {\bibfnamefont {C.}~\bibnamefont {Klempt}},\ and\ \bibinfo {author}
  {\bibfnamefont {G.}~\bibnamefont {T\'oth}},\ }\bibfield  {title} {\bibinfo
  {title} {Detecting metrologically useful entanglement in the vicinity of
  dicke states},\ }\href {https://doi.org/10.1088/1367-2630/17/8/083027}
  {\bibfield  {journal} {\bibinfo  {journal} {New J. Phys.}\ }\textbf {\bibinfo
  {volume} {17}},\ \bibinfo {pages} {083027} (\bibinfo {year}
  {2015})}\BibitemShut {NoStop}%
\bibitem [{\citenamefont {Wang}\ \emph {et~al.}(2020)\citenamefont {Wang},
  \citenamefont {Bai}, \citenamefont {Liu}, \citenamefont {Zhang},\ and\
  \citenamefont {Wang}}]{Wang:20}%
  \BibitemOpen
  \bibfield  {author} {\bibinfo {author} {\bibfnamefont {D.-Y.}\ \bibnamefont
  {Wang}}, \bibinfo {author} {\bibfnamefont {C.-H.}\ \bibnamefont {Bai}},
  \bibinfo {author} {\bibfnamefont {S.}~\bibnamefont {Liu}}, \bibinfo {author}
  {\bibfnamefont {S.}~\bibnamefont {Zhang}},\ and\ \bibinfo {author}
  {\bibfnamefont {H.-F.}\ \bibnamefont {Wang}},\ }\bibfield  {title} {\bibinfo
  {title} {Dissipative bosonic squeezing via frequency modulation and its
  application in optomechanics},\ }\href {https://doi.org/10.1364/OE.399687}
  {\bibfield  {journal} {\bibinfo  {journal} {Opt. Express}\ }\textbf {\bibinfo
  {volume} {28}},\ \bibinfo {pages} {28942} (\bibinfo {year}
  {2020})}\BibitemShut {NoStop}%
\bibitem [{\citenamefont {Jung}(1993)}]{JUNG1993175}%
  \BibitemOpen
  \bibfield  {author} {\bibinfo {author} {\bibfnamefont {P.}~\bibnamefont
  {Jung}},\ }\bibfield  {title} {\bibinfo {title} {Periodically driven
  stochastic systems},\ }\href
  {https://doi.org/https://doi.org/10.1016/0370-1573(93)90022-6} {\bibfield
  {journal} {\bibinfo  {journal} {Physics Reports}\ }\textbf {\bibinfo {volume}
  {234}},\ \bibinfo {pages} {175} (\bibinfo {year} {1993})}\BibitemShut
  {NoStop}%
\bibitem [{\citenamefont {Novi\ifmmode~\check{c}\else \v{c}\fi{}enko}\ \emph
  {et~al.}(2017)\citenamefont {Novi\ifmmode~\check{c}\else \v{c}\fi{}enko},
  \citenamefont {Anisimovas},\ and\ \citenamefont {Juzeli\ifmmode~\bar{u}\else
  \={u}\fi{}nas}}]{PhysRevA.95.023615}%
  \BibitemOpen
  \bibfield  {author} {\bibinfo {author} {\bibfnamefont {V.}~\bibnamefont
  {Novi\ifmmode~\check{c}\else \v{c}\fi{}enko}}, \bibinfo {author}
  {\bibfnamefont {E.}~\bibnamefont {Anisimovas}},\ and\ \bibinfo {author}
  {\bibfnamefont {G.}~\bibnamefont {Juzeli\ifmmode~\bar{u}\else
  \={u}\fi{}nas}},\ }\bibfield  {title} {\bibinfo {title} {Floquet analysis of
  a quantum system with modulated periodic driving},\ }\href
  {https://doi.org/10.1103/PhysRevA.95.023615} {\bibfield  {journal} {\bibinfo
  {journal} {Phys. Rev. A}\ }\textbf {\bibinfo {volume} {95}},\ \bibinfo
  {pages} {023615} (\bibinfo {year} {2017})}\BibitemShut {NoStop}%
\bibitem [{\citenamefont {Yu}\ \emph {et~al.}(2022)\citenamefont {Yu},
  \citenamefont {Wang}, \citenamefont {Sun},\ and\ \citenamefont
  {Wang}}]{2022Yu}%
  \BibitemOpen
  \bibfield  {author} {\bibinfo {author} {\bibfnamefont {N.}~\bibnamefont
  {Yu}}, \bibinfo {author} {\bibfnamefont {S.}~\bibnamefont {Wang}}, \bibinfo
  {author} {\bibfnamefont {C.}~\bibnamefont {Sun}},\ and\ \bibinfo {author}
  {\bibfnamefont {G.}~\bibnamefont {Wang}},\ }\bibfield  {title} {\bibinfo
  {title} {Switchable selective interactions in a dicke model with driven
  biased term},\ }\href {https://doi.org/10.1103/PhysRevE.105.034125}
  {\bibfield  {journal} {\bibinfo  {journal} {Phys. Rev. E}\ }\textbf {\bibinfo
  {volume} {105}},\ \bibinfo {pages} {034125} (\bibinfo {year}
  {2022})}\BibitemShut {NoStop}%
\bibitem [{\citenamefont {Cong}\ \emph {et~al.}(2020)\citenamefont {Cong},
  \citenamefont {Felicetti}, \citenamefont {Casanova}, \citenamefont {Lamata},
  \citenamefont {Solano},\ and\ \citenamefont {Arrazola}}]{2020CongL}%
  \BibitemOpen
  \bibfield  {author} {\bibinfo {author} {\bibfnamefont {L.}~\bibnamefont
  {Cong}}, \bibinfo {author} {\bibfnamefont {S.}~\bibnamefont {Felicetti}},
  \bibinfo {author} {\bibfnamefont {J.}~\bibnamefont {Casanova}}, \bibinfo
  {author} {\bibfnamefont {L.}~\bibnamefont {Lamata}}, \bibinfo {author}
  {\bibfnamefont {E.}~\bibnamefont {Solano}},\ and\ \bibinfo {author}
  {\bibfnamefont {I.}~\bibnamefont {Arrazola}},\ }\bibfield  {title} {\bibinfo
  {title} {Selective interactions in the quantum rabi model},\ }\href
  {https://doi.org/10.1103/PhysRevA.101.032350} {\bibfield  {journal} {\bibinfo
   {journal} {Phys. Rev. A}\ }\textbf {\bibinfo {volume} {101}},\ \bibinfo
  {pages} {032350} (\bibinfo {year} {2020})}\BibitemShut {NoStop}%
\bibitem [{\citenamefont {Shao}(2020)}]{2020Shao}%
  \BibitemOpen
  \bibfield  {author} {\bibinfo {author} {\bibfnamefont {X.-Q.}\ \bibnamefont
  {Shao}},\ }\bibfield  {title} {\bibinfo {title} {Selective rydberg pumping
  via strong dipole blockade},\ }\href
  {https://doi.org/10.1103/PhysRevA.102.053118} {\bibfield  {journal} {\bibinfo
   {journal} {Phys. Rev. A}\ }\textbf {\bibinfo {volume} {102}},\ \bibinfo
  {pages} {053118} (\bibinfo {year} {2020})}\BibitemShut {NoStop}%
\bibitem [{\citenamefont {Neuman}\ \emph {et~al.}(2020)\citenamefont {Neuman},
  \citenamefont {Trusheim},\ and\ \citenamefont {Narang}}]{2020Neuman}%
  \BibitemOpen
  \bibfield  {author} {\bibinfo {author} {\bibfnamefont {T.}~\bibnamefont
  {Neuman}}, \bibinfo {author} {\bibfnamefont {M.}~\bibnamefont {Trusheim}},\
  and\ \bibinfo {author} {\bibfnamefont {P.}~\bibnamefont {Narang}},\
  }\bibfield  {title} {\bibinfo {title} {Selective acoustic control of
  photon-mediated qubit-qubit interactions},\ }\href
  {https://doi.org/10.1103/PhysRevA.101.052342} {\bibfield  {journal} {\bibinfo
   {journal} {Phys. Rev. A}\ }\textbf {\bibinfo {volume} {101}},\ \bibinfo
  {pages} {052342} (\bibinfo {year} {2020})}\BibitemShut {NoStop}%
\bibitem [{\citenamefont {Mu}\ \emph {et~al.}(2020)\citenamefont {Mu},
  \citenamefont {Gao}, \citenamefont {Yin},\ and\ \citenamefont
  {Wang}}]{Mu:20}%
  \BibitemOpen
  \bibfield  {author} {\bibinfo {author} {\bibfnamefont {F.}~\bibnamefont
  {Mu}}, \bibinfo {author} {\bibfnamefont {Y.}~\bibnamefont {Gao}}, \bibinfo
  {author} {\bibfnamefont {H.}~\bibnamefont {Yin}},\ and\ \bibinfo {author}
  {\bibfnamefont {G.}~\bibnamefont {Wang}},\ }\bibfield  {title} {\bibinfo
  {title} {Dicke state generation via selective interactions in a dicke-stark
  model},\ }\href {https://doi.org/10.1364/OE.412914} {\bibfield  {journal}
  {\bibinfo  {journal} {Opt. Express}\ }\textbf {\bibinfo {volume} {28}},\
  \bibinfo {pages} {39574} (\bibinfo {year} {2020})}\BibitemShut {NoStop}%
\bibitem [{\citenamefont {Wu}\ \emph {et~al.}(2017)\citenamefont {Wu},
  \citenamefont {Guo}, \citenamefont {Wang}, \citenamefont {Wang},
  \citenamefont {Feng},\ and\ \citenamefont {Chen}}]{2017Chunfeng}%
  \BibitemOpen
  \bibfield  {author} {\bibinfo {author} {\bibfnamefont {C.}~\bibnamefont
  {Wu}}, \bibinfo {author} {\bibfnamefont {C.}~\bibnamefont {Guo}}, \bibinfo
  {author} {\bibfnamefont {Y.}~\bibnamefont {Wang}}, \bibinfo {author}
  {\bibfnamefont {G.}~\bibnamefont {Wang}}, \bibinfo {author} {\bibfnamefont
  {X.-L.}\ \bibnamefont {Feng}},\ and\ \bibinfo {author} {\bibfnamefont
  {J.-L.}\ \bibnamefont {Chen}},\ }\bibfield  {title} {\bibinfo {title}
  {Generation of dicke states in the ultrastrong-coupling regime of circuit qed
  systems},\ }\href {https://doi.org/10.1103/PhysRevA.95.013845} {\bibfield
  {journal} {\bibinfo  {journal} {Phys. Rev. A}\ }\textbf {\bibinfo {volume}
  {95}},\ \bibinfo {pages} {013845} (\bibinfo {year} {2017})}\BibitemShut
  {NoStop}%
\bibitem [{\citenamefont {Wu}\ \emph {et~al.}(2019)\citenamefont {Wu},
  \citenamefont {Wang}, \citenamefont {Guo}, \citenamefont {Ouyang},
  \citenamefont {Wang},\ and\ \citenamefont {Feng}}]{2019Chunfeng}%
  \BibitemOpen
  \bibfield  {author} {\bibinfo {author} {\bibfnamefont {C.}~\bibnamefont
  {Wu}}, \bibinfo {author} {\bibfnamefont {Y.}~\bibnamefont {Wang}}, \bibinfo
  {author} {\bibfnamefont {C.}~\bibnamefont {Guo}}, \bibinfo {author}
  {\bibfnamefont {Y.}~\bibnamefont {Ouyang}}, \bibinfo {author} {\bibfnamefont
  {G.}~\bibnamefont {Wang}},\ and\ \bibinfo {author} {\bibfnamefont {X.-L.}\
  \bibnamefont {Feng}},\ }\bibfield  {title} {\bibinfo {title} {Initializing a
  permutation-invariant quantum error-correction code},\ }\href
  {https://doi.org/1103/PhysRevA.99.012335} {\bibfield  {journal} {\bibinfo
  {journal} {Phys. Rev. A}\ }\textbf {\bibinfo {volume} {99}},\ \bibinfo
  {pages} {012335} (\bibinfo {year} {2019})}\BibitemShut {NoStop}%
\bibitem [{\citenamefont {Chu}\ and\ \citenamefont {Telnov}(2004)}]{PR2004Chu}%
  \BibitemOpen
  \bibfield  {author} {\bibinfo {author} {\bibfnamefont {S.-I.}\ \bibnamefont
  {Chu}}\ and\ \bibinfo {author} {\bibfnamefont {D.~A.}\ \bibnamefont
  {Telnov}},\ }\bibfield  {title} {\bibinfo {title} {Beyond the floquet
  theorem: generalized floquet formalisms and quasienergy methods for atomic
  and molecular multiphoton processes in intense laser fields},\ }\href
  {https://doi.org/https://doi.org/10.1016/j.physrep.2003.10.001} {\bibfield
  {journal} {\bibinfo  {journal} {Physics Reports}\ }\textbf {\bibinfo {volume}
  {390}},\ \bibinfo {pages} {1} (\bibinfo {year} {2004})}\BibitemShut {NoStop}%
\bibitem [{\citenamefont {Son}\ \emph {et~al.}(2009)\citenamefont {Son},
  \citenamefont {Han},\ and\ \citenamefont {Chu}}]{PRA2009Son}%
  \BibitemOpen
  \bibfield  {author} {\bibinfo {author} {\bibfnamefont {S.-K.}\ \bibnamefont
  {Son}}, \bibinfo {author} {\bibfnamefont {S.}~\bibnamefont {Han}},\ and\
  \bibinfo {author} {\bibfnamefont {S.-I.}\ \bibnamefont {Chu}},\ }\bibfield
  {title} {\bibinfo {title} {Floquet formulation for the investigation of
  multiphoton quantum interference in a superconducting qubit driven by a
  strong ac field},\ }\href {https://doi.org/10.1103/PhysRevA.79.032301}
  {\bibfield  {journal} {\bibinfo  {journal} {Phys. Rev. A}\ }\textbf {\bibinfo
  {volume} {79}},\ \bibinfo {pages} {032301} (\bibinfo {year}
  {2009})}\BibitemShut {NoStop}%
\bibitem [{\citenamefont {Luo}\ \emph {et~al.}(2013)\citenamefont {Luo},
  \citenamefont {Huang}, \citenamefont {Zhong}, \citenamefont {Qin},
  \citenamefont {Xie}, \citenamefont {Kivshar},\ and\ \citenamefont
  {Lee}}]{PRL2013Luo}%
  \BibitemOpen
  \bibfield  {author} {\bibinfo {author} {\bibfnamefont {X.}~\bibnamefont
  {Luo}}, \bibinfo {author} {\bibfnamefont {J.}~\bibnamefont {Huang}}, \bibinfo
  {author} {\bibfnamefont {H.}~\bibnamefont {Zhong}}, \bibinfo {author}
  {\bibfnamefont {X.}~\bibnamefont {Qin}}, \bibinfo {author} {\bibfnamefont
  {Q.}~\bibnamefont {Xie}}, \bibinfo {author} {\bibfnamefont {Y.~S.}\
  \bibnamefont {Kivshar}},\ and\ \bibinfo {author} {\bibfnamefont
  {C.}~\bibnamefont {Lee}},\ }\bibfield  {title} {\bibinfo {title}
  {Pseudo-parity-time symmetry in optical systems},\ }\href
  {https://doi.org/10.1103/PhysRevLett.110.243902} {\bibfield  {journal}
  {\bibinfo  {journal} {Phys. Rev. Lett.}\ }\textbf {\bibinfo {volume} {110}},\
  \bibinfo {pages} {243902} (\bibinfo {year} {2013})}\BibitemShut {NoStop}%
\bibitem [{\citenamefont {H\"anggi}\ \emph {et~al.}(1990)\citenamefont
  {H\"anggi}, \citenamefont {Talkner},\ and\ \citenamefont
  {Borkovec}}]{RMP1990H}%
  \BibitemOpen
  \bibfield  {author} {\bibinfo {author} {\bibfnamefont {P.}~\bibnamefont
  {H\"anggi}}, \bibinfo {author} {\bibfnamefont {P.}~\bibnamefont {Talkner}},\
  and\ \bibinfo {author} {\bibfnamefont {M.}~\bibnamefont {Borkovec}},\
  }\bibfield  {title} {\bibinfo {title} {Reaction-rate theory: fifty years
  after kramers},\ }\href {https://doi.org/10.1103/RevModPhys.62.251}
  {\bibfield  {journal} {\bibinfo  {journal} {Rev. Mod. Phys.}\ }\textbf
  {\bibinfo {volume} {62}},\ \bibinfo {pages} {251} (\bibinfo {year}
  {1990})}\BibitemShut {NoStop}%
\bibitem [{\citenamefont {Eckardt}\ and\ \citenamefont
  {Anisimovas}(2015)}]{NJP2015Eckardt}%
  \BibitemOpen
  \bibfield  {author} {\bibinfo {author} {\bibfnamefont {A.}~\bibnamefont
  {Eckardt}}\ and\ \bibinfo {author} {\bibfnamefont {E.}~\bibnamefont
  {Anisimovas}},\ }\bibfield  {title} {\bibinfo {title} {High-frequency
  approximation for periodically driven quantum systems from a floquet-space
  perspective},\ }\href {https://doi.org/10.1088/1367-2630/17/9/093039}
  {\bibfield  {journal} {\bibinfo  {journal} {New J. Phys.}\ }\textbf {\bibinfo
  {volume} {17}},\ \bibinfo {pages} {093039} (\bibinfo {year}
  {2015})}\BibitemShut {NoStop}%
\bibitem [{\citenamefont {Haase}\ \emph {et~al.}(2018)\citenamefont {Haase},
  \citenamefont {Wang}, \citenamefont {Casanova},\ and\ \citenamefont
  {Plenio}}]{PhysRevLett.121.050402}%
  \BibitemOpen
  \bibfield  {author} {\bibinfo {author} {\bibfnamefont {J.~F.}\ \bibnamefont
  {Haase}}, \bibinfo {author} {\bibfnamefont {Z.-Y.}\ \bibnamefont {Wang}},
  \bibinfo {author} {\bibfnamefont {J.}~\bibnamefont {Casanova}},\ and\
  \bibinfo {author} {\bibfnamefont {M.~B.}\ \bibnamefont {Plenio}},\ }\bibfield
   {title} {\bibinfo {title} {Soft quantum control for highly selective
  interactions among joint quantum systems},\ }\href
  {https://doi.org/10.1103/PhysRevLett.121.050402} {\bibfield  {journal}
  {\bibinfo  {journal} {Phys. Rev. Lett.}\ }\textbf {\bibinfo {volume} {121}},\
  \bibinfo {pages} {050402} (\bibinfo {year} {2018})}\BibitemShut {NoStop}%
\bibitem [{\citenamefont {Holstein}\ and\ \citenamefont
  {Primakoff}(1940{\natexlab{a}})}]{1940Holstein}%
  \BibitemOpen
  \bibfield  {author} {\bibinfo {author} {\bibfnamefont {T.}~\bibnamefont
  {Holstein}}\ and\ \bibinfo {author} {\bibfnamefont {H.}~\bibnamefont
  {Primakoff}},\ }\bibfield  {title} {\bibinfo {title} {Field dependence of the
  intrinsic domain magnetization of a ferromagnet},\ }\href
  {https://doi.org/https://doi.org/10.1103/PhysRev.58.1098} {\bibfield
  {journal} {\bibinfo  {journal} {Phys. Rev.}\ }\textbf {\bibinfo {volume}
  {58}},\ \bibinfo {pages} {1098} (\bibinfo {year}
  {1940}{\natexlab{a}})}\BibitemShut {NoStop}%
\bibitem [{\citenamefont {Kitagawa}\ and\ \citenamefont
  {Ueda}(1993)}]{PhysRevA.47.5138}%
  \BibitemOpen
  \bibfield  {author} {\bibinfo {author} {\bibfnamefont {M.}~\bibnamefont
  {Kitagawa}}\ and\ \bibinfo {author} {\bibfnamefont {M.}~\bibnamefont
  {Ueda}},\ }\bibfield  {title} {\bibinfo {title} {Squeezed spin states},\
  }\href {https://doi.org/10.1103/PhysRevA.47.5138} {\bibfield  {journal}
  {\bibinfo  {journal} {Phys. Rev. A}\ }\textbf {\bibinfo {volume} {47}},\
  \bibinfo {pages} {5138} (\bibinfo {year} {1993})}\BibitemShut {NoStop}%
\bibitem [{\citenamefont {Huang}\ \emph {et~al.}(2017)\citenamefont {Huang},
  \citenamefont {Liao}, \citenamefont {Tian},\ and\ \citenamefont
  {Kuang}}]{PhysRevA.96.043849}%
  \BibitemOpen
  \bibfield  {author} {\bibinfo {author} {\bibfnamefont {J.-F.}\ \bibnamefont
  {Huang}}, \bibinfo {author} {\bibfnamefont {J.-Q.}\ \bibnamefont {Liao}},
  \bibinfo {author} {\bibfnamefont {L.}~\bibnamefont {Tian}},\ and\ \bibinfo
  {author} {\bibfnamefont {L.-M.}\ \bibnamefont {Kuang}},\ }\bibfield  {title}
  {\bibinfo {title} {Manipulating counter-rotating interactions in the quantum
  rabi model via modulation of the transition frequency of the two-level
  system},\ }\href {https://doi.org/10.1103/PhysRevA.96.043849} {\bibfield
  {journal} {\bibinfo  {journal} {Phys. Rev. A}\ }\textbf {\bibinfo {volume}
  {96}},\ \bibinfo {pages} {043849} (\bibinfo {year} {2017})}\BibitemShut
  {NoStop}%
\bibitem [{\citenamefont {Zhou}\ \emph {et~al.}(2011)\citenamefont {Zhou},
  \citenamefont {Hu}, \citenamefont {Zou},\ and\ \citenamefont
  {Guo}}]{PhysRevA.84.042324}%
  \BibitemOpen
  \bibfield  {author} {\bibinfo {author} {\bibfnamefont {J.}~\bibnamefont
  {Zhou}}, \bibinfo {author} {\bibfnamefont {Y.}~\bibnamefont {Hu}}, \bibinfo
  {author} {\bibfnamefont {X.-B.}\ \bibnamefont {Zou}},\ and\ \bibinfo {author}
  {\bibfnamefont {G.-C.}\ \bibnamefont {Guo}},\ }\bibfield  {title} {\bibinfo
  {title} {Ground-state preparation of arbitrarily multipartite dicke states in
  the one-dimensional ferromagnetic spin-$\frac{1}{2}$ chain},\ }\href
  {https://doi.org/10.1103/PhysRevA.84.042324} {\bibfield  {journal} {\bibinfo
  {journal} {Phys. Rev. A}\ }\textbf {\bibinfo {volume} {84}},\ \bibinfo
  {pages} {042324} (\bibinfo {year} {2011})}\BibitemShut {NoStop}%
\bibitem [{\citenamefont {Hume}\ \emph {et~al.}(2009)\citenamefont {Hume},
  \citenamefont {Chou}, \citenamefont {Rosenband},\ and\ \citenamefont
  {Wineland}}]{PhysRevA.80.052302}%
  \BibitemOpen
  \bibfield  {author} {\bibinfo {author} {\bibfnamefont {D.~B.}\ \bibnamefont
  {Hume}}, \bibinfo {author} {\bibfnamefont {C.~W.}\ \bibnamefont {Chou}},
  \bibinfo {author} {\bibfnamefont {T.}~\bibnamefont {Rosenband}},\ and\
  \bibinfo {author} {\bibfnamefont {D.~J.}\ \bibnamefont {Wineland}},\
  }\bibfield  {title} {\bibinfo {title} {Preparation of dicke states in an ion
  chain},\ }\href {https://doi.org/10.1103/PhysRevA.80.052302} {\bibfield
  {journal} {\bibinfo  {journal} {Phys. Rev. A}\ }\textbf {\bibinfo {volume}
  {80}},\ \bibinfo {pages} {052302} (\bibinfo {year} {2009})}\BibitemShut
  {NoStop}%
\bibitem [{\citenamefont {Tameshtit}\ and\ \citenamefont
  {Sipe}(1994)}]{PhysRevA.49.89}%
  \BibitemOpen
  \bibfield  {author} {\bibinfo {author} {\bibfnamefont {A.}~\bibnamefont
  {Tameshtit}}\ and\ \bibinfo {author} {\bibfnamefont {J.~E.}\ \bibnamefont
  {Sipe}},\ }\bibfield  {title} {\bibinfo {title} {Evolution of coherences and
  populations in the secular approximation},\ }\href
  {https://doi.org/10.1103/PhysRevA.49.89} {\bibfield  {journal} {\bibinfo
  {journal} {Phys. Rev. A}\ }\textbf {\bibinfo {volume} {49}},\ \bibinfo
  {pages} {89} (\bibinfo {year} {1994})}\BibitemShut {NoStop}%
\bibitem [{\citenamefont {Creffield}(2003)}]{PhysRevB.67.165301}%
  \BibitemOpen
  \bibfield  {author} {\bibinfo {author} {\bibfnamefont {C.}~\bibnamefont
  {Creffield}},\ }\bibfield  {title} {\bibinfo {title} {Location of crossings
  in the floquet spectrum of a driven two-level system},\ }\href
  {https://doi.org/10.1103/PhysRevB.67.165301} {\bibfield  {journal} {\bibinfo
  {journal} {Phys. Rev. B}\ }\textbf {\bibinfo {volume} {67}},\ \bibinfo
  {pages} {165301} (\bibinfo {year} {2003})}\BibitemShut {NoStop}%
\bibitem [{\citenamefont {Schlegel}\ \emph {et~al.}(2022)\citenamefont
  {Schlegel}, \citenamefont {Minganti},\ and\ \citenamefont
  {Savona}}]{PhysRevA.106.022431}%
  \BibitemOpen
  \bibfield  {author} {\bibinfo {author} {\bibfnamefont {D.~S.}\ \bibnamefont
  {Schlegel}}, \bibinfo {author} {\bibfnamefont {F.}~\bibnamefont {Minganti}},\
  and\ \bibinfo {author} {\bibfnamefont {V.}~\bibnamefont {Savona}},\
  }\bibfield  {title} {\bibinfo {title} {Quantum error correction using
  squeezed schr\"odinger cat states},\ }\href
  {https://doi.org/10.1103/PhysRevA.106.022431} {\bibfield  {journal} {\bibinfo
   {journal} {Phys. Rev. A}\ }\textbf {\bibinfo {volume} {106}},\ \bibinfo
  {pages} {022431} (\bibinfo {year} {2022})}\BibitemShut {NoStop}%
\bibitem [{\citenamefont {Izumi}\ \emph {et~al.}(2018)\citenamefont {Izumi},
  \citenamefont {Takeoka}, \citenamefont {Wakui}, \citenamefont {Fujiwara},
  \citenamefont {Ema},\ and\ \citenamefont {Sasaki}}]{Izumi2018}%
  \BibitemOpen
  \bibfield  {author} {\bibinfo {author} {\bibfnamefont {S.}~\bibnamefont
  {Izumi}}, \bibinfo {author} {\bibfnamefont {M.}~\bibnamefont {Takeoka}},
  \bibinfo {author} {\bibfnamefont {K.}~\bibnamefont {Wakui}}, \bibinfo
  {author} {\bibfnamefont {M.}~\bibnamefont {Fujiwara}}, \bibinfo {author}
  {\bibfnamefont {K.}~\bibnamefont {Ema}},\ and\ \bibinfo {author}
  {\bibfnamefont {M.}~\bibnamefont {Sasaki}},\ }\bibfield  {title} {\bibinfo
  {title} {Projective measurement onto arbitrary superposition of weak coherent
  state bases},\ }\href {https://doi.org/10.1038/s41598-018-21092-8} {\bibfield
   {journal} {\bibinfo  {journal} {Scientific Reports}\ }\textbf {\bibinfo
  {volume} {8}},\ \bibinfo {pages} {2999} (\bibinfo {year} {2018})}\BibitemShut
  {NoStop}%
\bibitem [{\citenamefont {Shao}\ \emph {et~al.}(2023)\citenamefont {Shao},
  \citenamefont {Liu}, \citenamefont {Xue}, \citenamefont {Mu},\ and\
  \citenamefont {Li}}]{PhysRevApplied.20.014014}%
  \BibitemOpen
  \bibfield  {author} {\bibinfo {author} {\bibfnamefont {X.}~\bibnamefont
  {Shao}}, \bibinfo {author} {\bibfnamefont {F.}~\bibnamefont {Liu}}, \bibinfo
  {author} {\bibfnamefont {X.}~\bibnamefont {Xue}}, \bibinfo {author}
  {\bibfnamefont {W.}~\bibnamefont {Mu}},\ and\ \bibinfo {author}
  {\bibfnamefont {W.}~\bibnamefont {Li}},\ }\bibfield  {title} {\bibinfo
  {title} {High-fidelity interconversion between greenberger-horne-zeilinger
  and $w$ states through floquet-lindblad engineering in rydberg atom arrays},\
  }\href {https://doi.org/10.1103/PhysRevApplied.20.014014} {\bibfield
  {journal} {\bibinfo  {journal} {Phys. Rev. Appl.}\ }\textbf {\bibinfo
  {volume} {20}},\ \bibinfo {pages} {014014} (\bibinfo {year}
  {2023})}\BibitemShut {NoStop}%
\bibitem [{\citenamefont {Holstein}\ and\ \citenamefont
  {Primakoff}(1940{\natexlab{b}})}]{PhysRev.58.1098}%
  \BibitemOpen
  \bibfield  {author} {\bibinfo {author} {\bibfnamefont {T.}~\bibnamefont
  {Holstein}}\ and\ \bibinfo {author} {\bibfnamefont {H.}~\bibnamefont
  {Primakoff}},\ }\bibfield  {title} {\bibinfo {title} {Field dependence of the
  intrinsic domain magnetization of a ferromagnet},\ }\href
  {https://doi.org/10.1103/PhysRev.58.1098} {\bibfield  {journal} {\bibinfo
  {journal} {Phys. Rev.}\ }\textbf {\bibinfo {volume} {58}},\ \bibinfo {pages}
  {1098} (\bibinfo {year} {1940}{\natexlab{b}})}\BibitemShut {NoStop}%
\bibitem [{\citenamefont {Dowling}(2008)}]{doi:10.1080/00107510802091298}%
  \BibitemOpen
  \bibfield  {author} {\bibinfo {author} {\bibfnamefont {J.~P.}\ \bibnamefont
  {Dowling}},\ }\bibfield  {title} {\bibinfo {title} {Quantum optical
  metrology–the lowdown on high-n00n states},\ }\href
  {https://doi.org/10.1080/00107510802091298} {\bibfield  {journal} {\bibinfo
  {journal} {Contemporary Physics}\ }\textbf {\bibinfo {volume} {49}},\
  \bibinfo {pages} {125} (\bibinfo {year} {2008})}\BibitemShut {NoStop}%
\bibitem [{\citenamefont {Qi}\ and\ \citenamefont
  {Jing}(2020)}]{PhysRevA.101.033809}%
  \BibitemOpen
  \bibfield  {author} {\bibinfo {author} {\bibfnamefont {S.-f.}\ \bibnamefont
  {Qi}}\ and\ \bibinfo {author} {\bibfnamefont {J.}~\bibnamefont {Jing}},\
  }\bibfield  {title} {\bibinfo {title} {Generating noon states in circuit qed
  using a multiphoton resonance in the presence of counter-rotating
  interactions},\ }\href {https://doi.org/10.1103/PhysRevA.101.033809}
  {\bibfield  {journal} {\bibinfo  {journal} {Phys. Rev. A}\ }\textbf {\bibinfo
  {volume} {101}},\ \bibinfo {pages} {033809} (\bibinfo {year}
  {2020})}\BibitemShut {NoStop}%
\bibitem [{\citenamefont {Beaudoin}\ \emph {et~al.}(2011)\citenamefont
  {Beaudoin}, \citenamefont {Gambetta},\ and\ \citenamefont
  {Blais}}]{PhysRevA.84.043832}%
  \BibitemOpen
  \bibfield  {author} {\bibinfo {author} {\bibfnamefont {F.}~\bibnamefont
  {Beaudoin}}, \bibinfo {author} {\bibfnamefont {J.~M.}\ \bibnamefont
  {Gambetta}},\ and\ \bibinfo {author} {\bibfnamefont {A.}~\bibnamefont
  {Blais}},\ }\bibfield  {title} {\bibinfo {title} {Dissipation and ultrastrong
  coupling in circuit qed},\ }\href
  {https://doi.org/10.1103/PhysRevA.84.043832} {\bibfield  {journal} {\bibinfo
  {journal} {Phys. Rev. A}\ }\textbf {\bibinfo {volume} {84}},\ \bibinfo
  {pages} {043832} (\bibinfo {year} {2011})}\BibitemShut {NoStop}%
\bibitem [{\citenamefont {Settineri}\ \emph {et~al.}(2018)\citenamefont
  {Settineri}, \citenamefont {Macr\'{\i}}, \citenamefont {Ridolfo},
  \citenamefont {Di~Stefano}, \citenamefont {Kockum}, \citenamefont {Nori},\
  and\ \citenamefont {Savasta}}]{PhysRevA.98.053834}%
  \BibitemOpen
  \bibfield  {author} {\bibinfo {author} {\bibfnamefont {A.}~\bibnamefont
  {Settineri}}, \bibinfo {author} {\bibfnamefont {V.}~\bibnamefont
  {Macr\'{\i}}}, \bibinfo {author} {\bibfnamefont {A.}~\bibnamefont {Ridolfo}},
  \bibinfo {author} {\bibfnamefont {O.}~\bibnamefont {Di~Stefano}}, \bibinfo
  {author} {\bibfnamefont {A.~F.}\ \bibnamefont {Kockum}}, \bibinfo {author}
  {\bibfnamefont {F.}~\bibnamefont {Nori}},\ and\ \bibinfo {author}
  {\bibfnamefont {S.}~\bibnamefont {Savasta}},\ }\bibfield  {title} {\bibinfo
  {title} {Dissipation and thermal noise in hybrid quantum systems in the
  ultrastrong-coupling regime},\ }\href
  {https://doi.org/10.1103/PhysRevA.98.053834} {\bibfield  {journal} {\bibinfo
  {journal} {Phys. Rev. A}\ }\textbf {\bibinfo {volume} {98}},\ \bibinfo
  {pages} {053834} (\bibinfo {year} {2018})}\BibitemShut {NoStop}%
\bibitem [{\citenamefont {Song}\ and\ \citenamefont
  {Jin}(2023)}]{PhysRevB.108.054302}%
  \BibitemOpen
  \bibfield  {author} {\bibinfo {author} {\bibfnamefont {L.}~\bibnamefont
  {Song}}\ and\ \bibinfo {author} {\bibfnamefont {J.}~\bibnamefont {Jin}},\
  }\bibfield  {title} {\bibinfo {title} {Crossover from discontinuous to
  continuous phase transition in a dissipative spin system with collective
  decay},\ }\href {https://doi.org/10.1103/PhysRevB.108.054302} {\bibfield
  {journal} {\bibinfo  {journal} {Phys. Rev. B}\ }\textbf {\bibinfo {volume}
  {108}},\ \bibinfo {pages} {054302} (\bibinfo {year} {2023})}\BibitemShut
  {NoStop}%
\bibitem [{\citenamefont {Macr\`{\i}}\ \emph {et~al.}(2020)\citenamefont
  {Macr\`{\i}}, \citenamefont {Nori}, \citenamefont {Savasta},\ and\
  \citenamefont {Zueco}}]{PhysRevA.101.053818}%
  \BibitemOpen
  \bibfield  {author} {\bibinfo {author} {\bibfnamefont {V.}~\bibnamefont
  {Macr\`{\i}}}, \bibinfo {author} {\bibfnamefont {F.}~\bibnamefont {Nori}},
  \bibinfo {author} {\bibfnamefont {S.}~\bibnamefont {Savasta}},\ and\ \bibinfo
  {author} {\bibfnamefont {D.}~\bibnamefont {Zueco}},\ }\bibfield  {title}
  {\bibinfo {title} {Spin squeezing by one-photon--two-atom excitation
  processes in atomic ensembles},\ }\href
  {https://doi.org/10.1103/PhysRevA.101.053818} {\bibfield  {journal} {\bibinfo
   {journal} {Phys. Rev. A}\ }\textbf {\bibinfo {volume} {101}},\ \bibinfo
  {pages} {053818} (\bibinfo {year} {2020})}\BibitemShut {NoStop}%
\bibitem [{\citenamefont {Gelhausen}\ \emph {et~al.}(2017)\citenamefont
  {Gelhausen}, \citenamefont {Buchhold},\ and\ \citenamefont
  {Strack}}]{PhysRevA.95.063824}%
  \BibitemOpen
  \bibfield  {author} {\bibinfo {author} {\bibfnamefont {J.}~\bibnamefont
  {Gelhausen}}, \bibinfo {author} {\bibfnamefont {M.}~\bibnamefont
  {Buchhold}},\ and\ \bibinfo {author} {\bibfnamefont {P.}~\bibnamefont
  {Strack}},\ }\bibfield  {title} {\bibinfo {title} {Many-body quantum optics
  with decaying atomic spin states: ($\ensuremath{\gamma},
  \ensuremath{\kappa}$) dicke model},\ }\href
  {https://doi.org/10.1103/PhysRevA.95.063824} {\bibfield  {journal} {\bibinfo
  {journal} {Phys. Rev. A}\ }\textbf {\bibinfo {volume} {95}},\ \bibinfo
  {pages} {063824} (\bibinfo {year} {2017})}\BibitemShut {NoStop}%
\bibitem [{\citenamefont {Yoshihara}\ \emph
  {et~al.}(2017{\natexlab{a}})\citenamefont {Yoshihara}, \citenamefont {Fuse},
  \citenamefont {Ashhab}, \citenamefont {Kakuyanagi}, \citenamefont {Saito},\
  and\ \citenamefont {Semba}}]{PhysRevA.95.053824}%
  \BibitemOpen
  \bibfield  {author} {\bibinfo {author} {\bibfnamefont {F.}~\bibnamefont
  {Yoshihara}}, \bibinfo {author} {\bibfnamefont {T.}~\bibnamefont {Fuse}},
  \bibinfo {author} {\bibfnamefont {S.}~\bibnamefont {Ashhab}}, \bibinfo
  {author} {\bibfnamefont {K.}~\bibnamefont {Kakuyanagi}}, \bibinfo {author}
  {\bibfnamefont {S.}~\bibnamefont {Saito}},\ and\ \bibinfo {author}
  {\bibfnamefont {K.}~\bibnamefont {Semba}},\ }\bibfield  {title} {\bibinfo
  {title} {Characteristic spectra of circuit quantum electrodynamics systems
  from the ultrastrong- to the deep-strong-coupling regime},\ }\href
  {https://doi.org/10.1103/PhysRevA.95.053824} {\bibfield  {journal} {\bibinfo
  {journal} {Phys. Rev. A}\ }\textbf {\bibinfo {volume} {95}},\ \bibinfo
  {pages} {053824} (\bibinfo {year} {2017}{\natexlab{a}})}\BibitemShut
  {NoStop}%
\bibitem [{\citenamefont {Yoshihara}\ \emph
  {et~al.}(2017{\natexlab{b}})\citenamefont {Yoshihara}, \citenamefont {Fuse},
  \citenamefont {Ashhab}, \citenamefont {Kakuyanagi}, \citenamefont {Saito},\
  and\ \citenamefont {Semba}}]{Yoshihara2017}%
  \BibitemOpen
  \bibfield  {author} {\bibinfo {author} {\bibfnamefont {F.}~\bibnamefont
  {Yoshihara}}, \bibinfo {author} {\bibfnamefont {T.}~\bibnamefont {Fuse}},
  \bibinfo {author} {\bibfnamefont {S.}~\bibnamefont {Ashhab}}, \bibinfo
  {author} {\bibfnamefont {K.}~\bibnamefont {Kakuyanagi}}, \bibinfo {author}
  {\bibfnamefont {S.}~\bibnamefont {Saito}},\ and\ \bibinfo {author}
  {\bibfnamefont {K.}~\bibnamefont {Semba}},\ }\bibfield  {title} {\bibinfo
  {title} {Superconducting qubit--oscillator circuit beyond the
  ultrastrong-coupling regime},\ }\href {https://doi.org/10.1038/nphys3906}
  {\bibfield  {journal} {\bibinfo  {journal} {Nature Physics}\ }\textbf
  {\bibinfo {volume} {13}},\ \bibinfo {pages} {44} (\bibinfo {year}
  {2017}{\natexlab{b}})}\BibitemShut {NoStop}%
\bibitem [{\citenamefont {Chen}\ \emph {et~al.}(2021)\citenamefont {Chen},
  \citenamefont {Qin}, \citenamefont {Stassi}, \citenamefont {Wang},\ and\
  \citenamefont {Nori}}]{PhysRevResearch.3.033275}%
  \BibitemOpen
  \bibfield  {author} {\bibinfo {author} {\bibfnamefont {Y.-H.}\ \bibnamefont
  {Chen}}, \bibinfo {author} {\bibfnamefont {W.}~\bibnamefont {Qin}}, \bibinfo
  {author} {\bibfnamefont {R.}~\bibnamefont {Stassi}}, \bibinfo {author}
  {\bibfnamefont {X.}~\bibnamefont {Wang}},\ and\ \bibinfo {author}
  {\bibfnamefont {F.}~\bibnamefont {Nori}},\ }\bibfield  {title} {\bibinfo
  {title} {Fast binomial-code holonomic quantum computation with ultrastrong
  light-matter coupling},\ }\href
  {https://doi.org/10.1103/PhysRevResearch.3.033275} {\bibfield  {journal}
  {\bibinfo  {journal} {Phys. Rev. Res.}\ }\textbf {\bibinfo {volume} {3}},\
  \bibinfo {pages} {033275} (\bibinfo {year} {2021})}\BibitemShut {NoStop}%
\bibitem [{\citenamefont {Niemczyk}\ \emph {et~al.}(2010)\citenamefont
  {Niemczyk}, \citenamefont {Deppe}, \citenamefont {Huebl}, \citenamefont
  {Menzel}, \citenamefont {Hocke}, \citenamefont {Schwarz}, \citenamefont
  {Garcia-Ripoll}, \citenamefont {Zueco}, \citenamefont {H{\"u}mmer},
  \citenamefont {Solano}, \citenamefont {Marx},\ and\ \citenamefont
  {Gross}}]{Niemczyk2010}%
  \BibitemOpen
  \bibfield  {author} {\bibinfo {author} {\bibfnamefont {T.}~\bibnamefont
  {Niemczyk}}, \bibinfo {author} {\bibfnamefont {F.}~\bibnamefont {Deppe}},
  \bibinfo {author} {\bibfnamefont {H.}~\bibnamefont {Huebl}}, \bibinfo
  {author} {\bibfnamefont {E.~P.}\ \bibnamefont {Menzel}}, \bibinfo {author}
  {\bibfnamefont {F.}~\bibnamefont {Hocke}}, \bibinfo {author} {\bibfnamefont
  {M.~J.}\ \bibnamefont {Schwarz}}, \bibinfo {author} {\bibfnamefont {J.~J.}\
  \bibnamefont {Garcia-Ripoll}}, \bibinfo {author} {\bibfnamefont
  {D.}~\bibnamefont {Zueco}}, \bibinfo {author} {\bibfnamefont
  {T.}~\bibnamefont {H{\"u}mmer}}, \bibinfo {author} {\bibfnamefont
  {E.}~\bibnamefont {Solano}}, \bibinfo {author} {\bibfnamefont
  {A.}~\bibnamefont {Marx}},\ and\ \bibinfo {author} {\bibfnamefont
  {R.}~\bibnamefont {Gross}},\ }\bibfield  {title} {\bibinfo {title} {Circuit
  quantum electrodynamics in the ultrastrong-coupling regime},\ }\href
  {https://doi.org/10.1038/nphys1730} {\bibfield  {journal} {\bibinfo
  {journal} {Nature Physics}\ }\textbf {\bibinfo {volume} {6}},\ \bibinfo
  {pages} {772} (\bibinfo {year} {2010})}\BibitemShut {NoStop}%
\bibitem [{\citenamefont {Romero}\ \emph {et~al.}(2012)\citenamefont {Romero},
  \citenamefont {Ballester}, \citenamefont {Wang}, \citenamefont {Scarani},\
  and\ \citenamefont {Solano}}]{PhysRevLett.108.120501}%
  \BibitemOpen
  \bibfield  {author} {\bibinfo {author} {\bibfnamefont {G.}~\bibnamefont
  {Romero}}, \bibinfo {author} {\bibfnamefont {D.}~\bibnamefont {Ballester}},
  \bibinfo {author} {\bibfnamefont {Y.~M.}\ \bibnamefont {Wang}}, \bibinfo
  {author} {\bibfnamefont {V.}~\bibnamefont {Scarani}},\ and\ \bibinfo {author}
  {\bibfnamefont {E.}~\bibnamefont {Solano}},\ }\bibfield  {title} {\bibinfo
  {title} {Ultrafast quantum gates in circuit qed},\ }\href
  {https://doi.org/10.1103/PhysRevLett.108.120501} {\bibfield  {journal}
  {\bibinfo  {journal} {Phys. Rev. Lett.}\ }\textbf {\bibinfo {volume} {108}},\
  \bibinfo {pages} {120501} (\bibinfo {year} {2012})}\BibitemShut {NoStop}%
\bibitem [{\citenamefont {Forn-D\'{\i}az}\ \emph {et~al.}(2019)\citenamefont
  {Forn-D\'{\i}az}, \citenamefont {Lamata}, \citenamefont {Rico}, \citenamefont
  {Kono},\ and\ \citenamefont {Solano}}]{RevModPhys.91.025005}%
  \BibitemOpen
  \bibfield  {author} {\bibinfo {author} {\bibfnamefont {P.}~\bibnamefont
  {Forn-D\'{\i}az}}, \bibinfo {author} {\bibfnamefont {L.}~\bibnamefont
  {Lamata}}, \bibinfo {author} {\bibfnamefont {E.}~\bibnamefont {Rico}},
  \bibinfo {author} {\bibfnamefont {J.}~\bibnamefont {Kono}},\ and\ \bibinfo
  {author} {\bibfnamefont {E.}~\bibnamefont {Solano}},\ }\bibfield  {title}
  {\bibinfo {title} {Ultrastrong coupling regimes of light-matter
  interaction},\ }\href {https://doi.org/10.1103/RevModPhys.91.025005}
  {\bibfield  {journal} {\bibinfo  {journal} {Rev. Mod. Phys.}\ }\textbf
  {\bibinfo {volume} {91}},\ \bibinfo {pages} {025005} (\bibinfo {year}
  {2019})}\BibitemShut {NoStop}%
\bibitem [{\citenamefont {Stern}\ \emph {et~al.}(2014)\citenamefont {Stern},
  \citenamefont {Catelani}, \citenamefont {Kubo}, \citenamefont {Grezes},
  \citenamefont {Bienfait}, \citenamefont {Vion}, \citenamefont {Esteve},\ and\
  \citenamefont {Bertet}}]{PhysRevLett.113.123601}%
  \BibitemOpen
  \bibfield  {author} {\bibinfo {author} {\bibfnamefont {M.}~\bibnamefont
  {Stern}}, \bibinfo {author} {\bibfnamefont {G.}~\bibnamefont {Catelani}},
  \bibinfo {author} {\bibfnamefont {Y.}~\bibnamefont {Kubo}}, \bibinfo {author}
  {\bibfnamefont {C.}~\bibnamefont {Grezes}}, \bibinfo {author} {\bibfnamefont
  {A.}~\bibnamefont {Bienfait}}, \bibinfo {author} {\bibfnamefont
  {D.}~\bibnamefont {Vion}}, \bibinfo {author} {\bibfnamefont {D.}~\bibnamefont
  {Esteve}},\ and\ \bibinfo {author} {\bibfnamefont {P.}~\bibnamefont
  {Bertet}},\ }\bibfield  {title} {\bibinfo {title} {Flux qubits with long
  coherence times for hybrid quantum circuits},\ }\href
  {https://doi.org/10.1103/PhysRevLett.113.123601} {\bibfield  {journal}
  {\bibinfo  {journal} {Phys. Rev. Lett.}\ }\textbf {\bibinfo {volume} {113}},\
  \bibinfo {pages} {123601} (\bibinfo {year} {2014})}\BibitemShut {NoStop}%
\bibitem [{\citenamefont {Orgiazzi}\ \emph {et~al.}(2016)\citenamefont
  {Orgiazzi}, \citenamefont {Deng}, \citenamefont {Layden}, \citenamefont
  {Marchildon}, \citenamefont {Kitapli}, \citenamefont {Shen}, \citenamefont
  {Bal}, \citenamefont {Ong},\ and\ \citenamefont
  {Lupascu}}]{PhysRevB.93.104518}%
  \BibitemOpen
  \bibfield  {author} {\bibinfo {author} {\bibfnamefont {J.-L.}\ \bibnamefont
  {Orgiazzi}}, \bibinfo {author} {\bibfnamefont {C.}~\bibnamefont {Deng}},
  \bibinfo {author} {\bibfnamefont {D.}~\bibnamefont {Layden}}, \bibinfo
  {author} {\bibfnamefont {R.}~\bibnamefont {Marchildon}}, \bibinfo {author}
  {\bibfnamefont {F.}~\bibnamefont {Kitapli}}, \bibinfo {author} {\bibfnamefont
  {F.}~\bibnamefont {Shen}}, \bibinfo {author} {\bibfnamefont {M.}~\bibnamefont
  {Bal}}, \bibinfo {author} {\bibfnamefont {F.~R.}\ \bibnamefont {Ong}},\ and\
  \bibinfo {author} {\bibfnamefont {A.}~\bibnamefont {Lupascu}},\ }\bibfield
  {title} {\bibinfo {title} {Flux qubits in a planar circuit quantum
  electrodynamics architecture: Quantum control and decoherence},\ }\href
  {https://doi.org/10.1103/PhysRevB.93.104518} {\bibfield  {journal} {\bibinfo
  {journal} {Phys. Rev. B}\ }\textbf {\bibinfo {volume} {93}},\ \bibinfo
  {pages} {104518} (\bibinfo {year} {2016})}\BibitemShut {NoStop}%
\bibitem [{\citenamefont {Paik}\ \emph {et~al.}(2011)\citenamefont {Paik},
  \citenamefont {Schuster}, \citenamefont {Bishop}, \citenamefont {Kirchmair},
  \citenamefont {Catelani}, \citenamefont {Sears}, \citenamefont {Johnson},
  \citenamefont {Reagor}, \citenamefont {Frunzio}, \citenamefont {Glazman},
  \citenamefont {Girvin}, \citenamefont {Devoret},\ and\ \citenamefont
  {Schoelkopf}}]{PhysRevLett.107.240501}%
  \BibitemOpen
  \bibfield  {author} {\bibinfo {author} {\bibfnamefont {H.}~\bibnamefont
  {Paik}}, \bibinfo {author} {\bibfnamefont {D.~I.}\ \bibnamefont {Schuster}},
  \bibinfo {author} {\bibfnamefont {L.~S.}\ \bibnamefont {Bishop}}, \bibinfo
  {author} {\bibfnamefont {G.}~\bibnamefont {Kirchmair}}, \bibinfo {author}
  {\bibfnamefont {G.}~\bibnamefont {Catelani}}, \bibinfo {author}
  {\bibfnamefont {A.~P.}\ \bibnamefont {Sears}}, \bibinfo {author}
  {\bibfnamefont {B.~R.}\ \bibnamefont {Johnson}}, \bibinfo {author}
  {\bibfnamefont {M.~J.}\ \bibnamefont {Reagor}}, \bibinfo {author}
  {\bibfnamefont {L.}~\bibnamefont {Frunzio}}, \bibinfo {author} {\bibfnamefont
  {L.~I.}\ \bibnamefont {Glazman}}, \bibinfo {author} {\bibfnamefont {S.~M.}\
  \bibnamefont {Girvin}}, \bibinfo {author} {\bibfnamefont {M.~H.}\
  \bibnamefont {Devoret}},\ and\ \bibinfo {author} {\bibfnamefont {R.~J.}\
  \bibnamefont {Schoelkopf}},\ }\bibfield  {title} {\bibinfo {title}
  {Observation of high coherence in josephson junction qubits measured in a
  three-dimensional circuit qed architecture},\ }\href
  {https://doi.org/10.1103/PhysRevLett.107.240501} {\bibfield  {journal}
  {\bibinfo  {journal} {Phys. Rev. Lett.}\ }\textbf {\bibinfo {volume} {107}},\
  \bibinfo {pages} {240501} (\bibinfo {year} {2011})}\BibitemShut {NoStop}%
\bibitem [{\citenamefont {Megrant}\ \emph {et~al.}(2012)\citenamefont
  {Megrant}, \citenamefont {Neill}, \citenamefont {Barends}, \citenamefont
  {Chiaro}, \citenamefont {Chen}, \citenamefont {Feigl}, \citenamefont {Kelly},
  \citenamefont {Lucero}, \citenamefont {Mariantoni}, \citenamefont
  {O’Malley}, \citenamefont {Sank}, \citenamefont {Vainsencher},
  \citenamefont {Wenner}, \citenamefont {White}, \citenamefont {Yin},
  \citenamefont {Zhao}, \citenamefont {Palmstrøm}, \citenamefont {Martinis},\
  and\ \citenamefont {Cleland}}]{10.1063/1.3693409}%
  \BibitemOpen
  \bibfield  {author} {\bibinfo {author} {\bibfnamefont {A.}~\bibnamefont
  {Megrant}}, \bibinfo {author} {\bibfnamefont {C.}~\bibnamefont {Neill}},
  \bibinfo {author} {\bibfnamefont {R.}~\bibnamefont {Barends}}, \bibinfo
  {author} {\bibfnamefont {B.}~\bibnamefont {Chiaro}}, \bibinfo {author}
  {\bibfnamefont {Y.}~\bibnamefont {Chen}}, \bibinfo {author} {\bibfnamefont
  {L.}~\bibnamefont {Feigl}}, \bibinfo {author} {\bibfnamefont
  {J.}~\bibnamefont {Kelly}}, \bibinfo {author} {\bibfnamefont
  {E.}~\bibnamefont {Lucero}}, \bibinfo {author} {\bibfnamefont
  {M.}~\bibnamefont {Mariantoni}}, \bibinfo {author} {\bibfnamefont {P.~J.~J.}\
  \bibnamefont {O’Malley}}, \bibinfo {author} {\bibfnamefont
  {D.}~\bibnamefont {Sank}}, \bibinfo {author} {\bibfnamefont {A.}~\bibnamefont
  {Vainsencher}}, \bibinfo {author} {\bibfnamefont {J.}~\bibnamefont {Wenner}},
  \bibinfo {author} {\bibfnamefont {T.~C.}\ \bibnamefont {White}}, \bibinfo
  {author} {\bibfnamefont {Y.}~\bibnamefont {Yin}}, \bibinfo {author}
  {\bibfnamefont {J.}~\bibnamefont {Zhao}}, \bibinfo {author} {\bibfnamefont
  {C.~J.}\ \bibnamefont {Palmstrøm}}, \bibinfo {author} {\bibfnamefont
  {J.~M.}\ \bibnamefont {Martinis}},\ and\ \bibinfo {author} {\bibfnamefont
  {A.~N.}\ \bibnamefont {Cleland}},\ }\bibfield  {title} {\bibinfo {title}
  {{Planar superconducting resonators with internal quality factors above one
  million}},\ }\href {https://doi.org/10.1063/1.3693409} {\bibfield  {journal}
  {\bibinfo  {journal} {Applied Physics Letters}\ }\textbf {\bibinfo {volume}
  {100}},\ \bibinfo {pages} {113510} (\bibinfo {year} {2012})}\BibitemShut
  {NoStop}%
\bibitem [{\citenamefont {Qiu}\ \emph {et~al.}(2014)\citenamefont {Qiu},
  \citenamefont {Xiong}, \citenamefont {Tian},\ and\ \citenamefont
  {You}}]{PhysRevA.89.042321}%
  \BibitemOpen
  \bibfield  {author} {\bibinfo {author} {\bibfnamefont {Y.}~\bibnamefont
  {Qiu}}, \bibinfo {author} {\bibfnamefont {W.}~\bibnamefont {Xiong}}, \bibinfo
  {author} {\bibfnamefont {L.}~\bibnamefont {Tian}},\ and\ \bibinfo {author}
  {\bibfnamefont {J.~Q.}\ \bibnamefont {You}},\ }\bibfield  {title} {\bibinfo
  {title} {Coupling spin ensembles via superconducting flux qubits},\ }\href
  {https://doi.org/10.1103/PhysRevA.89.042321} {\bibfield  {journal} {\bibinfo
  {journal} {Phys. Rev. A}\ }\textbf {\bibinfo {volume} {89}},\ \bibinfo
  {pages} {042321} (\bibinfo {year} {2014})}\BibitemShut {NoStop}%
\end{thebibliography}%
\vspace{8pt}
\end{document}